\PassOptionsToPackage{pdfpagelabels=false}{hyperref}
\PassOptionsToPackage{nonamebreak}{natbib}   
\documentclass[fleqn,usenatbib]{mnras}
\usepackage{newtxtext,newtxmath}

\usepackage[T1]{fontenc}
\usepackage{ae,aecompl}
\pdfoutput=1

\usepackage{graphicx}	
\usepackage{amsmath}	
\usepackage{amssymb}	
\usepackage{subfigure}

\newcommand{\epsurl}{\href{https://github.com/ylu2010/mergertree}{\url{https://github.com/ylu2010/mergertree}}}
\defcitealias{Grudic_2016}{G18}

\newcommand\altaffilmark[1]{$^{#1}$}
\newcommand\altaffiltext[1]{$^{#1}$}

\title[GC formation and hierarchical assembly]{The formation and hierarchical assembly of globular cluster populations}

\author[El-Badry et al.]{
\parbox[t]{\textwidth}{ 
Kareem El-Badry\thanks{E-mail: kelbadry@berkeley.edu}\altaffilmark{1,2},
Eliot Quataert\altaffilmark{1}, 
Daniel R. Weisz\altaffilmark{1},
Nick Choksi\altaffilmark{1},
and Michael Boylan-Kolchin\altaffilmark{3}
} 
\vspace*{6pt} \\
\altaffiltext{1}{Department of Astronomy and Theoretical Astrophysics Center, University of California Berkeley, Berkeley, CA 94720} \\
\altaffiltext{2}{Max Planck Institute for Astronomy, D-69117 Heidelberg, Germany} \\
\altaffiltext{3}{Department of Astronomy, The University of Texas at Austin, Austin, TX 78712} \\
}

\date{Accepted to MNRAS}
\pubyear{2018}

\begin{document}
\label{firstpage}
\pagerange{\pageref{firstpage}--\pageref{lastpage}}
\maketitle

\begin{abstract}
We use a semi-analytic model for globular cluster (GC) formation built on dark matter merger trees to explore the relative role of formation physics and hierarchical assembly in determining the properties of GC populations.
Many previous works have argued that the observed linear relation between total GC mass and halo mass points to a fundamental GC -- dark matter connection or indicates that GCs formed at very high redshift before feedback processes introduced nonlinearity in the baryon-to-dark matter mass relation.
We demonstrate that at $M_{\rm vir}(z=0) \gtrsim 10^{11.5} M_{\odot}$, a constant ratio between halo mass and total GC mass is in fact an almost inevitable consequence of hierarchical assembly: by the central limit theorem, it is expected at $z=0$ independent of the GC-to-halo mass relation at the time of GC formation. 
The GC-to-halo mass relation at $M_{\rm vir}(z=0) < 10^{11.5} M_{\odot}$ is more sensitive to the details of the GC formation process.
In our fiducial model, GC formation occurs in galaxies when the gas surface density exceeds a critical value. This model naturally predicts bimodal GC color distributions similar to those observed in nearby galaxies and reproduces the observed relation between GC system metallicity and halo mass. It predicts that the cosmic GC formation rate peaked at $z$ $\sim$ 4, too late for GCs to contribute significantly to the UV luminosity density during reionization. 
\end{abstract}




\section{Introduction}
Globular clusters (GCs) are relics of star formation under extreme conditions in the early Universe. Although it may soon become feasible to observe young GCs at high redshift as they form \citep{Carlberg_2002, Katz_2013, BoylanKolchin_2017b, Vanzella_2017, Renzini_2017, Zick_2018}, at present, most of what we know about GCs comes from observations of the old GC populations of nearby galaxies. 

Studies of GCs in the local Universe have highlighted striking differences between galaxies' GC and field star populations. While the galaxy stellar-to-halo mass relation is strongly nonlinear, the total mass of globular clusters within a dark matter halo is a constant fraction of halo mass over almost 5 decades in mass \citep[e.g.][]{Blakeslee_1997, Harris_2013, Hudson_2014, Durrell_2014, Harris_2015, Harris_2017}. The GC populations of most individual halos exhibit bimodality in color and/or metallicity \citep{Zepf_1993, Harris_2006, Peng_2006, Brodie_2012}, in contrast to the stars in the central galaxy or the stellar halo. And although GCs were once thought to be simple, uniform stellar populations formed in a single burst, detailed observations reveal evidence of multiple stellar populations and anomalous abundance patterns that remain poorly understood \citep[e.g.][]{Piotto_2015, Bastian_2017}.

The observed constant GC-to-halo mass ratio and the old ages measured for Milky Way (MW) GCs have led many authors to suggest that most GCs, particularly those that are blue and metal-poor, formed at very early times, before feedback processes introduced nonlinearity in the baryon-to-dark matter mass relation \citep{Blakeslee_1997, Kavelaars_1999, Diemand_2005, Moore_2006, Bekki_2008, Spitler_2008, Spitler_2009, Hudson_2014, Moran_2014, Harris_2015, Katz_2014, Trenti_2015, BoylanKolchin_2017}. Such a formation scenario most directly implies a constant relation between GC mass and halo mass {\it at the time of GC formation}, but \citet{BoylanKolchin_2017} showed that if a constant GC-to-halo mass ratio was set at high redshift, it would be preserved to $z=0$ during hierarchical assembly.

Star formation is a local process. If GC formation did occur proportional to dark matter halo mass at high redshift, a successful GC formation theory must attempt to tie the mass of a dark matter halo at high redshift to local gas conditions conducive to GC formation. Doing so is challenging both because the properties of a DM halo do not uniquely determine the baryonic conditions in its central galaxy, even at high redshift \citep[e.g.][]{Wise_2012, OShea_2015}, and because the local gas conditions required for the formation of massive bound clusters remain imperfectly understood \citep[e.g][]{McKee_2007, Krumholz_2014, Skinner_2015, Grudic_2016, Tsang_2017}.

Some numerical studies have begun to resolve aspects of the GC formation process in a cosmological context \citep{Kravtsov_2005, Boley_2009, Trenti_2015, Kimm_2016, Kim_2016, Mandelker_2017}. Because GCs are much smaller than the scales typically resolved in cosmological zoom-in simulations, such works face strong trade-offs between resolution, simulation volume, and final redshift. Simulations reaching the sub-parsec scale resolution required to study details of the GC formation process have therefore to date focused on small volumes and have been terminated at high redshift, making comparison with observations difficult.

A complementary approach, which we take in this work, is to adopt simple prescriptions to predict the GC formation rate and/or the dynamical evolution of GCs in a halo as a function of galaxy-scale gas conditions or the properties of the dark matter halo. This approach, which has been fruitfully employed in a number of previous studies \citep{Ashman_1992, Cote_1998, Beasley_2002, Prieto_2008, Muratov_2010, Tonini_2013, Katz_2014, Li_2014, Kruijssen_2015, Choksi_2018, Pfeffer_2018}, makes it possible to efficiently predict the observable GC populations of galaxies at $z=0$ for a wide range of GC formation models. Such ``semi-analytic'' models cannot predict the internal properties of GCs with high fidelity and are not guaranteed to capture all the physical processes relevant to GC formation. However, their simplicity aids their interpretability: because such models have only a few free parameters, they make it straightforward to gauge the sensitivity of observables to different aspects of the GC formation model.

This work models GC formation as the product of ``normal'' star formation in the high-density disks of gas-rich galaxies. Motivated by simulations of molecular cloud collapse, we use the ansatz that massive bound clusters form preferentially when the gas surface density exceeds a critical threshold. We apply this ansatz to a semi-analytic gas model built on dark matter merger trees in order to predict the globular cluster populations of halos at $z=0$. We then explore how varying different aspects of the GC formation prescription changes the epoch at which GCs form, the GC-to-halo mass relation, and the $z=0$ color distributions of individual galaxies' GCs.  In contrast to some previous works \citep[e.g.][]{Beasley_2002, Tonini_2013, Amorisco_2018}, our model does not explicitly assume separate formation modes for blue and red GCs; we simply predict the metallicity and color of each GC based on the conditions in the galaxy in which it formed. Because the model is built on dark matter merger trees, it does not allow for straightforward predictions of GC formation divorced from dark matter \citep[e.g.][]{Zhao_2005, vanDokkum_2018}. However, we emphasize that GC formation in our model is most directly tied to baryonic conditions: the merger trees serve primarily to keep track of the gas conditions throughout the formation history of a GC population.

In agreement with previous work, we find that semi-analytic GC formation models can broadly reproduce many aspects of the observed GC population. However, one of our main results is that some observed GC scaling relations, particularly the constant GC-to-halo mass ratio, are primarily consequences of hierarchical assembly and are thus insensitive to details of the GC formation process. 

The rest of this paper is organized as follows. In Section~\ref{sec:methods}, we describe the assumptions and implementation of our semi-analytic model. We present the $z=0$ globular cluster populations predicted by our model and explore the model's sensitivity to several free parameters in Section~\ref{sec:results}. We summarize and discuss our findings in Section~\ref{sec:discussion}. We provide additional details of the underlying model in the appendices.

\section{Model}
\label{sec:methods}

\begin{figure*}
\includegraphics[width=\textwidth]{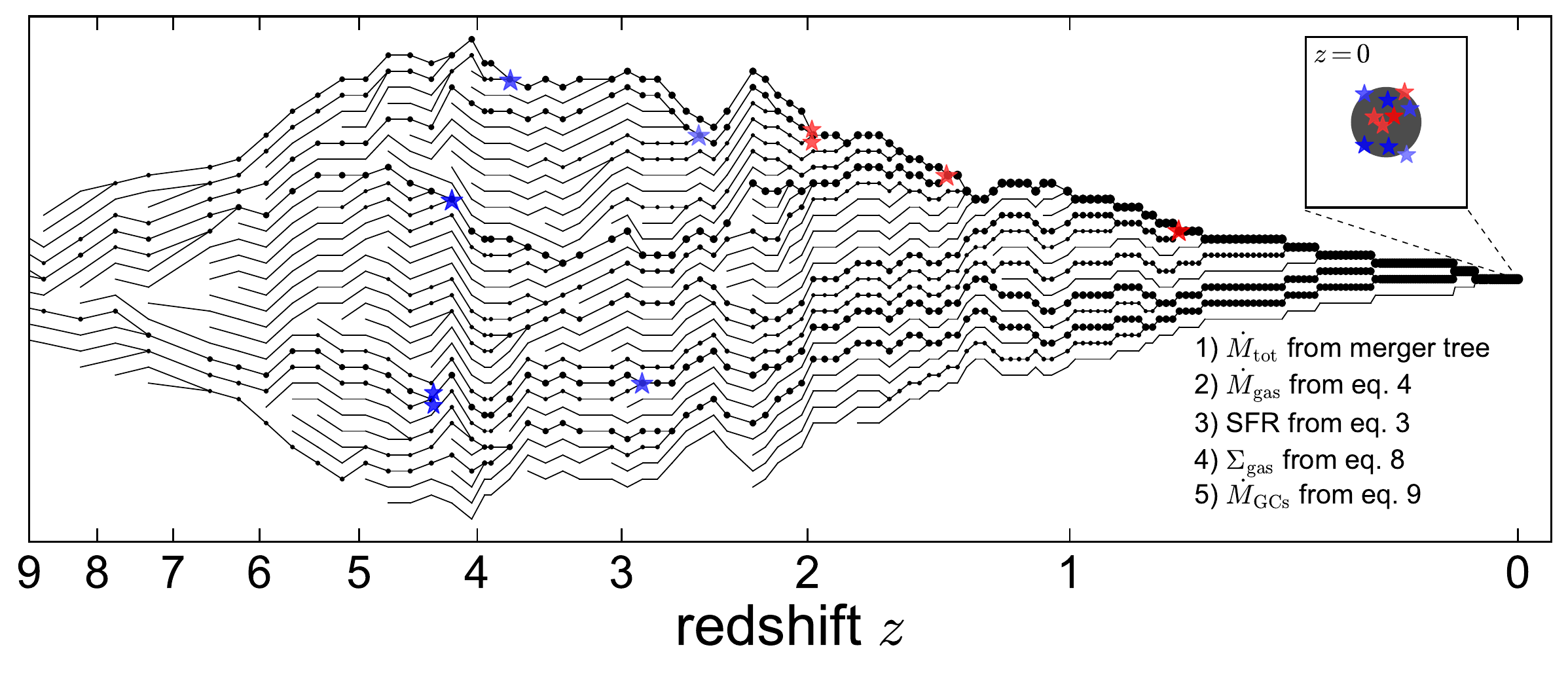}
\caption{Schematic illustration of our model. Colored stars represent GC formation events. At each node in the merger tree, we estimate the gas accretion rate from the total accretion rate (Equation~\ref{eq:mdot_gas_in}), the SFR from the gas accretion rate (Equation~\ref{eq:sfr}), and the gas surface density from the SFR (Equation~\ref{eq:KS_scaling}). Star formation in massive bound clusters occurs when the gas surface density exceeds a critical value (Equation~\ref{eq:Gamma}). These clusters are propagated through the merger tree to $z=0$, where they represent the observable GC population. We assume GCs form with the same metallicity as the gas in the galaxy in which they form, which is estimated from an analytic mass-metallicity relation (Equation~\ref{eq:mass_met}); as a result, most early-forming GCs have low metallicity (blue), while later-forming GCs have higher metallicity (red). }
\label{fig:schematic}
\end{figure*}

The basic idea of our model is that massive bound clusters, including the progenitors of GCs, form primarily when the gas surface density exceeds a critical value. We motivate this ansatz in Section~\ref{sec:motivation} and discuss the model's implementation in Section~\ref{sec:implementation}.

\subsection{GC formation at high surface density}
\label{sec:motivation}

Massive bound clusters like the progenitors of GCs are not a typical outcome of star formation under normal ISM conditions at low redshift. Although a large fraction of stars in nearby galaxies form in clusters, most clusters become gravitationally unbound and disrupted within a few dynamical times \citep[for a review, see][]{Lada_2003}. This cluster ``infant mortality'' owes to the fact that the star formation efficiency of giant molecular clouds (GMCs) in galaxies with MW-like gas surface densities is low ($\sim$\,1\%), so initially bound clusters become unbound when stellar feedback expels most of a cluster's gas mass and shallows the gravitational potential \citep{Tutukov_1978, Geyer_2001, Bastian_2006, Baumgardt_2007}. Clusters are more likely to remain bound if the star formation efficiency is high. 

Massive young star clusters \textit{are} observed in nearby galaxies with higher gas densities than the MW \citep[e.g.][]{Zwart_2010}, and the fraction of stars formed in long-lived clusters is observed to be higher in high-density environments \citep{Larsen_2000, Keto_2005, Goddard_2010, Johnson_2016}. Theoretical star formation models suggest that environments of high density and pressure are conducive to the formation of proto-GC-like clusters \citep{Elmegreen_1997, Murray_2010, Kruijssen_2012a, Kruijssen_2015}, because (a) the free-fall time becomes shorter than the few-Myr massive stellar evolution timescale, meaning that a large fraction of a gas cloud can turn into stars before the first supernovae explode \citep[and references therein]{Elmegreen_2017}, and (b) the self-gravity of a cloud increases more steeply with density than the energy injected by stellar feedback \citep{Murray_2010, Thompson_2016}, so that at sufficiently high densities, the feedback energy budget is insufficient to prevent runaway star formation.

\citet[][hereafter G18]{Grudic_2016} recently argued that the formation of massive bound star clusters depends most directly on high gas {\it surface density}, as opposed to volume density, escape velocity, pressure, or  other GMC properties. Using idealized cloud-collapse simulations of individual GMCs, they studied how the star formation efficiency, $\epsilon$ (i.e, the fraction of gas in a collapsing cloud that is converted to stars), and the cluster formation efficiency, $\Gamma$ (the fraction of stars formed in bound clusters), scale with the structural parameters of a GMC. \citetalias{Grudic_2016} found that at fixed cloud geometry, $\epsilon$ and $\Gamma$ are primarily functions of the gas surface density, $\Sigma_{\rm GMC}$, independent of the cloud mass and size.\footnote{\citetalias{Grudic_2016} did not identify any unique relation between $\epsilon$ and other integrated cloud properties, suggesting that $\epsilon$ depends most directly on gas surface density. This is expected if the star formation efficiency is set by the balance of the force self-gravity, which scales as $F_{\rm gravity}\sim M^2/R^2$, where $M$ and $R$ are the mass and radius of a GMC, and that of stellar feedback, which scales as the stellar mass formed: $F_{\rm feedback}\sim M_{\rm star} \sim M$. For any fixed cloud geometry, the ratio of these quantities scales as $\Sigma_{\rm GMC}\sim M/R^2$.} In particular, \citetalias{Grudic_2016} found $\epsilon$ to plateau, at a maximum value of order unity, for $\Sigma_{\rm GMC} \gg \Sigma_{\rm crit}$, and to fall off as $\epsilon \sim \Sigma_{\rm GMC}^{-\alpha}$, where $\alpha\sim 1$, at $\Sigma_{\rm GMC} \ll \Sigma_{\rm crit}$. A qualitatively similar scaling with $\Sigma_{\rm GMC}$ was found for $\Gamma$ (Grudic et al., private communication). \citetalias{Grudic_2016} found $\Sigma_{\rm crit}\approx 3000\,M_{\odot}\,{\rm pc^{-2}}$; several other works have predicted critical densities for GC formation in the range $\Sigma_{\rm crit} = 10^{3-4} M_{\odot}\,{\rm pc^{-2}}$ \citep{Beasley_2002, Elmegreen_2008, Fall_2010, Kruijssen_2012a, Kim_2016, Raskutti_2016, Li_2017}.

\subsection{Fiducial Model Implementation}
\label{sec:implementation}

To predict the GC population of a dark matter halo at $z=0$, we estimate the GC formation rate, based on estimates of the star formation rate (SFR) and gas surface density, throughout its assembly history. We propagate GCs formed at each point in the merger tree to $z=0$, assuming that at each merger, the descendant halo inherits the GC populations of both its progenitors. The model is illustrated schematically in Figure~\ref{fig:schematic}.

\subsubsection{Merger Trees}
\label{sec:merger_trees}
We generate merger trees based on extended Press Schechter theory \citep{Bond_1991}, using the Monte Carlo algorithm described in \citet{Parkinson_2008}.\footnote{We use an implementation of the algorithm provided by Yu Lu; it is available at \epsurl.}  Merger trees generated with this method have been shown to reproduce the statistical properties and mass accretion histories of merger trees extracted from N-body simulations with high fidelity \citep{Jiang_2014}. Using cosmological parameters from \citet{Plank_2016}, we generate merger trees for halos with $z=0$ masses $M_{\rm vir} = 10^{10-14}M_{\odot}$, with 0.1 dex spacing in $z=0$ mass. Here $M_{\rm vir}$ is the halo mass within the evolving virial overdensity from \citet{Bryan_1998}. We use a mass resolution of $m_{\rm res} = 10^{7} M_{\odot}$ for $M_{\rm vir} \leq 10^{13} M_{\odot}$ and $m_{\rm res} = 10^{8} M_{\odot}$ for $M_{\rm vir} > 10^{13} M_{\odot}$. 

\subsubsection{Populating merger trees with GCs}
We express the GC formation rate as 
\begin{align}
\label{eq:mdot_gcs}
\dot{M}_{{\rm GCs}}=\Gamma_{\rm GCs}\times{\rm SFR},
\end{align}
where $\Gamma_{\rm GCs}$ is the fraction of stars forming in bound clusters that are sufficiently massive to survive until $z=0$ (corresponding roughly to cluster birth masses $\gtrsim 10^{5} M_{\odot}$;  see \citealt{Muratov_2010}, and references therein). Motivated by the results of cloud-collapse simulations, $\Gamma_{\rm GCs}$ depends only on the mean surface density of GMCs. We estimate SFR and $\Sigma_{\rm GMC}$ throughout a merger tree as follows. 

We assume an equilibrium model wherein the gas content of a galaxy is set by a balance between cosmological inflow and stellar feedback-driven outflow \citep[e.g.][]{Dave_2012, Lilly_2013, Rodriguez_2016b}. In such models, the mass of a galaxy's cold gas reservoir is determined by 
\begin{align}
\label{eq:inflow}
\dot{M}_{{\rm gas,\,in}}=\dot{M}_{{\rm gas,\,out}}+ {\rm SFR},
\end{align}
where $\dot{M}_{{\rm gas,\,in}}$ and  $\dot{M}_{{\rm gas,\,out}}$ are inflow and outflow rates of cold gas. 
If stellar feedback expels gas from the galaxy in proportion to the mass of stars formed with mass loading factor $\eta=\dot{M}_{{\rm gas,\,out}}/{\rm SFR}$, the star formation rate can be written as
\begin{align}
\label{eq:sfr}
{\rm SFR}=\dot{M}_{{\rm gas,\,in}}/\left(1+\eta\right).
\end{align}
The SFR at a given point in the merger tree thus depends on the cold gas accretion rate, $\dot{M}_{{\rm gas,\,in}}$, and the mass-loading factor, $\eta$. 

At sufficiently high redshift, we expect $\dot{M}_{{\rm gas,\,in}} \approx f_{\rm b}\dot{M}_{{\rm tot,\,in}}$, where $f_{\rm b} = 0.165$ is the cosmic baryon fraction, and $\dot{M}_{{\rm tot,\,in}}$ is the total (dark matter plus baryon) accretion rate \citep[e.g.][]{Dekel_2009}. At later times, the fraction of all baryons that are in cold gas drops due to a combination of star formation and heating by the UV background, virial shocks, and stellar and AGN feedback. Following \citet{Dave_2012}, we approximate this suppression in the cold gas accretion rate as 
\begin{align}
\label{eq:mdot_gas_in}
\dot{M}_{{\rm gas,\,in}}=f_{{\rm b}}\dot{M}_{{\rm tot,\,in}}\times\zeta,
\end{align}
where $\zeta \leq 1$ is a function that represents the mass fraction of accreted material that is in cold gas relative to the cosmic baryon fraction. In practice, $\zeta$ is calculated as the product of several terms representing heating due to various sources; it varies with redshift and with the masses of the primary and accreted halo. Details on our adopted form of $\zeta$, which is largely phenomenological, can be found in Appendix~\ref{sec:prev_fdbk}. The basic effect of $\zeta$ is that at high $z$, $\dot{M}_{{\rm gas,\,in}}\approx f_{{\rm b}}\dot{M}_{{\rm tot,\,in}}$ over a wide range of masses, but at late times, the fraction of baryons in cold gas is suppressed at all masses, especially outside the range $10.5 \lesssim \log(M_{\rm vir}/M_{\odot}) \lesssim 12$.

We calculate $\dot{M}_{\rm tot,\,in}$ directly from the dark matter merger trees. If a merger occurs in a given timestep (i.e., if a halo has more than one progenitor), we calculate $\dot{M}_{\rm tot,\,in}$ as the mass of the accreted satellite divided by the merger timescale, which we describe in Section~\ref{sec:timescale}. We do not trace the accretion of halos below the resolution limit (``smooth'' accretion).

The SFR also depends on the mass loading factor, $\eta$. This parameter is both observationally and theoretically poorly constrained (see \citealt{Schroetter_2015}, and references therein). A generic prediction of many theoretical studies is that $\eta$ is higher in lower-mass halos, since less energy is required to eject gas from a shallow potential well than from a deep one. We parameterize $\eta$ as 
\begin{align}
\label{eq:eta}
\eta=\alpha_{\eta}\left(\frac{M_{{\rm vir}}}{10^{12}M_{\odot}}\right)^{-\beta_{\eta}},
\end{align}
where $\alpha_{\eta}$ and $\beta_{\eta}$ are free parameters. As a fiducial estimate, we choose $\beta_{\eta} = 1/3$ and $\alpha_{\eta} = 1.0$, which matches the prediction for momentum-driven winds \citep{Murray_2005, Dave_2011} and is typical of the scalings predicted by simulations (see \citealt{Schroetter_2015}, their Figure 10).

The parameters $\alpha_{\eta}$ and $\beta_{\eta}$ are not independent, because  $\eta$ determines the SFR, and thus, the integrated stellar mass of a halo at $z=0$. We treat $\beta_{\eta}$ as the free parameter; for a given value of $\beta_{\eta}$, we choose $\alpha_{\eta}$ such that the total $z=0$ stellar mass implied by the model matches the $z=0$ total observed stellar-to-halo mass relation at the high-mass end (see Figure~\ref{fig:mstar_mhalo}). Increasing $\beta_{\eta}$ increases the slope of the stellar-to-halo mass relation at low masses. Because a change in $\beta_{\eta}$ causes $\eta(M_{\rm vir})$ to ``pivot'' around $M_{\rm vir}=10^{12}M_{\odot}$, $\alpha_{\eta}$ (as required to produce the correct normalization of the stellar-to-halo mass relation) is only  weakly dependent on $\beta_{\eta}$, varying by $\sim$\,50\% over $0 \lesssim \beta_{\eta} \lesssim 1/2$.

A smaller value of $\beta_{\eta}$ causes a larger fraction of stars and GCs to form in low-mass halos. Given the fixed stellar mass-metallicity relation we adopt (Equation~\ref{eq:mass_met}), changing $\beta_{\eta}$ also changes the predicted metallicity distributions of GCs and field stars; this is discussed further in Appendix~\ref{sec:stellar_mdf}.

To determine the gas surface density in a given halo, we use a KS-like relation that relates it to the star formation rate surface density. We first calculate the average star formation rate surface density as $\Sigma_{{\rm SFR}}\approx {\rm SFR}/(\pi R_{d}^{2})$, where $R_d$ is the scale length of the gas disk. We estimate $R_{d}$ using a model wherein the scale length of the disk is set by the specific angular momentum of the halo, \citep[e.g.][]{Fall_1980, Mo_1998},
\begin{align}
\label{eq:mmw}
R_{d}=\frac{\lambda}{\sqrt{2}}R_{{\rm vir}}\approx 0.025 R_{{\rm vir}},
\end{align}
where $\lambda$ is the halo spin parameter, which is fixed at a typical value of 0.035 \citep{Bullock_2001}. Although the scaling of $R_d$ with $\lambda$ implied by Equation~\ref{eq:mmw} likely does not hold in detail \citep{Desmond_2017, ElBadry_2018, GarrisonKimmel_2017}, the prediction of a constant scaling between disk size and $R_{\rm vir}$ has been found to hold within a factor of $\sim$2 over redshifts $0 < z < 8$ \citep{Shibuya_2015} and over nearly eight decades of stellar mass \citep{Kravtsov_2013, Huang_2017}.\footnote{\citet{Kravtsov_2013} found the normalization constant 0.025 in Equation~\ref{eq:mmw} to be closer to 0.01 at low redshift for stellar disks, but noted that this is expected if the \citet{Mo_1998} normalization held at the epoch of disk formation and halos subsequently grew by pseudo-evolution.}

Given $\Sigma_{\rm SFR}$, we estimate the corresponding gas surface density as described in \citet{FG_2013}. In their model, gravity is balanced by feedback-driven turbulence such that disks self-regulate to a Toomre parameter $Q\sim 1$. This leads to a KS-like relation of the form 
\begin{align}
\label{eq:KS_FG}
\Sigma_{{\rm gas}}^{2}=\frac{\left(P_{\star}/m_{\star}\right)\mathcal{F}}{2\sqrt{2}\pi QG}\Sigma_{{\rm SFR}}.
\end{align}
Here $(P_{\star}/m_{\star})$ is the momentum ultimately injected into the ISM by supernovae per stellar mass formed \citep{Cioffi_1988, Ostriker_2011}, and $\mathcal{F}$ is a factor of order unity encapsulating various uncertainties in the model; \citet{FG_2013} found $\mathcal{F}=2$ to provide a good match to observations. We set $Q=1$, $(P_{\star}/m_{\star})= 3000\,{\rm km\,s^{-1}}$, and $\mathcal{F}=2$, yielding 
\begin{align}
\label{eq:KS_scaling}
\frac{\Sigma_{{\rm gas}}}{M_{\odot}\,{\rm pc}^{-2}}=1.2\times10^{3}\left(\frac{\Sigma_{{\rm SFR}}}{10\,M_{\odot}\,{\rm kpc^{-2}}\,{\rm yr}^{-1}}\right)^{1/2}.
\end{align}

Here $\Sigma_{\rm gas}$ represents the disk-averaged surface density of cold gas. The  surface densities of individual molecular clouds are expected to higher than the disk-averaged surface density. We adopt $\Sigma_{\rm GMC} = 5 \times \Sigma_{\rm gas}$, which is roughly the mean relation found in observations of nearby galaxies and in simulations over a wide range of gas densities \citep[e.g.][]{Bolatto_2008, Hopkins_2012}. The factor of 5 is of course uncertain, but we do not leave it as a free parameter because varying it has exactly the same effect as varying $\Sigma_{\rm crit}$.

Given $\Sigma_{\rm gas}$, we can relate the SFR to the GC formation rate, $\dot{M}_{\rm GCs}$. Motivated by the theoretical prediction that the cluster formation efficiency plateaus at $\Sigma_{\rm GMC} \gg \Sigma_{\rm crit}$, we parameterize the GC formation efficiency as 

\begin{align}
\label{eq:Gamma}
\Gamma_{\rm GCs}\equiv \frac{\dot{M}_{{\rm GCs}}}{{\rm SFR}}=\frac{\alpha_{\Gamma}}{1+\left(\Sigma_{\rm GMC}/\Sigma_{{\rm crit}}\right)^{-\beta_{\Gamma}}}.
\end{align}
This parameterization causes the GC formation rate to approach $\alpha_{\Gamma} \times \rm SFR$ when $\Sigma_{\rm GMCs} \gg \Sigma_{\rm crit}$ and to be suppressed at $\Sigma_{\rm GMCs} \ll \Sigma_{\rm crit}$. How strongly GC formation is suppressed at $\Sigma_{\rm GMC} \ll \Sigma_{\rm crit}$ is set by $\beta_{\Gamma}$ (see Figure~\ref{fig:sup_funcs}). The case of $\beta_{\Gamma}=0$ corresponds to {\it no} dependence on surface density; in this case, the GC formation rate will be a constant multiple of the total star formation rate implied by the model. 
Following \citetalias{Grudic_2016}, we set fiducial values of $\Sigma_{\rm crit}=3000\,\rm M_{\odot}\,{\rm pc^{-2}}$ and $\beta_{\Gamma} = 1$. For particular values of $\Sigma_{\rm crit}$ and $\beta_{\Gamma}$, we set $\alpha_{\Gamma}$ such that the model reproduces the normalization of the observed GC-to-halo mass relation at the high-mass end. This implies $\alpha_{\Gamma} = 2.1\times 10^{-3}$ for the fiducial model. 
In reality, a large fraction of star clusters born in high-density gas disks are disrupted shortly after their formation \citep[e.g.][]{Fall_2005, Fall_2012}. This causes the effective formation efficiency of surviving clusters, represented by $\alpha_\Gamma$ in our model, to be small.

\subsubsection{Merger timescale}
\label{sec:timescale}
Each time a halo in the merger tree is accreted, our model requires an estimate of the accretion timescale to determine the implied $\dot{M}_{\rm tot,\,in}$ (Equation~\ref{eq:mdot_gas_in}). One possibility is to use the output timestep of the merger trees. In this case, the value of $\Sigma_{\rm gas}$, and thus, the properties of the predicted GC population, will depend on the time resolution of the merger tree. We therefore instead define the merger timescale to be roughly the dynamical time of the galaxy: 
\begin{align}
\label{eq:tau_merger}
\tau_{{\rm merger}}=\frac{0.05 R_{{\rm vir}}}{V_{{\rm vir}}},
\end{align}
where $0.05 R_{\rm vir}$ approximates the size of the galaxy and $V_{\rm}=\sqrt{GM_{\rm vir}/R_{\rm vir}}$. Because the relative velocity of galaxies during a merger is of order $V_{\rm vir}$, and the merging galaxies travel a distance of order their size during a merger, this timescale roughly represents how long the gas density is elevated during a merger. It depends only on redshift, varying from $\sim$10\,Myr at $z=5$ to $\sim$25\,Myr at $z=2$ to $\sim$100\,Myr at $z=0$.

When a halo of mass $M_{\rm vir,\,acc}$ is accreted, the resulting total mass accretion rate is 
\begin{align}
\label{eq:mdot_total}
\dot{M}_{{\rm tot,\,in}}=\frac{M_{{\rm vir,\,acc}}}{\tau_{{\rm merger}}}.
\end{align}
The total GC mass formed in a GC formation event is then 
\begin{align}
\label{eq:dm_gcs}
\Delta M_{{\rm GCs}}=\dot{M}_{{\rm GCs}}\times\tau_{{\rm merger}}, 
\end{align}
where $\dot{M}_{{\rm GCs}}$ is calculated from Equation~\ref{eq:mdot_gcs}. The adopted $\tau_{\rm merger}$ has no effect on the total stellar mass formed, since the resulting change in the implied mass accretion rate (Equation~\ref{eq:mdot_total}) is exactly balanced by the change in the timescale over which stars and GCs form (Equation~\ref{eq:dm_gcs}). However, the merger timescale does affect the total GC mass formed, because a decrease in $\tau_{\rm merger}$ implies an increase in the SFR, which implies an increase in $\Sigma_{\rm gas}$ and a higher fraction of stars formed in GCs. Increasing $\tau_{\rm merger}$ has the same effect as decreasing $\Sigma_{\rm crit}$.

\subsubsection{GC masses}
\label{sec:sampling}
We draw the masses of individual clusters for each GC formation event from an $m^{-2}$ power law with $m_{\rm min} = 10^{5} M_{\odot}$, assuming that the majority of lower-mass clusters would be disrupted by $z=0$ \citep[e.g.][]{Fall_2001, Muratov_2010}. Following \citet[][their equations 11 \& 12]{Muratov_2010}, we determine the mass of the most-massive cluster formed in a given event using ``optimal sampling'' \citep{Kroupa_2013}. Because we do not predict GC mass functions and do not implement mass-dependent GC disruption or evaporation in our fiducial model, these choices have little effect on our primary conclusions. We consider the effects of mass- and age-dependent GC disruption and evaporation in Appendix~\ref{sec:disrupt}; there, the GC mass spectrum does affect how much disruption occurs.

For a given minimum and maximum GC mass, we then compute $\overline{m}$, the mean mass of the GC mass function, and the predicted number of GCs formed, $\Delta N_{{\rm GCs}}=\Delta M_{{\rm GCs}}/\overline{m}.$ The number of GCs formed in a single event must always be an integer. In order to ensure that our procedure for stochastically drawing GC masses on average forms the correct total $\Delta M_{\rm GCs}$ as predicted by Equation~\ref{eq:dm_gcs}, we use another random draw to determine the number of GCs formed. For example, if $\Delta N_{{\rm GCs}}$ is 2.7, we form 3 GCs with 70\% probability and 2 GCs with 30\% probability. If $\Delta N_{{\rm GCs}} < 0.5$, no GCs are formed. This limit prevents spurious GC formation in accretion events in which the mass of accreted cold gas is insufficient to form a GC. 

\subsubsection{Metallicities and Colors}
\label{sec:metallicity}
We assume that GCs inherit the gas-phase metallicity  of the galaxy in which they formed, which we calculate using the mass-metallicity relation from \citet{Ma_2016}:\footnote{Ma et al. provide a fitting function for the mass-weighted total metallicty, $\log(Z_{\rm gas}/Z_{\odot})$. Following their convention, we then estimate $[\rm Fe/H]_{\rm gas}=\log(Z_{\rm gas}/Z_{\odot}) - 0.2$, where $[\rm Fe/H]_{\rm gas}$ represents the logarithmic iron abundance relative to the Solar value. The mass and redshift-evolution predicted by this relation is similar to that predicted in the model of \citet{Choksi_2018}. However, the normalization is lower at all masses and redshifts, by $0.2-0.3$ dex on average. We note that Ma et al. assumed $Z_{\odot} = 0.02$; assuming $Z_{\odot} = 0.014$ would increase all [Fe/H] values by 0.15 dex.}
\begin{align}
\label{eq:mass_met}
\left[{\rm Fe/H}\right]_{{\rm gas}}=0.35\left(\log\frac{M_{{\rm star}}}{M_{\odot}}-10\right)+0.93\exp\left(-0.43z\right)-1.25.
\end{align}
When calculating metallicities, we assign a stellar mass to each halo using the median stellar-to-halo mass relation from \citep{Behroozi_2013}.\footnote{We do not use the stellar mass calculated directly from our model because it includes the mass of unmerged satellites while the mass-metallicity relation is for individual galaxies (see Section~\ref{sec:disc_mstar_mhalo}).}
Following \citet{Tremonti_2004}, we assume an intrinsic Gaussian scatter in metallicity at fixed $M_{\rm star}$ of $\sigma_{\rm [Fe/H]} = 0.1$\,dex. We calculate colors for model GCs using PARSEC isochrones \citep[v1.2S; ][]{Bressan_2012, Tang_2014, Chen_2014, Chen_2015}. We treat each GC as a simple stellar population with a \citet{Kroupa_2001} initial mass function. A model GC's color at a given time thus depends only on its age and metallicty. 

\subsubsection{GC disruption}
Our fiducial model does not include any GC disruption, tidal stripping, or mass loss due to two-body evaporation, besides the assumption that clusters with birth masses below $10^5 M_{\odot}$ will be disrupted by $z=0$. Although disruption likely does have non-negligible effects on some observable properties of the GC population \citep{Spitzer_1987,Gnedin_1999,Fall_2001,McLaughlin_2008, Carlberg_2017}, we do not believe that a model such as ours can capture disruption with much fidelity. The efficiency of disruption is highly dependent on the spatial distribution of GCs: GCs are subjected to strong tidal forces and can be rapidly destroyed as long as they reside in the disks within which they formed \citep{Kruijssen_2012b}. Tidal effects become much weaker once GCs migrate into the halo due to mergers \citep[e.g.][]{Kruijssen_2015} or feedback-driven fluctuations in the gravitational potential \citep[e.g.][]{ElBadry_2016, ElBadry_2018b}. Because our model does not include information about the spatial distribution of GCs, it cannot account for these effects.

We consider the effects of a simple analytical model for GC disruption and stripping, which has also been employed in other recent semi-analytic works, in Appendix~\ref{sec:disrupt}. We find that the primary effect of disruption as implemented in this model is to change the normalization of $\Gamma_{\rm GCs}$; i.e., the $\alpha_{\Gamma}$ parameter in Equation~\ref{eq:Gamma}) required to match the observed GC-to-halo mass relation. 

\subsection{Random GC Formation Model}
\label{sec:random}
We also construct a pathological random model for GC formation in which GC mass and halo mass are uncorrelated at the time of GC formation. We use this model to explore what aspects of the $z=0$ GC population are expected purely due to hierarchical assembly.

In the random model, we select a random subset of all halos in the merger tree at $z>2$ as GC formation sites. Each node in the merger tree at $z>2$ is assigned the same probability of hosting a GC formation event. We then randomly assign each of these halos a GC mass to form, $\Delta M_{\rm GCs}$, which we drawn from a log-uniform distribution over $10^{5-7} M_{\odot}$. The absolute probability of hosting a GC formation event is set such that the normalization of the resulting $z=0$ GC-to-halo mass relation matches the observed relation at the high-mass end. As in the fiducial model, masses of individual GCs for each formation event are sampled as described in Section~\ref{sec:sampling}. Both the halos in which GCs form and the GC mass formed in each halo are chosen without any consideration of halo mass, the mass accretion rate, or the implied SFR or gas density. 

In the random model, the mean GC mass formed {\it per halo} is independent of halo mass. Because low-mass halos are more abundant than high-mass halos, the mean GC mass formed {\it per unit halo mass} decreases with $M_{\rm vir}$, roughly as $M_{\rm vir}^{-1}$. We experimented with an alternate random model in which the probability of hosting a GC formation event is instead proportional to halo mass, such that the mean GC mass formed {\it per unit halo mass} is independent of halo mass. All the results we present are unchanged under this alternate model.

\section{Results}
\label{sec:results}

\subsection{GC system mass -- halo mass relation}
\label{sec:mgc_mhalo}

\begin{figure*}
\includegraphics[width=\textwidth]{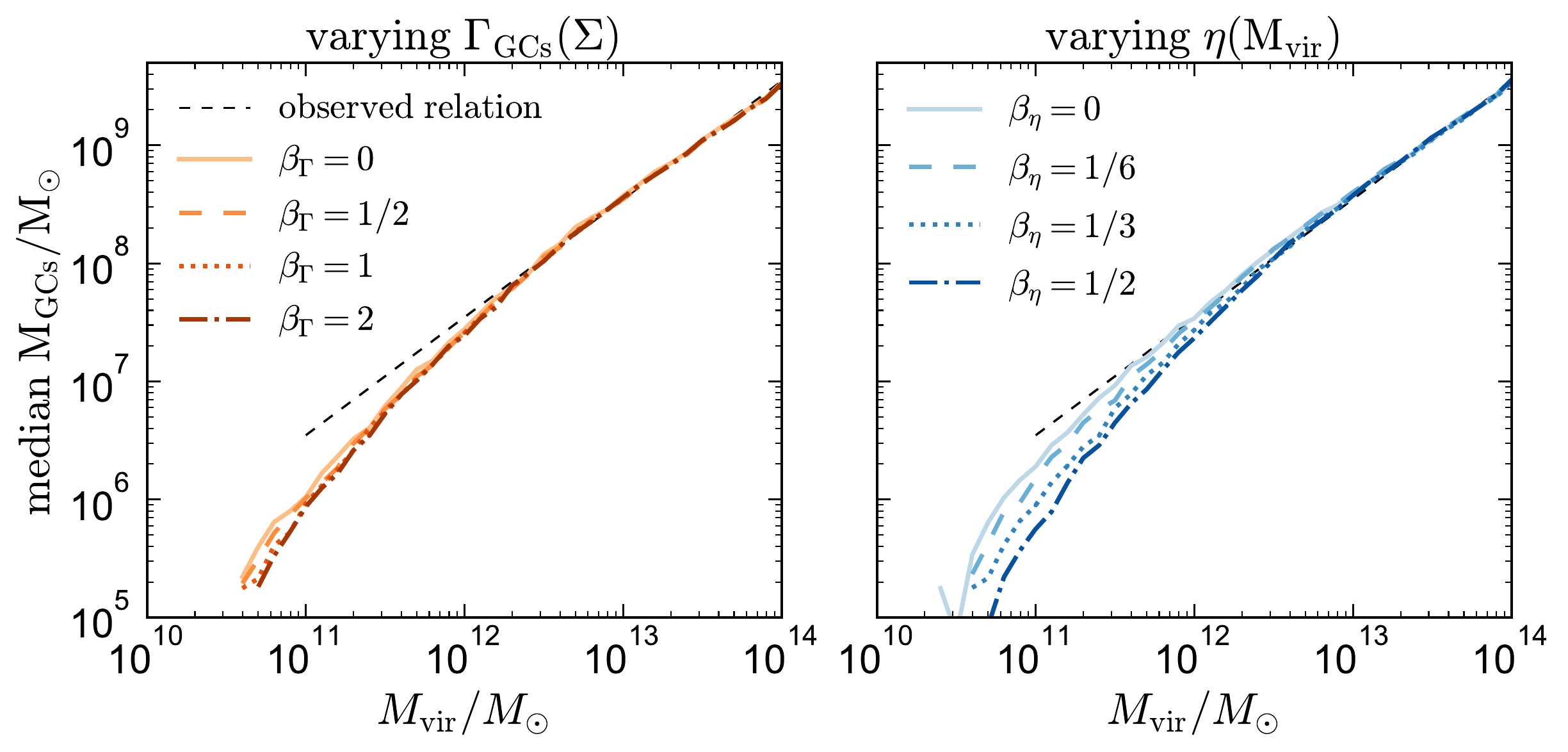}
\caption{Median total GC-to-halo mass relation predicted by our model for different values of $\beta_{\Gamma}$ (left) and $\beta_{\eta}$ (right). The normalization of the relation is forced to match the observed value at the high-mass end. In the left (right) panel we fix, $\beta_{\eta} = 1/3$ ($\beta_{\Gamma} = 1$). The black dashed line represents the observed linear relation, $M_{\rm GCs} \approx 3.5\times 10^{-5} M_{\rm vir}$. {\bf Left}: the relation is not sensitive to $\beta_{\Gamma}$, which parameterizes how steeply the cluster formation efficiency falls off at low $\Sigma_{\rm GMC}$ (Equation~\ref{eq:Gamma}). The model reproduces the observed linear behavior at $M_{\rm vir} \gtrsim 10^{12} M_{\odot}$, but the prediction falls somewhat below linear at $M_{\rm vir}\lesssim 10^{11.5}M_{\odot}$. {\bf Right}: Increasing $\beta_{\eta}$ leads to a decrease in $M_{\rm GCs}$ at low halo masses. At high halo masses, {\it all} models predict a linear GC-to-halo mass relation.}
\label{fig:mgc_mhalo}
\end{figure*}

We first consider the $z=0$ relation between halo mass and the total mass of all GCs in a halo ($M_{\rm GCs}$, which in our default model without GC disruption is simply the sum of the masses of all GCs formed in the merger tree). Figure~\ref{fig:mgc_mhalo} shows the effect of varying $\beta_{\Gamma}$ (left) and $\beta_{\eta}$ (right). The black dashed line represents the observed relation, which is well-fit by a constant GC-to-halo mass ratio, $M_{{\rm GCs}}=3.5\times10^{-5}M_{{\rm vir}}$, over $M_{\rm vir} \simeq 10^{11-15} M_{\odot}$ \citep{Harris_1996, Hudson_2014, Harris_2015}. We plot the median total GC mass predicted by the model for four choices of $\beta_{\Gamma}$ (left, while keeping $\beta_{\eta} = 1/3$ fixed) and four choices of $\beta_{\eta}$ (right, while keeping $\beta_{\Gamma}=1$ fixed). The median relation is calculated from 20 Monte Carlo merger trees at each 0.1 dex interval in $M_{\rm vir}$, which we find to be sufficient for all quantities to be converged. 

The left panel shows that the shape of the global GC-to-halo mass relation is not sensitive to $\beta_{\Gamma}$. The overall normalization of the relation {\it does} vary somewhat with $\beta_{\Gamma}$, but we always set the parameter $\alpha_{\Gamma}$ in Equation~\ref{eq:Gamma} such that the normalization of the GC-to-halo mass matches the observed value at the high-mass end. With this constraint in place, the GC-to-halo mass ratio is essentially independent of $\beta_{\Gamma}$ over all halo masses and is completely linear at high halo masses. As we will discuss in Section~\ref{sec:disc_mgc_mhalo}, this owes to the self-similar assembly histories of dark matter halos. At low masses, the fiducial model with $\beta_{\eta} = 1/3$ produces lower total GC mass on average than predicted by a constant GC-to-halo mass relation. Halos with $M_{\rm vir} \lesssim 3\times 10^{10} M_{\odot}$ on average do not form any GCs. 

The right panel of Figure~\ref{fig:mgc_mhalo} shows that the shape of the GC-to-halo mass relation does depend somewhat on $\beta_{\eta}$, which determines how the mass loading factor varies with halo mass. A larger value of $\beta_{\eta}$ leads to larger $\eta$, lower SFR, and thus,  lower $\Sigma_{\rm GMC}$ at $M_{\rm vir} < 10^{12}M_{\odot}$ (Equation~\ref{eq:eta}), so higher values of $\beta_{\eta}$ lead to fewer GCs forming in low-mass halos. Interpreted at face value, Figure~\ref{fig:mgc_mhalo} would appear to suggest that a value of $\beta_{\eta}$ close to 0 provides the best match the observed constant GC-to-halo mass relation; however, we caution that there is significant scatter in the observed relation at low halo masses and some indication that observed GC systems on average also fall below the constant ratio at $M_{\rm vir} \lesssim 10^{12}M_{\odot}$; see \citet[][their Figure 3]{Choksi_2018}. On the other hand, varying $\beta_{\eta}$ has little effect on the shape of the GC-to-halo mass relation at higher masses, $M_{\rm vir} \gtrsim 10^{11.5} M_{\odot}$. As we will now show, a constant GC-to-halo mass relation at high halo masses is a generic consequence of hierarchical assembly.

\subsubsection{GC-to-halo mass relation with random GC formation}

\begin{figure*}
\includegraphics[width=\textwidth]{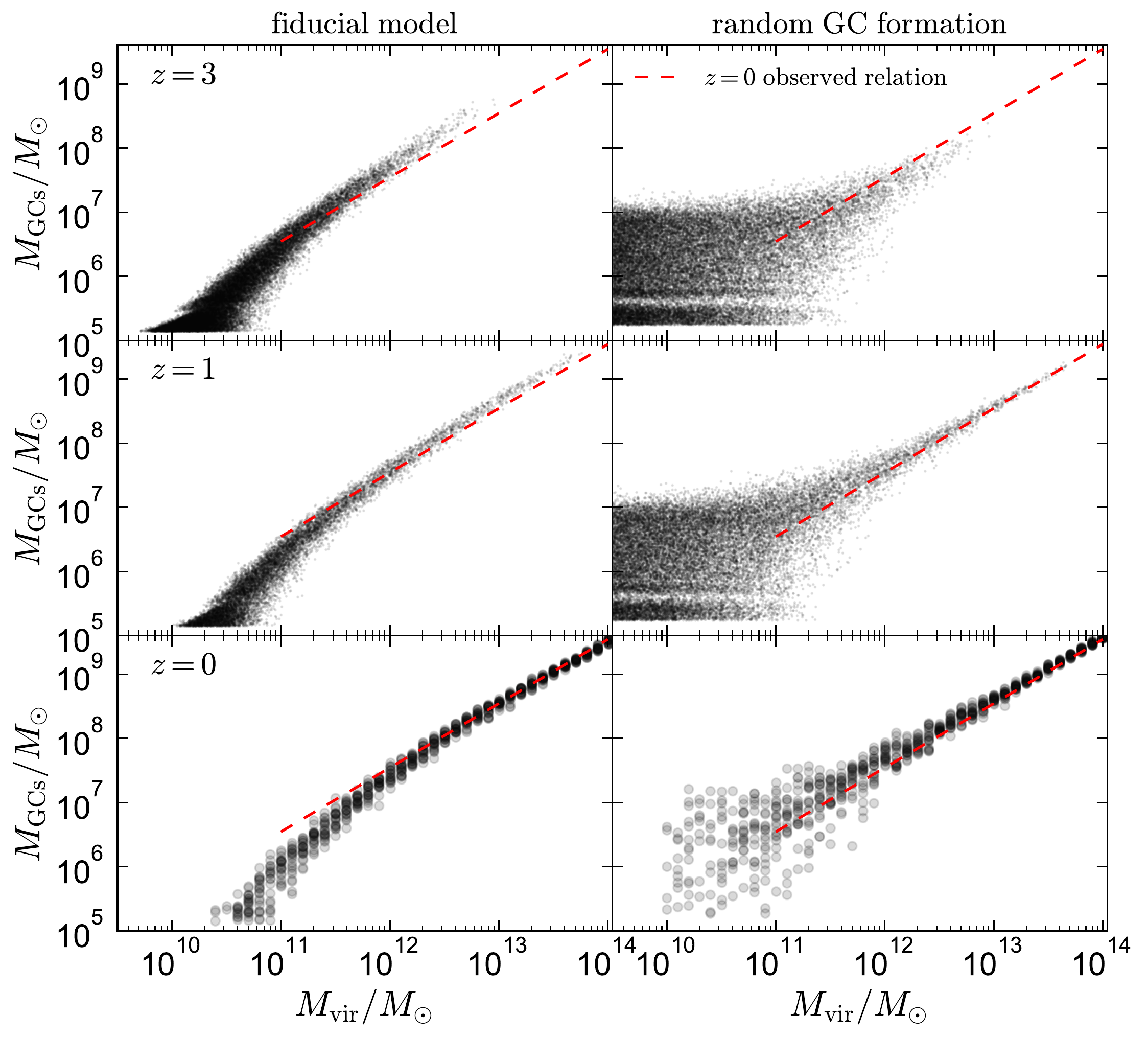}
\caption{Total GC system mass vs. halo mass relation. Each point represents a single halo at $z=3$ (top), $z=1$ (middle) and $z=0$ (bottom); red line represents the observed $z=0$ relation. {\bf Left}: Fiducial model, in which GC formation occurs at high surface density (see Section~\ref{sec:implementation}). {\bf Right}: Random model (Section~\ref{sec:random}), in which GCs form in an entirely random subset of all halos at $z>2$. A tight, linear GC-to-halo mass relation is predicted by $z=0$ at high halo masses purely as a result of the central limit theorem: high-mass halos form through mergers of low-mass halos, so the total GC-to-halo mass ratio tends to average out by $z=0$. The linearity of the GC-to-halo mass relation at $M_{\rm vir}\gtrsim 10^{11.5}M_{\odot}$ thus does not contain much information about GC formation beyond the fact that GCs are old.}
\label{fig:random_model}
\end{figure*}

Figure~\ref{fig:random_model} compares the evolution of the GC-to-halo mass relation in the fiducial and random models. Each gray point represents a single halo at $z=3$ (top), $z=1$ (middle) and $z=0$ (bottom). The left panels show the fiducial model, with $\beta_{\eta} = 1/3$ and $\beta_{\Gamma} = 1$, for which the median relation was shown in Figure~\ref{fig:mgc_mhalo}. For the fiducial model, an approximately constant GC-to-halo mass ratio is already in place at high redshift, but the relation is offset to higher $M_{\rm GCs}$ at fixed $M_{\rm vir}$ relative to the $z=0$ relation. Because the fraction of the total mass that is assembled at early times is larger for GCs than for halos, halos move rightward in the $M_{\rm GCs}$ -- $M_{\rm vir}$ plane as they evolve. By $z=0$, the total GC-to-halo mass relation matches the observed constant ratio at high halo masses but falls somewhat below linearity at low halo masses. 

In the right panels, we show the relation predicted by the random model. There is no correlation between GC mass and halo mass at the time the GCs form,\footnote{The substructure at high redshift and low $M_{\rm GCs}$ is an artifact of the GC mass sampling procedure for GC formation events in which only a single GC is formed (Section~\ref{sec:sampling}).} but a linear relation emerges at later times as both GC system masses and halo masses grow through mergers. At $z=0$, the random model -- which does not attempt to model any of the physics of GC formation, except that GCs form at $z>2$ -- produces a tight, constant relation at $M_{\rm vir} \gtrsim 10^{11.5} M_{\odot}$, with scatter comparable to that predicted by the fiducial model. At lower masses, the scatter grows larger, but the median GC mass does not drop off as it does for the fiducial model.  

In the random model, mergers alone drive the GC-to-halo mass ratio towards a constant value. This is a manifestation of the central limit theorem. The total GC mass at $z=0$ is essentially the result of adding together a long list of random numbers (the GC masses formed in each GC formation event). The same is true for the total halo mass at $z=0$, which is the sum of all progenitor halo masses. At higher halo masses, the number of random numbers is larger, driving down the scatter in their sum. Irrespective of the GC formation model and the initial relation between GC mass and halo mass, a constant GC-to-halo mass ratio is expected at late times as long as GCs form relatively early, such that there are enough mergers after the majority of GCs form to drive the ratio towards a constant value. 

The scatter in the GC-to-halo mass relation for the random model is large for halo masses at which the maximum $\Delta M_{\rm GCs}$ per formation event ($10^7 M_{\odot}$ in our implementation) exceeds the value of $M_{\rm GCs}$ implied by the observed relation. This corresponds to $M_{\rm vir} \lesssim 3\times 10^{11} M_{\odot}$ in Figure~\ref{fig:random_model}. At higher masses, the $z=0$ GC population is always the result of several GC formation events, driving the total GC population toward the mean relation. Increasing the maximum GC mass formed per formation event in the random model increases the scatter in the GC-to-halo mass relation at low masses.

\begin{figure}
\includegraphics[width=\columnwidth]{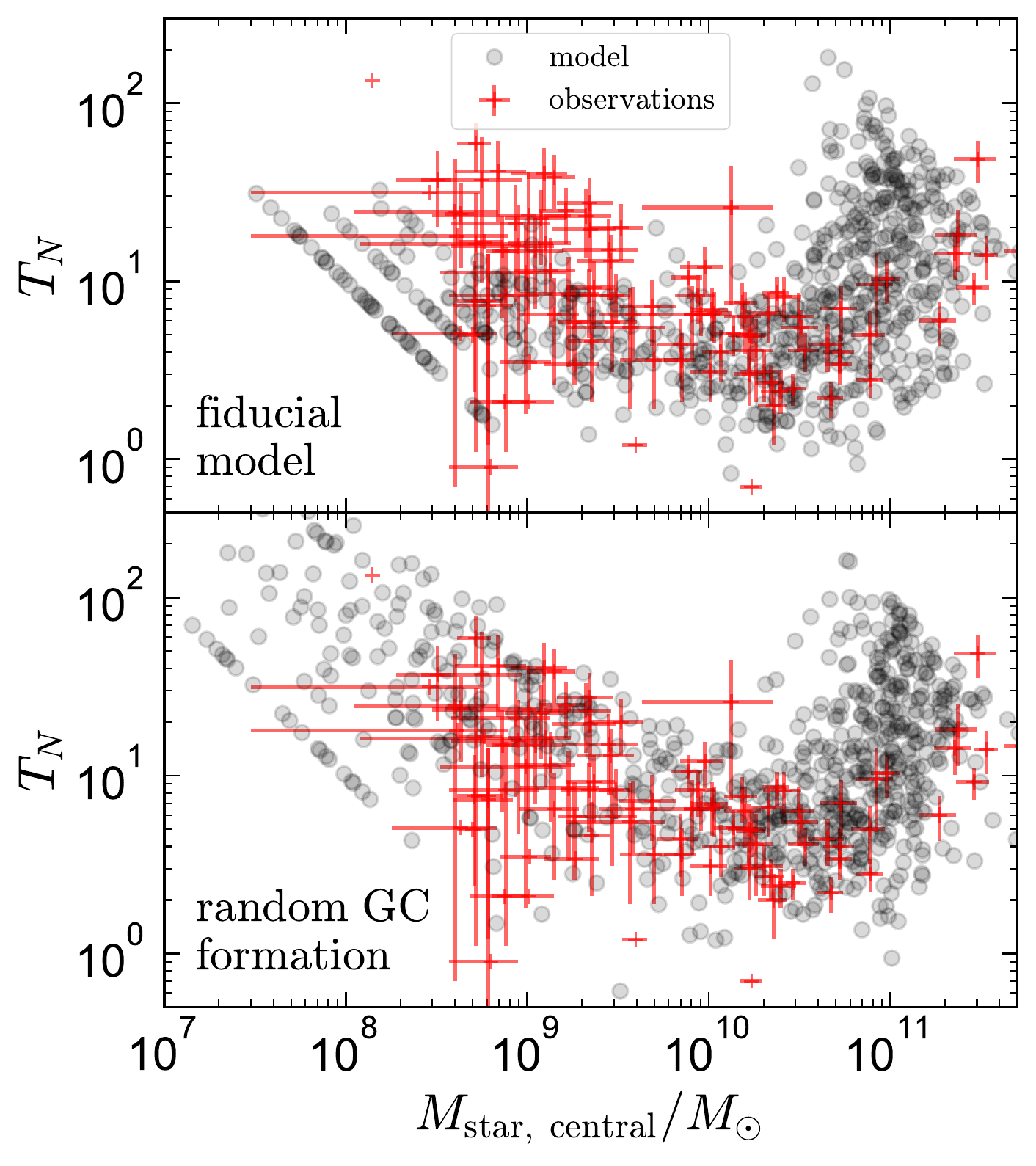}
\caption{GC frequency, $T_{N}=N_{{\rm GCs}}/(M_{{\rm star}}/10^{9}M_{\odot})$, vs. stellar mass of the host galaxy. Observational data are from the ACS Virgo Cluster Survey \citep{Peng_2008}. Model values of $M_{\rm star,\,central}$ are computed assuming the stellar-to-halo mass relation from \citet{Behroozi_2013}. Predictions for the fiducial model and random GC formation scenario are very similar at $M_{\rm star,\,central} \gtrsim 10^{9} M_{\odot}$ and are in good agreement with observations. GC formation in the fiducial model is suppressed in lower-mass galaxies, leading to lower $T_N$ than in the random formation scenario.}
\label{fig:spec_freq}
\end{figure}

In Figure~\ref{fig:spec_freq}, we show the stellar mass-normalized GC frequency, $T_{N}=N_{{\rm GCs}}/(M_{{\rm star}}/10^{9}M_{\odot}),$ predicted for the random and fiducial models. We compute the stellar mass of the host galaxy using the stellar-to-halo mass relation from \citet{Behroozi_2013}, including 0.22 dex scatter. We compare the predictions of both models to observations from the ACS Virgo Cluster Survey \citep{Peng_2008}. Both the fiducial and random models match the shape of the observed relation; this is expected if the observed and model galaxies follow a similar stellar-to-halo mass relation. In low-mass galaxies, ($M_{\rm star} \ll 10^9 M_{\odot}$), the random formation scenario predicts significantly higher values of $T_N$, as the fiducial model suppresses GC formation in low-mass halos. The observed scatter in $T_N$ increases markedly at low masses, as is predicted if the constant GC-to-halo mass relation is primarily a consequence of the central limit theorem in hierarchical assembly. 

Observational data are sparse in the mass range where the predictions of the random and fiducial models strongly differ but agree somewhat better with the random formation model. We caution, however, that because $T_N$ depends on the {\it number} of GCs rather than on their total mass, neglecting GC disruption and evaporation decreases $T_N$ predicted by the model at fixed $M_{\rm GCs}$ (see Appendix~\ref{sec:disrupt}). We discuss the GC populations of low-mass galaxies further in Section~\ref{sec:low_mass}.

We have thus far neglected the possible effects of GC evaporation, stripping, and disruption on the GC-to-halo mass relation. In Appendix~\ref{sec:disrupt}, we show that although these processes are expected to change the normalization of the GC-to-halo mass relation at fixed $\alpha_{\Gamma}$, they do not change the conclusion that a constant GC-to-halo mass ratio is expected at high masses purely due to the effects of mergers. 

We also emphasize that the random model implemented here is not designed to produce a realistic GC population at $z=0$, and it can be ruled out by comparing its predictions to observables beside the GC-to-halo mass relation. For example, because randomly selecting nodes in the merger tree as GC formation sites preferentially selects low-mass halos, the random model predicts a large majority of GCs to be metal-poor. Our contention is simply that mergers alone will drive the GC-to-halo mass relation toward a constant ratio for a large range of GC formation models. We return to this discussion in Section~\ref{sec:disc_mgc_mhalo}.

\subsubsection{GC-to-halo mass relation for red and blue GCs}
\label{sec:red_vs_blue}
\begin{figure}
\includegraphics[width = \columnwidth]{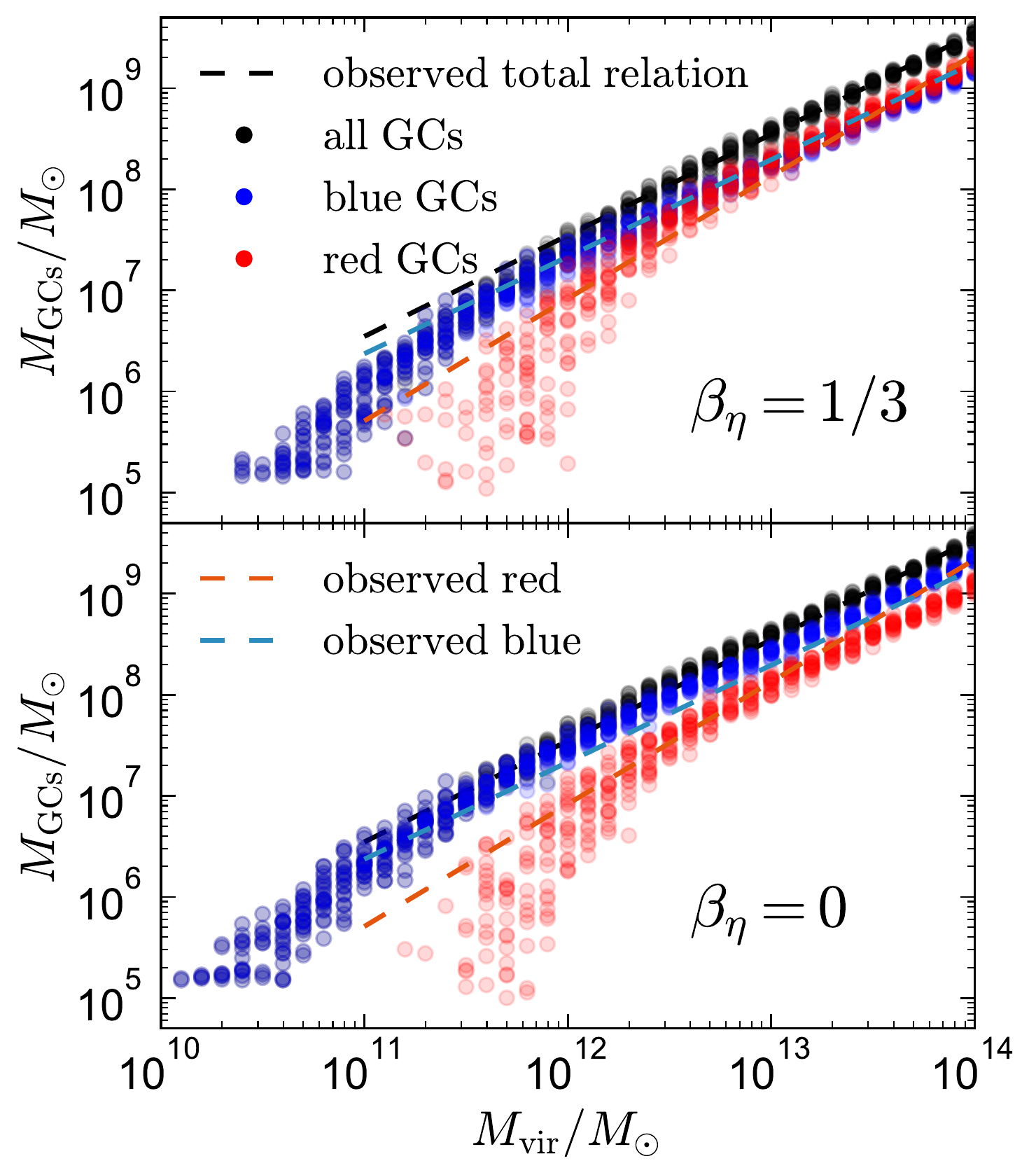}
\caption{Total GC-to-halo mass relation for red GCs, blue GCs, and all GCs. Dashed line show median observed relations. Top panel shows our fiducial model with $\beta_{\eta} = 1/3$; bottom panel shows the extreme case in which the galactic wind mass loading factor $\eta$ does not depend on halo mass. Because blue GCs form in lower-mass halos and at earlier times than red GCs, they fall on a tighter relation at lower masses, where they dominate the GC population. Red GCs make up an increasing fraction of the population at higher halo masses, so the GC-to-halo mass relation for red GCs is superlinear at intermediate halo masses. }
\label{fig:mass_scaling_red_blue}
\end{figure}

In Figure~\ref{fig:mass_scaling_red_blue}, we show the $z=0$ GC-to-halo mass relations predicted by our fiducial model for red and blue GCs separately. We divide red and blue clusters based on their $V-I$ color (Section~\ref{sec:metallicity}), with red GCs having $V-I > 1.0$. This corresponds roughly to the division between the two peaks in the GC color distributions predicted by our model, and to a metallicity of $\rm [Fe/H]=-1$ for typical GC ages (Section~\ref{sec:bimodality}). Blue GCs dominate the population at low halo masses. GC color is driven primarily by metallicity, and the metal-poor progenitors of low-mass halos form GCs that are metal-poor.

$\beta_{\Gamma} = 1$ is fixed in both panels. The top panel shows predictions for the fiducial mass loading factor scaling of $\beta_{\eta} = 1/3$. Slightly more red GCs than blue GCs are predicted at the high-mass end, with equal GC mass in the two populations near $M_{\rm vir} = 10^{13} M_{\odot}$. The bottom panel shows predictions for $\beta_{\eta} = 0$. In this case, $\eta$ is lower in low-mass, metal-poor galaxies, making their SFR and $\Sigma_{\rm GMC}$ higher. This results in a higher fraction of blue GCs at all $z=0$ halo masses. Although it is not shown in Figure~\ref{fig:mass_scaling_red_blue}, we find that varying $\beta_{\Gamma}$ also changes the relative numbers of red and blue GCs: at fixed $M_{\rm vir}$ and $\beta_{\eta}$, increasing $\beta_{\Gamma}$ decreases the fraction of GCs that are red because a larger fraction of GCs form at high redshift.

Because the fraction of GCs that are red increases with halo mass, the GC-to-halo mass relation is steeper for red GCs than for blue GCs up to halo masses of a few $\times\,10^{13}M_{\odot}$. This is also found observationally: the best power-law fit to the observed blue GC-to-halo mass relation has a slope of $\sim$0.96, similar to the constant ratio (slope 1) observed for all GCs, while the slope for red GCs is steeper, at $\sim$1.21 \citep{Harris_2015}. The red fraction predicted by our model flattens at the highest halo masses. Whether such flatting is also found for observed GC populations is unclear due to the small number of observed GC systems in high-mass halos; however, we note that there is substantial scatter in the observed $f_{\rm red}$ at fixed halo mass \citep[e.g.][]{Beasley_2018}.

The fraction of GCs that are red indeed decreases in low-mass halos in the local Universe \citep{Brodie_1991, Cote_1998, Larsen_2001}. However, the sharpness of the transition from red to blue GCs predicted by our model is steeper that what is observed: some red GCs {\it are} observed in halos with masses $M_{\rm vir} < M^{11} M_{\odot}$, where our model predicts all GCs to be blue. \citet{Harris_2015} find a red fraction of $\sim$30\% at $M_{\rm vir} = 10^{12} M_{\odot}$ and $\sim$20\% at $M_{\rm vir} = 10^{11} M_{\odot}$. The observed red fraction does eventually reach $\sim$\,0, but only at $M_{\rm vir} \lesssim 10^{10} M_{\odot}$ \citep{Georgiev_2010}. 

The mean metallicity predicted by our fiducial model agrees well with observations (see Section~\ref{sec:mass_metallicity}), so the dearth of red GCs primarily reflects the fact that the scatter in GC metallicity at fixed mass predicted by our model is lower than is observed. Our model assumes an intrinsic scatter in the galaxy mass-metallicity relation of only 0.1 dex. This value is consistent with what is observed for galaxies in the local Universe \citep{Tremonti_2004}, but the relation is uncertain at low masses and high redshifts, where its scatter may also be larger. The observed scatter in the {\it stellar} mass-metallicity relation at low masses is of order 0.2 dex in the Local Group \citep{Kirby_2013}. Our model also assigns the same metallicity to all GCs formed in a given GC formation event, implicitly treating the ISM as homogeneous. The ISM in real galaxies can exhibit spatial abundance fluctuations over a wide range of scales \citep[e.g.][]{Sanders_2012, Krumholz_2018}, and it is also possible that GCs self-enrich during formation \citep{Bailin_2009}. Increased scatter in the mass-metallicity relation due to such effects could also plausibly account for the red GCs observed in low-mass halos. 

\subsection{Cosmic GC formation rate}
\label{sec:cSFR}

\begin{figure}
\includegraphics[width=\columnwidth]{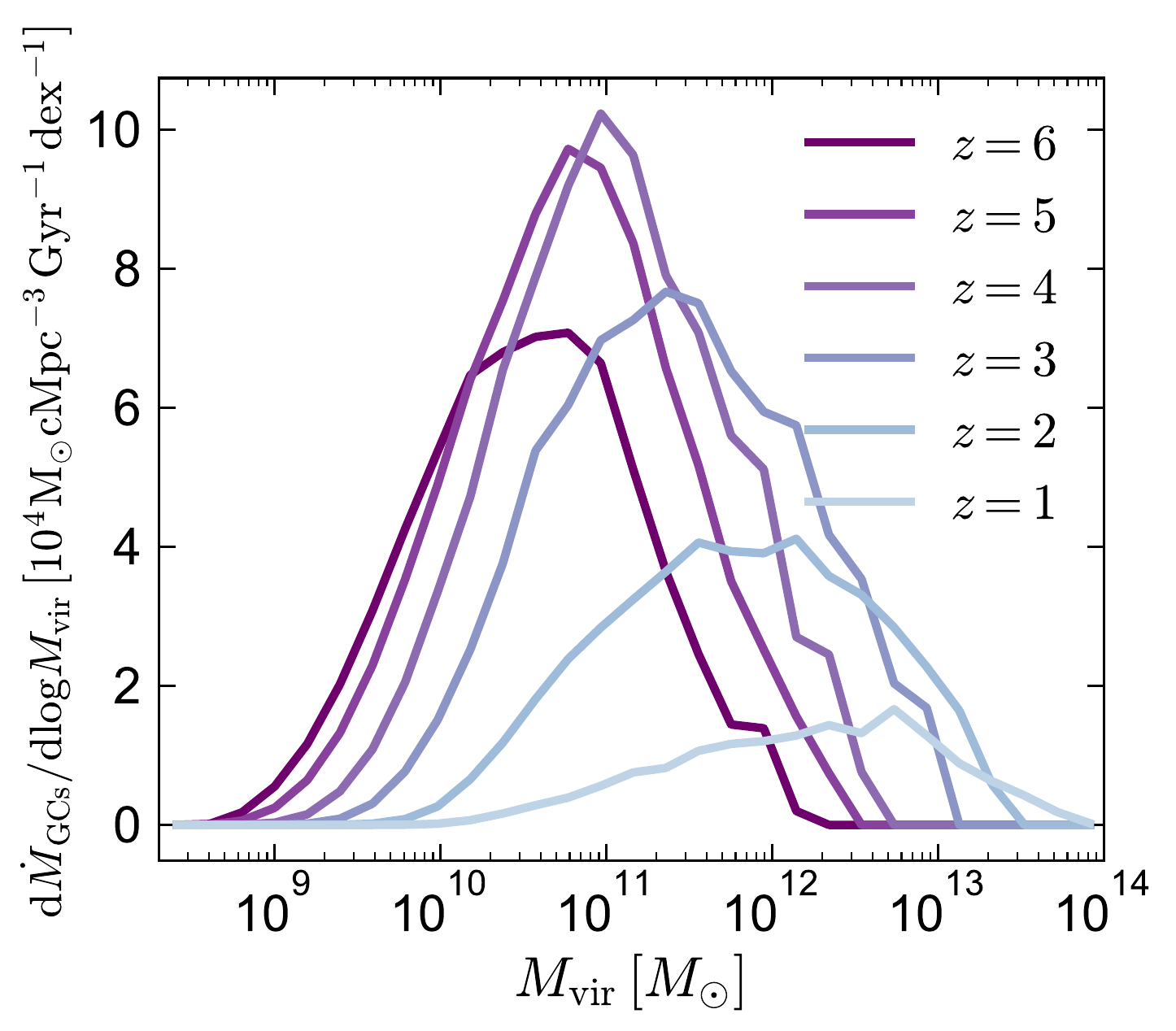}
\caption{Cosmically averaged distribution of halo masses hosting GC formation at different redshifts, for our fiducial model with $\beta_{\eta}=1/3$ and $\beta_{\Gamma}=1$. GCs form in progressively more massive halos at later times.}
\label{fig:mdot_gc_vs_mass}
\end{figure}

To calculate the cosmic mean GC formation rate at a given redshift, we calculate the mean GC formation rate per halo as a function of halo mass and redshift and then weight by the halo mass function: 
\begin{align}
\label{eq:mdot_GCs_vs_mass}
\frac{{\rm d}\dot{M}_{{\rm GCs}}}{{\rm d}M_{{\rm vir}}}=\frac{{\rm d}\dot{M}_{{\rm GCs}}}{{\rm d}n}\frac{{\rm d}n}{{\rm d}M_{{\rm vir}}}
\end{align}
Here ${\rm d}\dot{M}_{{\rm GCs}}/{\rm d}n=\left\langle \Delta M_{{\rm GCs}}/\Delta t\right\rangle \left(M_{{\rm vir}},z\right)$ is the mean GC formation rate per halo\footnote{This quantity is averaged over all halos, including those not forming any GCs. $\Delta M_{{\rm GCs}}$ represents the GC mass formed in a particular timestep, and $\Delta t$, the length of the timestep.} for halos of a particular mass and redshift, and ${\rm d}n/{\rm d}M_{{\rm vir}}$ is the halo mass function. We use the halo mass function measured from the Bolshoi-Planck and MultiDark-Planck simulations by \citet[][their Equation 23]{Rodriguez_2016}. Figure~\ref{fig:mdot_gc_vs_mass} shows the resulting cosmically-averaged distribution of GC formation sites at different redshifts. The peak of the distribution is set by the competing effects of a higher average GC formation rate in more massive halos and a larger absolute number of low mass halos; it moves to higher masses at late times as the Schechter mass increases and accretion of cold gas is suppressed in low-mass halos. 

\begin{figure}
\includegraphics[width=\columnwidth]{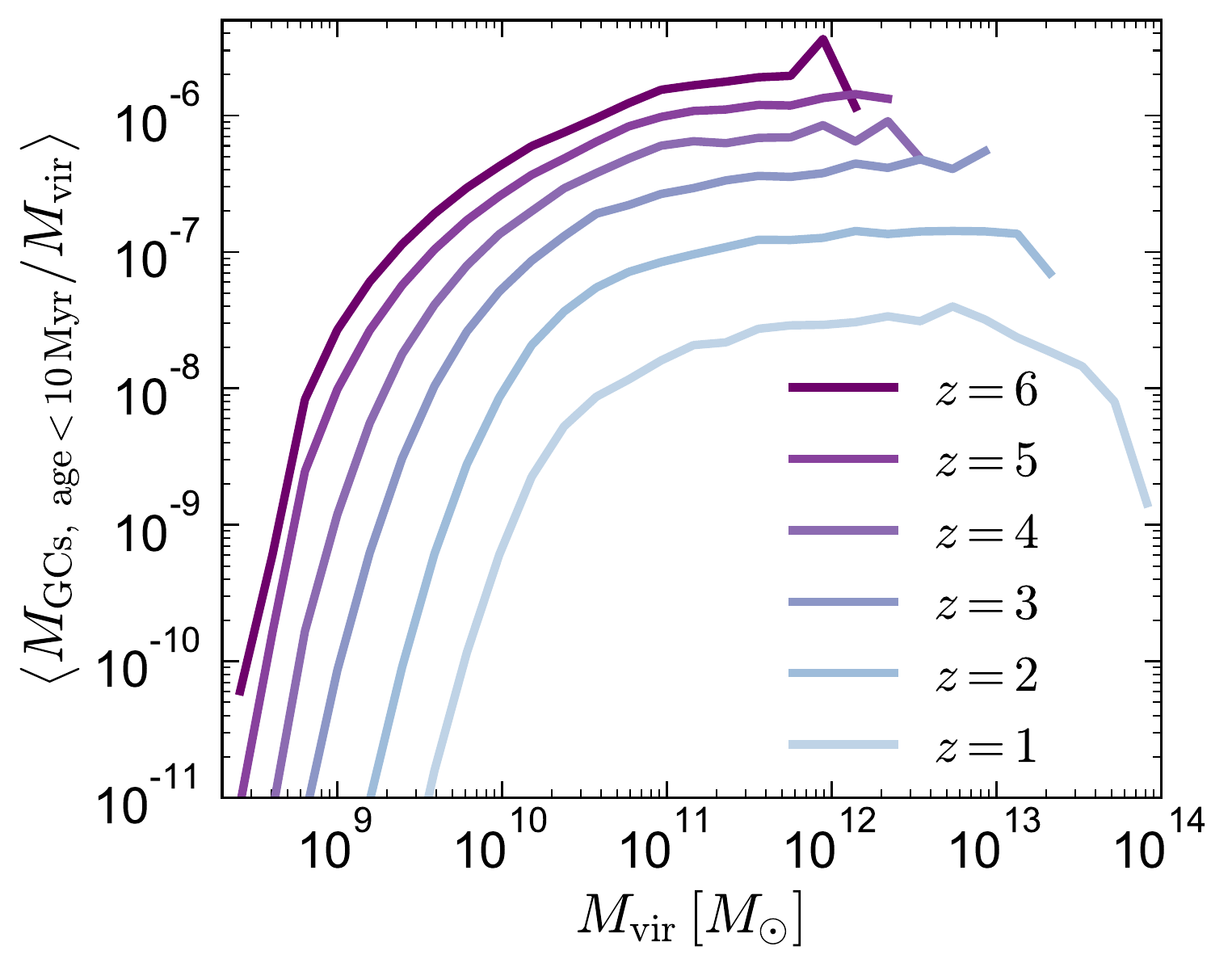}
\caption{Mean GC-to-halo mass ratio predicted by our fiducial model for young GCs only (age < 10\,Myr). GC formation is suppressed at low halo masses, and at high halo masses at late times. The ratio is highest at high $z$, but the cosmic abundance of massive halos is lower at high $z$.}
\label{fig:young}
\end{figure}

If the GC formation rate scaled linearly with halo mass, the distributions in Figure~\ref{fig:mdot_gc_vs_mass} would be flat below the Schechter mass because each decade in $M_{\rm vir}$ contributes the same total mass for a ${\rm d}n/{\rm d}M_{\rm vir} \sim M_{\rm vir}^{-2}$ halo mass function. The fact that this is {\it not} the case is primarily a consequence of the cold gas-to-dark matter relation adopted in our model (Appendix~\ref{sec:prev_fdbk} and Figure~\ref{fig:zeta2D}), which imprints a mass scale at which GC formation is most efficient. This can be seen explicitly in Figure~\ref{fig:young}, which shows the GC-to-halo mass ratio for young GCs only. This ratio is nearly constant at intermediate halo masses\footnote{This occurs because the gas accretion rate scales nearly linearly with halo mass \citep{Dekel_2009}.} but drops off sharply at low halo masses, and at later times, at high halo masses. Thus, although our fiducial model predicts the integrated GC mass in a halo at a given redshift to scale linearly with halo mass (Figure~\ref{fig:random_model}), the specific GC formation {\it rate} at any redshift varies with halo mass.

\begin{figure}
\includegraphics[width=\columnwidth]{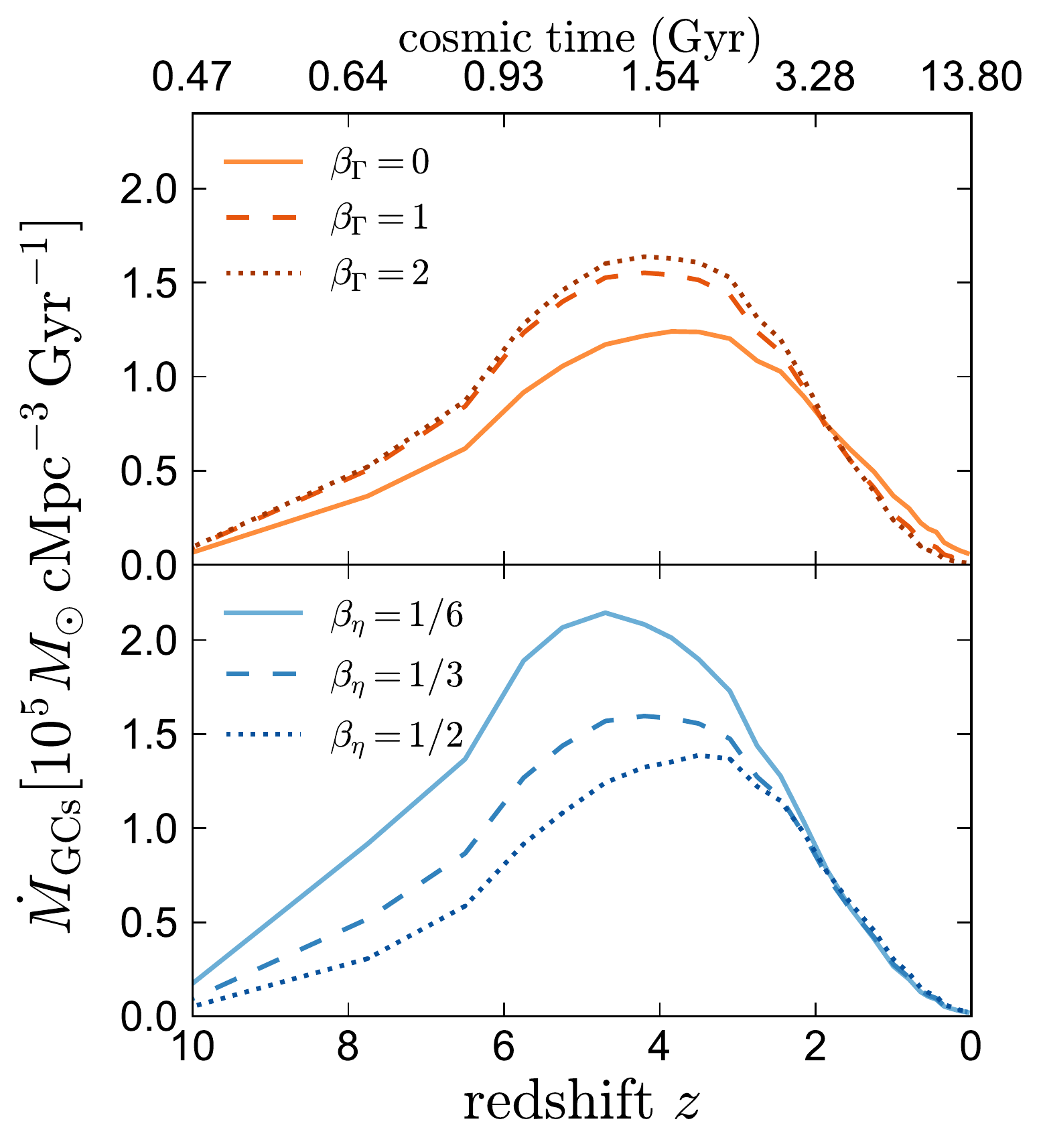}
\caption{Cosmic GC formation rate predicted by our model; i.e., the result of integrating the distributions in Figure~\ref{fig:mdot_gc_vs_mass} over all halo masses. Top panel varies $\beta_{\Gamma}$ while holding $\beta_{\eta}=1/3$ fixed; bottom panel holds $\beta_{\Gamma}=1$ fixed and varies $\beta_{\eta}$. The GC formation rate peaks at $3\lesssim z \lesssim 5$. As a result, GCs in these models contribute only a few percent of the UV luminosity during reionization.}
\label{fig:analytic}
\end{figure}

The total cosmic GC formation rate can be computed by integrating over all halo masses:
\begin{align}
\label{eq:cosmic_mdot_gc}
\dot{M}_{{\rm GCs}}\left(z\right)=\int\frac{{\rm d}\dot{M}_{{\rm GCs}}}{{\rm d}M_{{\rm vir}}}\left(M_{{\rm vir}},z\right)\,{\rm d}M_{{\rm vir}}.
\end{align}
The resulting GC formation rate per comoving volume is shown for three values of $\beta_{\Gamma}$ and $\beta_{\eta}$ in Figure~\ref{fig:analytic}. For typical model choices, the GC formation rate peaks at $z \sim 3-5$. This peak is set primarily by the balance between lower $\Sigma_{\rm GMC}$ at low redshift and a dearth of massive halos at high redshift. The peak moves toward higher $z$ for higher $\beta_{\Gamma}$ or lower $\beta_{\eta}$, both of which cause a larger fraction of GCs to form in low-mass halos at early times. 

Figures~\ref{fig:mdot_gc_vs_mass} and ~\ref{fig:analytic} imply that although the GC formation rate peaked at $z\sim 3-5$, some GCs should continue to form at late times in gas-rich galaxies with high SFRs over a wide range of halo masses. These GCs can be associated with massive star clusters observed forming in the nearby Universe \citep[e.g.][]{Zwart_2010}. The high GC formation rate predicted at $z\gtrsim 2$ is also in agreement with previous works \citep[e.g.][]{Shapiro_2010} that have associated GC formation with the bright star-forming clumps observed in galaxies at $z\sim 2$ \citep{Forster_2009, Forster_2011, Adamo_2013}, or with compact bright sources seen in lensed fields at higher redshifts \citep{Vanzella_2017, Bouwens_2017}.

We note that although the GC formation {\it rate} predicted by our model peaks at $z\sim 3-5$, the most common GC formation redshift is somewhat younger, typically corresponding to $z_{\rm form}\sim 2.5$ (see Figure~\ref{fig:GC_mmr}). There is more time for GCs to form at low redshifts, so the lower formation rate predicted by our model at late times still contributes significantly to the total GC population. Integrated over all formation redshifts, our fiducial model predicts a mean cosmic GC mass density of $\phi_{{\rm GCs}}=5\times10^{5}\,{\rm M_{\odot}\,Mpc^{-3}}$ at $z=0$, as is required to match the observed GC-to-halo mass relation. 

We calculate the contribution of GCs to the cosmic UV luminosity density using  approximations for the time evolution of the UV luminosity at 1500 Angstroms of simple stellar populations from \citet[][their Equations $17-18$]{BoylanKolchin_2017}. At $z=6$, our fiducial model predicts $\rho_{{\rm UV}}=1.95\times 10^{24}\,{\rm erg\,s^{-1}Hz^{-1}\,cMpc^{-3}}$. This is roughly 1\% of the total UV luminosity density found by \citet{Bouwens_2015b} at the same redshift when integrating the luminosity function down to $M_{\rm UV}=-13$. The predicted UV luminosity density due to GCs is less than 2\% of the total cosmic value over $4<z<9$, so in the fiducial model, GCs form too late to contribute substantially to reionization.

\subsection{GC populations}
\label{sec:mass_metallicity}
\subsubsection{Trends with halo mass}

\begin{figure}
\includegraphics[width=\columnwidth]{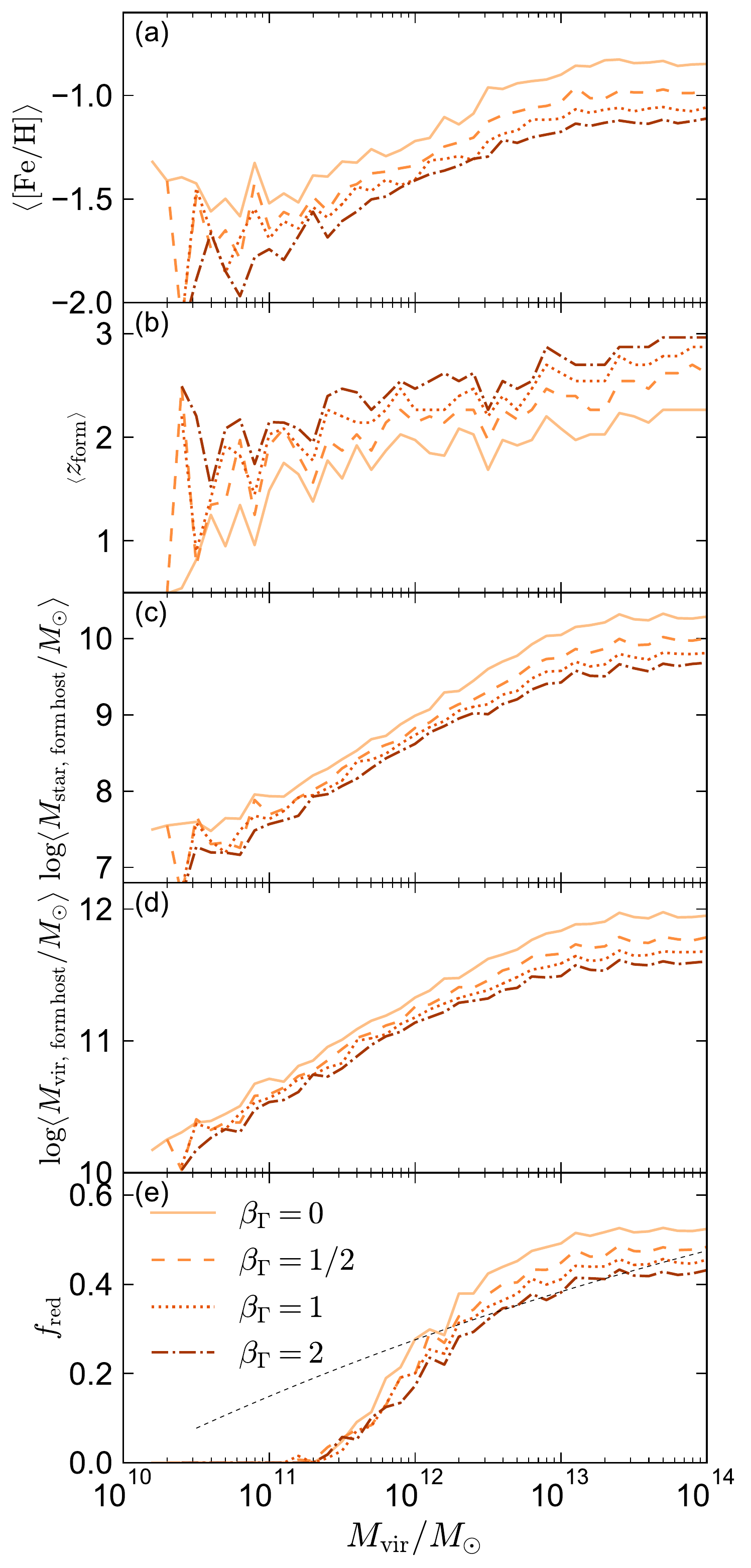}
\caption{Median GC properties as a function of halo mass at $z=0$. {\bf (a)}: GC metallicity. {\bf (b)}: GC formation redshift. {\bf (c)}: stellar mass of the host galaxy in which the GC formed, at the time of the GC's formation. {\bf (d)}: virial mass of the host halo in which the GC formed, at the time of the GC's formation. {\bf (e)}: fraction of GCs that are red ($\rm V-I$ > 1.0 mag); dashed black line shows a fit to observations.}
\label{fig:GC_mmr}
\end{figure}

\begin{figure*}
\includegraphics[width=\textwidth]{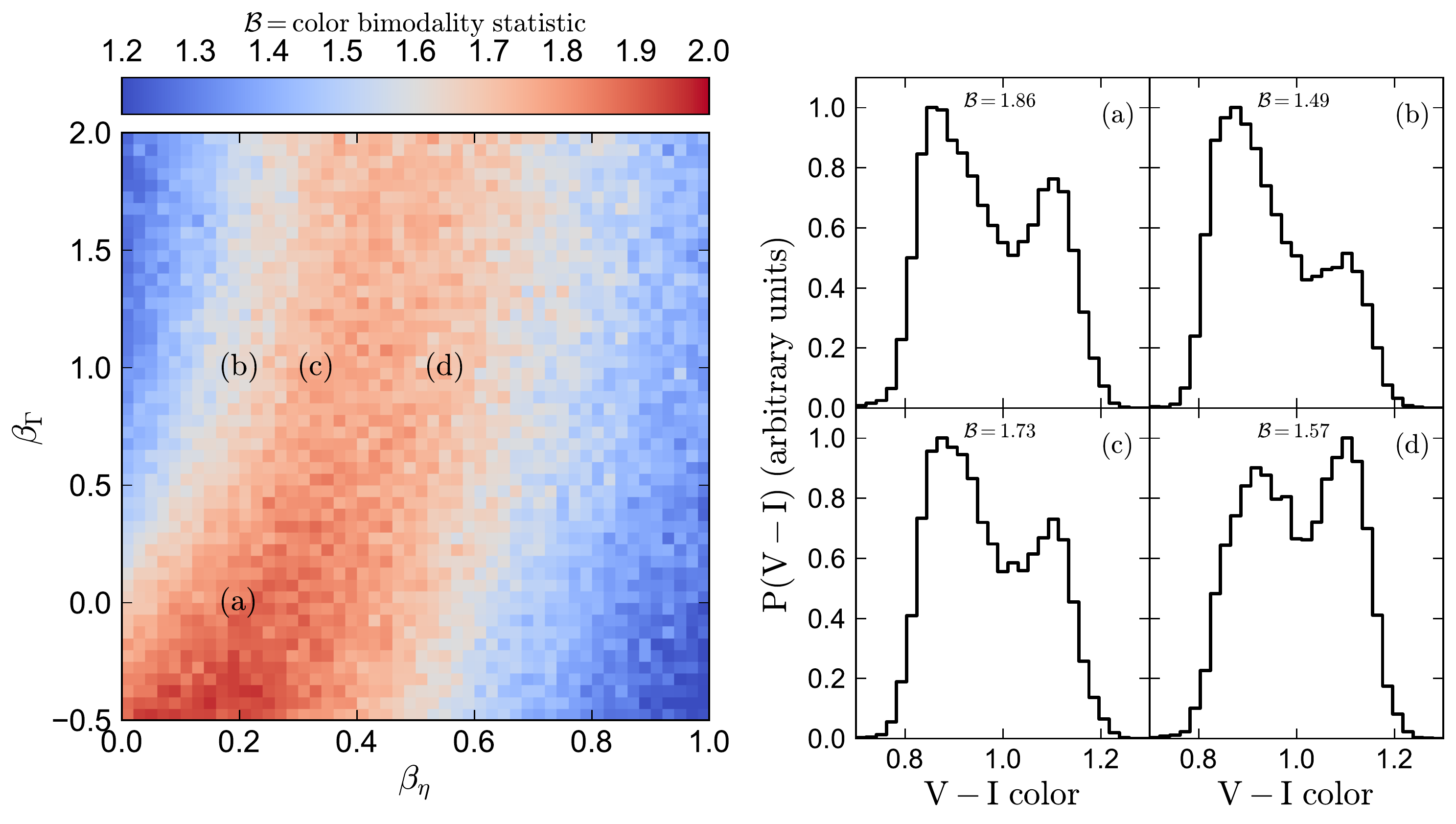}
\caption{{\bf Left}: GC $V-I$ color bimodality statistic (Equation~\ref{eq:bimodality}) for halos with $M_{\rm vir}=10^{13}M_{\odot}.$ We show bimodality values predicted for a wide range of model parameters. $\beta_{\Gamma}$ parameterizes how the cluster formation efficiency varies with $\Sigma_{\rm GMC}$ (Equation~\ref{eq:Gamma}). $\beta_{\eta}$ parameterizes how the mass-loading factor, which relates the SFR and gas accretion rate, varies with $M_{\rm vir}$ (Equation~\ref{eq:eta}). {\bf Right}: Example GC color distributions for different combinations of $\beta_{\eta}$ and $\beta_{\Gamma}$. Our model does not produce a bimodal GC color distribution for $\beta_{\eta}\gtrsim 2/3$, because then too few GCs are produced in low-mass, metal-poor galaxies; however, a wide range of GC formation efficiencies (i.e., different choices of $\beta_{\Gamma}$) predict bimodal GC color distributions at $z=0$.}
\label{fig:bimodality_surface}
\end{figure*}

We now examine how properties of the GC population predicted by our model scale with halo mass. Figure~\ref{fig:GC_mmr} shows the median GC metallicity, formation redshift, and birth galaxy and halo mass, as well as the fraction of GCs that are red. We fix $\beta_{\eta} = 1/3$ in all panels and show predictions for 4 different values of $\beta_{\Gamma}$, corresponding to the cluster formation efficiencies shown in Figure~\ref{fig:sup_funcs}. At fixed halo mass, increasing $\beta_{\Gamma}$ causes the GC population to form in lower-mass halos and become older, bluer, and more metal poor. A higher value of $\beta_{\Gamma}$ limits GC formation to galaxies with higher $\Sigma_{\rm GMC}$, and  $\Sigma_{\rm GMC}$ is on average higher at high $z$. We find that decreasing $\beta_{\eta}$ has qualitatively similar effects to increasing $\beta_{\Gamma}$.

The median metallicity of GC systems (panel a) and stellar and halo masses of GC formation sites (panels c and d), as well as the fraction of GCs that are red (panel e) all increase monotonically with halo mass at $M_{\rm vir} \lesssim 10^{13} M_{\odot}$ and then flatten off at high halo masses. The primary reason for this flattening is that our model suppresses cold gas accretion at high halo masses and late times (see Appendix~\ref{sec:prev_fdbk}). Thus, few GCs form in halos with $M_{\rm vir} \gtrsim 10^{13} M_{\odot}$. The $z=0$ GC populations of these halos consist primarily of GCs that formed in lower-mass halos that subsequently merged. Thus, halos with $M_{\rm vir} \gtrsim 10^{13} M_{\odot}$ all have similar GC population demographics, reflecting the average demographics of the lower-mass progenitors in which most of the GCs formed. The uniformity of GC systems predicted at high halo masses is a consequence of the fact that most GCs form relatively early, before high-mass halos assembled. 

At the highest $z=0$ halo masses, most GCs formed in halos with $M_{\rm vir} \simeq 10^{11-12} M_{\odot}$ and $M_{\rm star} \simeq 10^{9 - 10.5} M_{\odot}$ at the time of GC formation; this is the mass regime in which cold gas accretion is most efficient. Such halos form earlier on average in overdense regions that collapse into cluster-mass halos by $z=0$ than in underdense regions; this causes the median GC formation redshift to increase weakly with halo mass (panel b). 

Our model predicts no red GCs in halos with $M_{\rm vir} \lesssim 10^{11} M_{\odot}$ (panel e). The dashed black line shows a log-linear fit to the $f_{\rm red}$ values for nearby galaxies compiled in \citet{Harris_2015}, highlighting the discrepancy between the observed nonzero occurrence rate of blue GCs in low-mass halos and the predictions of the model. We note that although the observed red fraction is higher than what is predicted by our model at low halo masses, the mean metallicity predicted by our model at the low-mass end is in good agreement with observed values, which find $-2\lesssim\left\langle {\rm \left[Fe/H\right]}\right\rangle \lesssim-1.5$ in the lowest-mass galaxies hosting GCs \citep[e.g.][]{Georgiev_2010, deBoer_2016, Choksi_2018}. Thus, the tension between our model's predictions and observations relates to the large scatter in the colors and metallicities of observed GC systems in low-mass halos.

\subsubsection{GC population bimodality}
\label{sec:bimodality}

\begin{figure*}
\includegraphics[width=\textwidth]{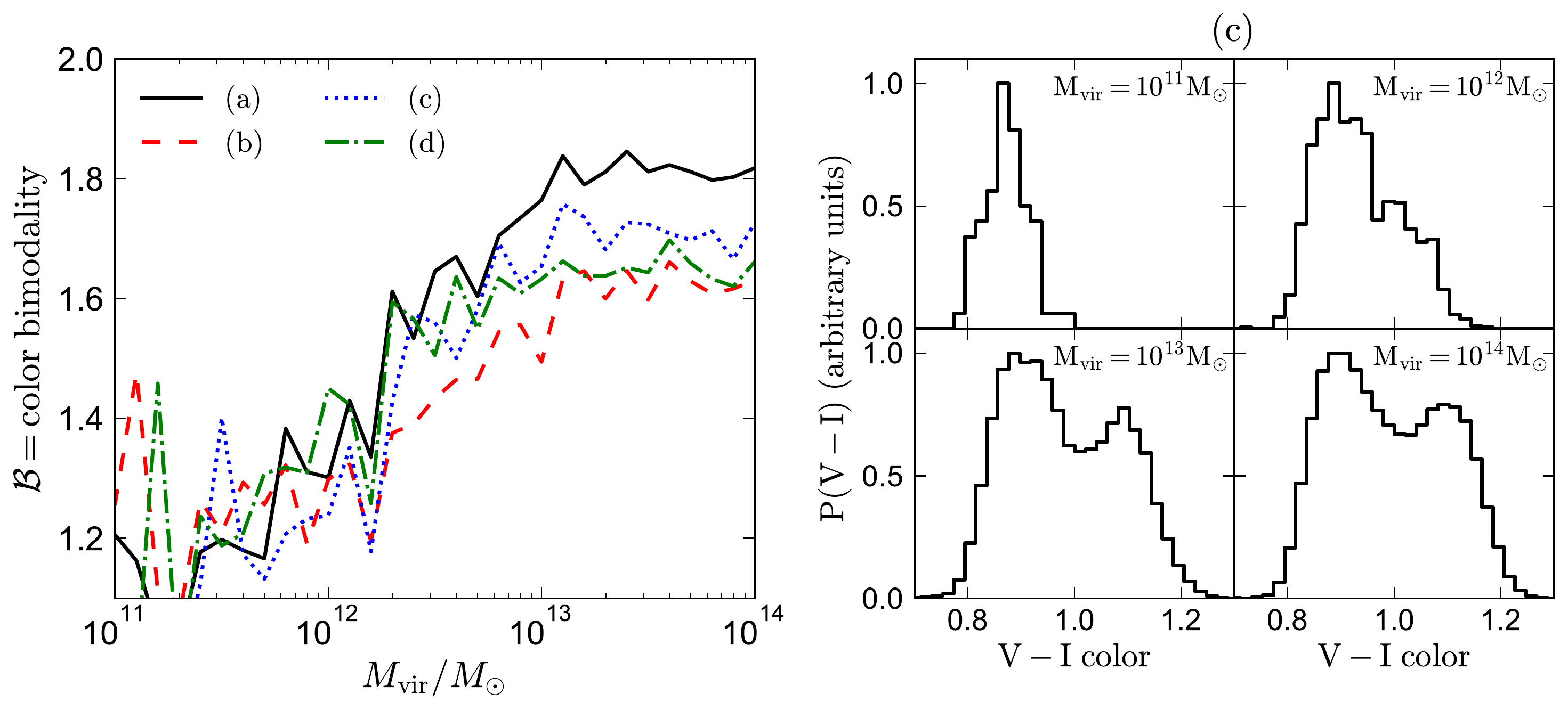}
\caption{{\bf Left}: GC color bimodality (Equation~\ref{eq:bimodality}) vs. host halo mass. Lines show the median of an ensemble of different merger tree realizations and correspond to the same choice of model parameters as in Figure~\ref{fig:bimodality_surface}. Strong color bimodality implies $\mathcal{B}\gtrsim 1.6$. {\bf Right}: GC color distribution for a range of halo masses, all for our fiducial parameters of $\beta_{\eta}=1/3$ and $\beta_{\Gamma}=1$; i.e., line (c).}
\label{fig:bm_vs_mass}
\end{figure*}

The color distributions of the GC populations of many nearby galaxies are bimodal, so GCs are often divided into ``red'' and ``blue'' subpopulations \citep{Ashman_1992, Zepf_1993, Harris_2006, Peng_2006, Brodie_2012}. Color bimodality has been interpreted as indicative of bimodality in GC metallicity \citep[e.g.][]{Brodie_2012} and possibly also GC age (e.g. \citealt{Woodley_2010, Dotter_2011, Leaman_2013}, but see \citealt{Strader_2005}). We now investigate what ranges of model parameters lead to bimodal color distributions in our model.

To quantify bimodality, we introduce a ``bimodality statistic'', $\mathcal{B}$. Given an array of values $x_i$, we define $x_{{\rm upper},\,i}$ and $x_{{\rm lower},\,i}$ as the upper and lower halves of the sorted array. We then compute 
\begin{align}
\label{eq:bimodality}
\mathcal{B}\equiv \frac{{\rm med}\left(x_{{\rm upper},\,i}\right)-{\rm med}\left(x_{{\rm lower},\,i}\right)}{{\rm std}\left(x_{{\rm upper},\,i}\right)+{\rm std}\left(x_{{\rm lower},\,i}\right)}.
\end{align}
$\mathcal{B}$ measures the separation of the ``upper'' and ``lower'' sub-populations relative to their internal dispersion. We find that a clearly bimodal distribution similar to the observed GC color distributions of many giant ellipticals has $\mathcal{B}\sim 1.8$, a marginally bimodal distribution without clear separation between the two peaks has $\mathcal{B}\sim 1.5$, and a Gaussian has $\mathcal{B} = 1.1$. Because the separation between the upper and lower sub-populations always occurs at the median value of the sample (not at a fixed color cut), a high value of $\mathcal{B}$ can only occur when the GC population contains a comparable number of red and blue GCs. To make $\mathcal{B}$ values more stable to stochastic fluctuations, we only compute $\mathcal{B}$ for clusters with $0.65 < {\rm V-I} < 1.35$. This includes the vast majority of GCs formed in our model but excludes GCs younger than $\sim$2 Gyr. 

We explore the range of model parameters $\beta_{\Gamma}$ and $\beta_{\eta}$ that produce a bimodal GC color distribution in Figure~\ref{fig:bimodality_surface}. For each point in $\beta_{\eta}$--$\beta_{\Gamma}$ parameter space, we predict the GC population for 20 merger tree realizations with $M_{\rm vir} = 10^{13} M_{\odot}$ (roughly the mass where the observed color bimodality is most pronounced; e.g. \citealt{Harris_2017b}) and then compute the median $\mathcal{B}$. We assume photometric uncertainties of 0.02 mag.  The color scale in the left panel shows the median $\mathcal{B}$ for 20 merger tree realizations; in the right panels, we show representative color distributions corresponding to several points in $\beta_{\eta}$--$\beta_{\Gamma}$ parameters space that are marked in the left panel. Point (c) corresponds to our fiducial model with $\beta_{\eta}=1/3$ and $\beta_{\Gamma}=1$.

\begin{figure*}
\includegraphics[width=\textwidth]{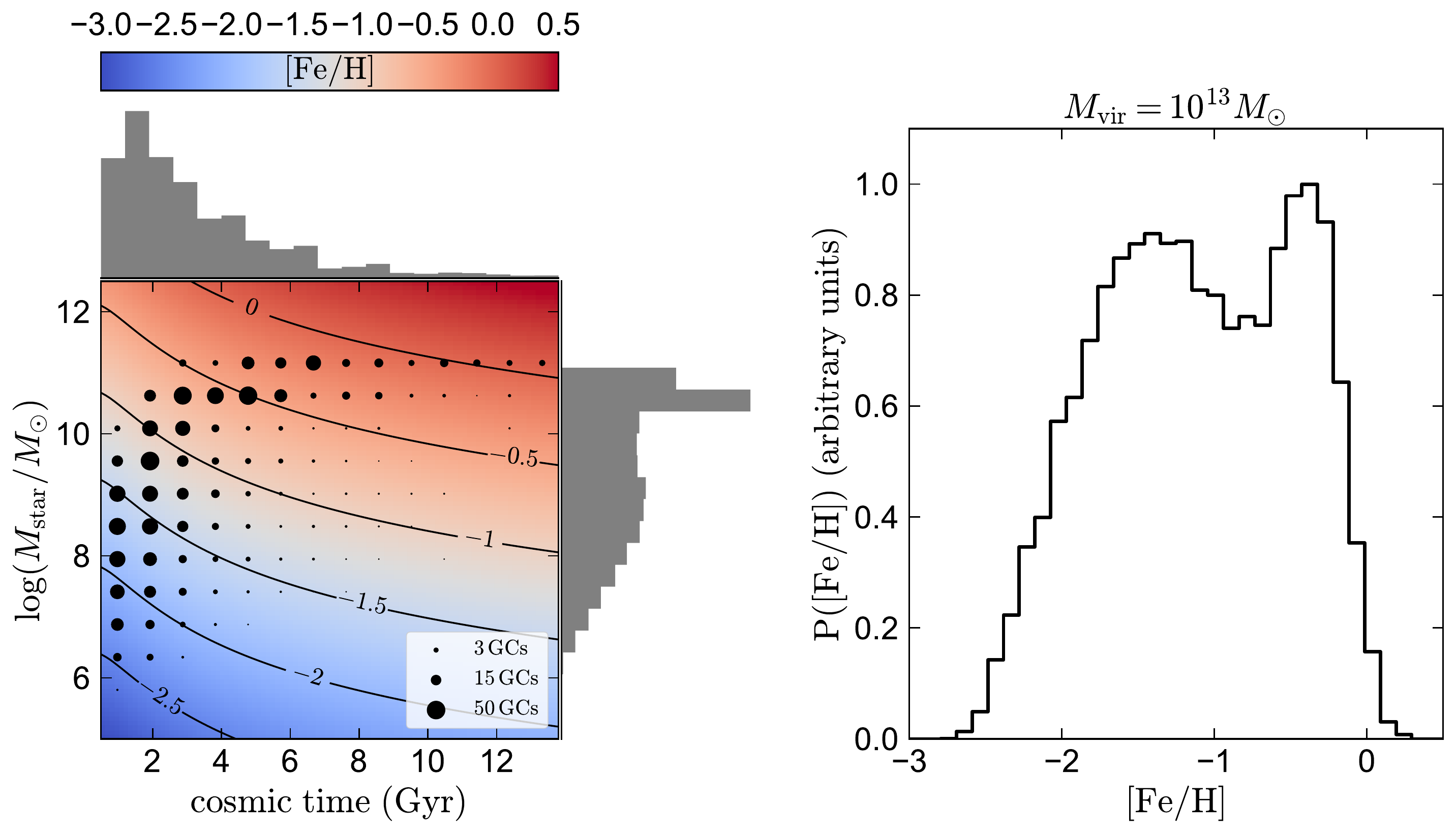}
\caption{{\bf Left}: Color scale shows the gas-phase mass-metallicity relation from \citet{Ma_2016}, which is built into our model. Black circles show the total number of GCs formed in each grid cell; i.e., in galaxies within each $M_{\rm star}$ interval at a given time interval, for a merger tree with $M_{\rm vir} = 10^{13} M_{\odot}$ at $z=0$. The area of each circle scales with the number of GCs formed. GCs form with the metallicity of the galaxy in which they form. {\bf Right}: Metallicity distribution for all GCs formed by $z=0$. Metallicity bimodality is driven primarily by bimodality in $M_{\rm star}$ of the galaxy from which the GC formed.}
\label{fig:bimodality_feh}
\end{figure*}

\begin{figure*}
\includegraphics[width=\textwidth]{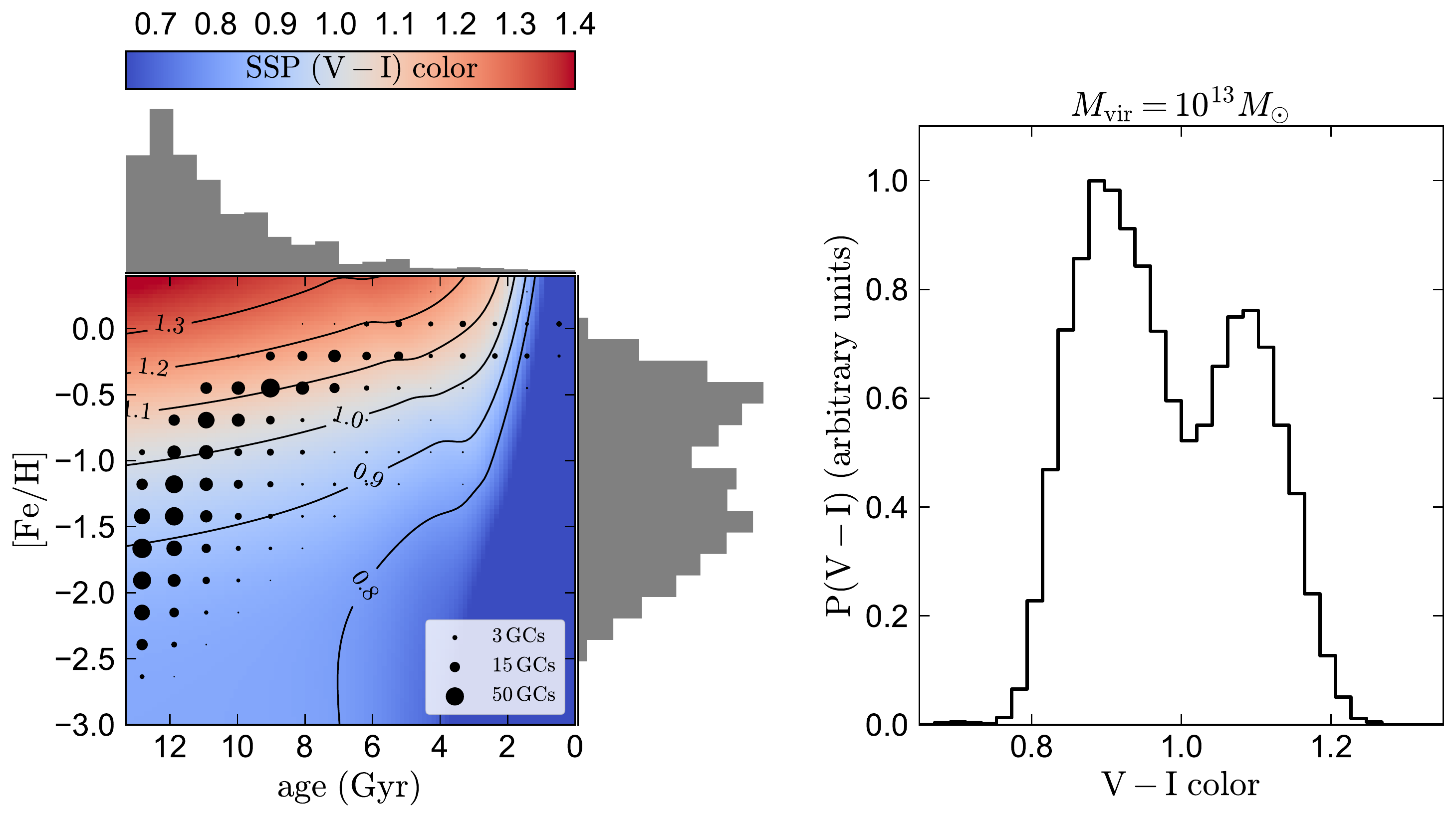}
\caption{{\bf Left}: IMF-integrated color-metallicity-age relation from Padova isochrones adopted by our model. Black circles show the total number of GCs formed in each age-metallicity grid cell for a merger tree with $M_{\rm vir} = 10^{13} M_{\odot}$ at $z=0$. The area of each circle scales with the number of GCs formed. {\bf Right}: Distribution of GC colors at $z=0$. The color bimodality is driven primarily by the metallicitity bimodality.}
\label{fig:bimodality}
\end{figure*}

Consistent with expectations from Figures~\ref{fig:mass_scaling_red_blue} and~\ref{fig:GC_mmr}, the bimodality of the population depends on both $\beta_{\eta}$ and $\beta_{\Gamma}$. A high value of $\beta_{\eta}$ or a low value of $\beta_{\Gamma}$ suppress GC formation at early times in low-mass halos, leading to a unimodal, red GC population. Conversely, low $\beta_{\eta}$ or high $\beta_{\Gamma}$ leads to a unimodal, blue GC population formed in low-mass halos at early times. However, models across a wide swath of $\beta_{\eta}$--$\beta_{\Gamma}$ parameter space produce roughly equal numbers of red and blue GCs, and the right panels show that the GC populations predicted by these models generally have two distinct peaks. 

Figure~\ref{fig:bimodality_surface} thus shows that requiring the $z=0$ GC population to exhibit color bimodality does not strongly constrain the GC formation process, at least in the absences of priors imposed from other observables. This is in some sense unsurprising, since a large number of other semi-analytic GC formation models \citep{Ashman_1992, Cote_1998, Beasley_2002, Tonini_2013, Muratov_2010, Li_2014, Kruijssen_2015, Choksi_2018, Pfeffer_2018} have predicted bimodal GC color distributions while employing a wide range of GC formation prescriptions, mass-metallicity relations, and assumptions regarding the origin of the red and blue GCs. However, some models that predict bimodal GC populations can be ruled out on other grounds. For example, models with $\beta_{\eta} \sim 0$ imply unrealistically metal-poor metallicity distributions for field stars (see Appendix~\ref{sec:stellar_mdf}). Models with $\beta_{\Gamma} \sim 0$ can be excluded because they produce GC populations with the same age and metallicity distribution as field stars.

Figure~\ref{fig:bm_vs_mass} shows how the color distributions predicted by our model vary with halo mass. The right panel shows color distributions for our fiducial model parameters (point (c) in Figure~\ref{fig:bimodality_surface}) for halos of different masses. At $M_{\rm vir} = 10^{11} M_{\odot}$, all GCs are blue. This is simply a consequence of our adopted stellar-to-halo mass and stellar mass-metallicity relations: a halo of mass $M_{\rm vir} = 10^{11} M_{\odot}$ hosts a galaxy with $z=0$ gas-phase metallicity $[\rm Fe/H]_{\rm gas} = -0.8$, which is barely metal-rich enough to form a red GC (see Figure~\ref{fig:bimodality}). Its lower-mass progenitors at higher redshift had even lower metallicities, so without GC self-enrichment or additional scatter in the mass-metallicity relation, there is no possibility of forming a red GC. 

At higher halo masses, red GCs make up an increasing fraction of the total population. A red mode is barely apparent in the GC population predicted for Milky Way-mass halos but is already pronounced at $M_{\rm vir} = 10^{13} M_{\odot}$. The GC populations predicted by our model do not change significantly at $M_{\rm vir} \gg 10^{13} M_{\odot}$, because the GCs in these systems almost all formed in lower-mass halos (see Section~\ref{sec:mass_metallicity}). The left panel of Figure~\ref{fig:bm_vs_mass} shows that this behavior is qualitatively similar across all sets of model parameters $\beta_{\eta}$ and $\beta_{\Gamma}$: the strength of the color bimodality increases with mass up to $M_{\rm vir} = 10^{13} M_{\odot}$ and then flattens off. 

Although the GC color distributions of most observed massive halos can be well-fit by a sum of two Gaussians \citep{Peng_2006, Harris_2016}, the color distributions of some massive systems appear more complex \citep{Strader_2011, Harris_2017b} and have been interpreted as exhibiting either unimodality or trimodality. Likely due to the simplicity of our model and limited sources of scatter, the color distributions we predict at high halo masses are fairly uniform and almost all have two peaks. 

\subsubsection{Origin of bimodality}

Figure~\ref{fig:bimodality_feh} illustrates the origin of the metallicity bimodality for the GC population of a typical massive elliptical galaxy. The left panel shows when and where the GCs in the halo at $z=0$ formed. More than half of the GCs formed in the first $2.5$ Gyr of cosmic history ($z_{\rm form} > 2.6$). These early-forming GCs form primarily in lower-mass galaxies with typical stellar masses of $M_{\rm star} < 10^{10} M_{\odot}$ at $z\sim 2$. On the other hand, most GCs with $z_{\rm form} < 2$ formed in galaxies with $M_{\rm star} > 10^{10} M_{\odot}$ and have $\rm [Fe/H] > -1$. Consistent with the GC age estimates of some works \citep[e.g.][]{Woodley_2010, Dotter_2011, VandenBerg_2013}, our model predicts the metal-rich GCs to be younger than the metal-poor GCs by $\sim$2 Gyr on average. The distribution of GC formation times is not bimodal but has a long tail toward late formation times. The distribution of $M_{\rm star}$ of the host galaxy at the time of formation is marginally bimodal. The distribution of GC metallicities is more strongly bimodal, because at fixed $M_{\rm star}$, later-forming GCs have higher metallicity. This scenario is consistent with the conclusions of \citet{Li_2014}, who identified the redshift-evolution of the galaxy mass-metallicity relation as an important factor in producing bimodal GC metallicity distributions. 

Figure~\ref{fig:bimodality} shows how the distribution of GC ages and metallicities predicted by our model translates to a color distribution at $z=0$. GC color is a stronger function of metallicity than of age for GCs older than $\sim$3\,Gyr. Most red GCs ($\rm V - I > 1$) have $[\rm Fe/H] \gtrsim -1$. For young GCs, color is more strongly dependent on age than on metallicity, but young GCs constitute a negligible fraction of the total GC population in most cases. 

For our fiducial model parameters, massive ellipticals are predicted to have bimodal distributions of both color and metallicity. However, this is not generically true: for some choices of $\beta_{\Gamma}$ and $\beta_{\eta}$, the model predicts single-peaked $[\rm Fe/H]$ distributions while still predicting double-peaked color distributions; see Appendix~\ref{sec:dists}. This can occur because GCs with a range of colors and ages can fall on a line of constant color, such that a unimodal $[\rm Fe/H]$ distribution transforms a bimodal color distribution. Because the GC populations of most giant ellipticals do not have spectroscopic metallicity measurements, some previous works \citep[e.g.][]{Yoon_2006, Richtler_2006} have proposed that effects similar to this are responsible for the observed bimodal color distributions. In the few cases where spectroscopic metallicity measurements are available \citep[e.g.][]{Brodie_2012}, the $[\rm Fe/H]$ distributions do also appear to be bimodal.

\section{Summary and Discussion}
\label{sec:discussion}

\subsection{Summary}
\label{sec:summary}
We have used a semi-analytic model for globular cluster (GC) formation to explore the sensitivity of the observable properties and scaling relations of low-redshift GC populations to details of the GC formation process. Our model uses dark matter merger trees to predict the GC populations of halos at $z=0$, treating GC formation as an extension of normal star formation that occurs at high surface densities. Our primary results are as follows.  

\begin{enumerate}
\item {\it GC system mass -- halo mass relation}: At $z=0$, all the models we consider produce a constant GC-to-halo mass relation at high halo masses, independent of the details of the GC formation model (Figure~\ref{fig:mgc_mhalo}). In fact, a tight GC-to-halo mass relation at $M_{\rm vir} \gtrsim 10^{11.5} M_{\odot}$ is predicted even when we adopt a pathological random model for GC formation in which the GC formation probability is not tied to any properties of the host halo (Figure~\ref{fig:random_model}). This remains true when we add an approximate treatment of GC disruption and mass loss to the model (Appendix~\ref{sec:disrupt} and Figure~\ref{fig:disrupt_effect}). The GC specific frequency predicted by both the fiducial and random models is U-shaped, reflecting the nonlinearity in the stellar-to-halo mass relation (Figure~\ref{fig:spec_freq}).

A constant GC-to-halo mass ratio is predicted for a wide range of models as a result of the central limit theorem. Large halos are formed through mergers of smaller halos, and both the halo masses and GC system masses are summed during mergers. After many mergers, the ratio of total GC mass to halo mass tends to average out, irrespective of the GC-to-halo mass relation when GCs formed. This holds true as long as GCs form relatively early ($z\gtrsim 2$), such that enough mergers occur after the bulk of the GC population forms to drive the population toward the mean relation. GC age constraints from stellar models suggest that most GCs are indeed ancient. We therefore conclude that the observed constant GC-to-halo mass relation does not necessarily imply any fundamental GC-dark matter connection. 

At low halo masses ($M_{\rm vir} \lesssim 10^{11.5} M_{\odot}$ in our fiducial model), mergers alone are insufficient to produce a tight, linear GC-to-halo mass relation. In this regime, the GC-to-halo mass relation predicted by our fiducial model falls below linear, and small number statistics drive up the scatter in the $z=0$ GC-to-halo mass relation in the absence of a correlation between GC and halo mass at the time of GC formation (Figure~\ref{fig:random_model}). The GC populations of low-mass halos thus retain the most information about the physical conditions under which GCs formed.  

\item {\it Cosmic GC formation rate}: Our model predicts the cosmically-averaged GC formation rate to peak at $z\simeq 3-5$ (Figure~\ref{fig:analytic}). Our fiducial model predicts that GCs contributed  1-2\% of the total UV luminosity density during reionization. Most GCs form in halos with $M_{\rm vir} \simeq 10^{10-12} M_{\odot}$, with the typical halo mass hosting GC formation increasing over cosmic time (Figure~\ref{fig:mdot_gc_vs_mass}). Although the integrated GC-to-halo mass relation predicted by the model is constant at $z=0$, the GC formation rate at a particular redshift falls off at low and high halo masses (Figure~\ref{fig:young}), largely due to our input model for the gas accretion rate (Appendix~\ref{sec:prev_fdbk}). Because there is more time for GCs to form at lower redshifts, the median GC formation redshift is $z \sim 2.5$ (Figure~\ref{fig:GC_mmr}). 

\item {\it GC color/metallicity bimodality}: Our model predicts the metallicity (Figure~\ref{fig:bimodality_feh}) and color (Figure~\ref{fig:bimodality}) distributions of GCs in massive galaxies to be bimodal at $z=0$ down to MW-mass halos (Figure~\ref{fig:bm_vs_mass}). The fraction of GCs predicted to be red increases with halo mass, with all GCs in halos with $M_{\rm vir} \lesssim 10^{11} M_{\odot}$ predicted to be blue (Figure~\ref{fig:mass_scaling_red_blue}). Bimodal GC color distributions are predicted for a wide range of model parameters (Figure~\ref{fig:bimodality_surface}). Red, metal-rich GCs are on average younger by $\sim$2 Gyr than blue, metal-poor GCs. The metallicity bimodality predicted by our model arises primarily due to bimodality in the masses of the galaxies in which GCs form and is strengthened by the redshift evolution of the mass-metallicity relation (Figure~\ref{fig:bimodality_feh}). The median formation redshifts of red and blue GCs are $z_{\rm form} \sim 1.9$ and $z_{\rm form} \sim 3.9$, respectively.

\end{enumerate}

\subsection{Discussion: The GC -- dark matter connection}
\label{sec:disc_mgc_mhalo}
Many previous works have proposed causal models for the origin of the constant observed GC-to-halo mass ratio. \citet{Peebles_1968} first suggested that GCs formed immediately following recombination with a characteristic scale set by the cosmological Jeans mass at $z\sim 1000$. \citet{Peebles_1984} revised this model in the context of the CDM paradigm, suggesting that GCs formed in the centers of dark matter minihalos at $z\sim 50-100$ \citep[see also][]{Fall_1985, Rosenblatt_1988}. Considerations of the inefficiency of cooling in primordial gas have pushed the preferred epoch of GC formation in similar, more recent models to $z\sim 10-12$, still in dark matter halos at the highest density peaks \citep{Mashchenko_2005, Moore_2006, Bekki_2008, Spitler_2009, Boley_2009, Moran_2014}. These and other works \citep[e.g.][]{Santos_2003, Bekki_2005} have suggested that GC formation was truncated by reionization at $z\gtrsim 6$, at least for metal-poor GCs. 

Such models are appealing because they explain the uniformity of GCs found in very different environments and because if reionization truncated GC formation at roughly the same time throughout the Universe, they predict the $z=0$ GC system mass to scale with halo mass. However, absolute GC age constraints from stellar models have systematic uncertainties of $\pm$1-2 Gyr \citep[e.g.][]{Chaboyer_2017} and thus cannot distinguish between scenarios in which GCs form at $z\gtrsim 2$ and those in which they form prior to reionization.
We also note that the apparent lack of dark matter halos around observed GCs \citep{Moore_1996, Baumgardt_2009, Conroy_2011, Ibata_2013} poses a challenge for dark matter minihalo GC formation models. 

Irrespective of whether GCs formed in individual dark matter halos or in galactic disks, a number of recent works \citep[e.g.][]{Harris_2013, Hudson_2014, Harris_2015, Harris_2017} have argued that (a) a constant GC-to-halo mass ratio at the time of GC formation implies that GC formation was largely unaffected by feedback from UV radiation, stellar winds, supernovae, and AGN, and (b) the constant GC-to-halo mass ratio observed at $z=0$ implies a constant GC-to-halo mass ratio at the time of formation. 

An alternative interpretation of the GC-to-halo mass relation was proposed by \citet{Kruijssen_2015}. In this model, the total GC mass at $z=0$ is determined primarily by the fraction of GCs that survive a ``rapid destruction'' phase in the disks of high-redshift galaxies. This fraction depends on the stellar mass of the host galaxy at the time of GC formation. Largely by coincidence, the mass-scaling of the stellar-to-halo mass relation and the surviving GC-to-stellar mass relation nearly cancel in this model, such that the GC-to-halo mass ratio after the rapid destruction phase is nearly constant. \citet{Kruijssen_2015} then argues that {\it once a constant GC-to-halo mass relation is established}, it is likely to be preserved and/or strengthened by hierarchical mergers, as was shown explicitly by \citet{BoylanKolchin_2017}.

In contrast to previous work, we find that the existence of a constant GC-to-halo mass ratio at $z=0$ does not imply the existence of such a relation at high redshift: it is predicted by all the models we consider, including the pathological case in which GC formation occurs at random (Figure~\ref{fig:random_model}). If GCs are relatively old and the $z=0$ GC population is viewed as the composite population of GCs formed in progenitor halos and assembled through mergers, no coupling between GCs and dark matter halos is needed to explain the observed relation, at least at high halo masses.\footnote{Our results of course do not {\it rule out} the possibility that GC formation is directly linked to properties of dark matter halos. They do imply, however, that the $z=0$ relation cannot strongly distinguish between different GC formation scenarios.}

The fact that a constant GC-to-halo mass relation is expected due to mergers alone is perhaps most obvious when one considers the GC populations of galaxy clusters. Because galaxy clusters have long dynamical friction timescales, their GC populations are -- unlike those of MW-mass halos -- often dominated by GCs bound to satellites, not to the central galaxy. When only the GCs associated with the central galaxy are accounted for, massive clusters are found to have lower GC system masses than predicted for a constant GC-to-halo mass ratio \citep{Spitler_2009}. On the other hand, massive clusters fall on the observed constant ratio when $M_{\rm GCs}$ also includes both GCs associated directly with the individual member galaxies and intracluster GCs that are not bound to any individual galaxy \citep[see, e.g.][]{Spitler_2009, Peng_2011, Durrell_2014, Harris_2015}. Given that the individual member galaxies in clusters are known to fall on a constant GC-to-halo mass relation and clusters are composed of individual member galaxies (some already tidally destroyed), it follows that the GC population of a whole cluster will have the same GC-to-halo mass ratio as the constituent galaxies. 

The mechanism that enforces a constant GC-to-halo mass ratio in our model does not apply uniquely to GCs: it is expected to create a constant ratio at late times between halo mass and any property that is set at relatively early times and is passed on through mergers. In fact, a non-causal, merger-driven scenario is widely recognized as a plausible explanation for the observed constant black hole-to-bulge mass ratio \citep{Peng_2007, Hirschmann_2010, Jahnke_2011}: because mergers are expected to cause both the bulges and central black holes of merging galaxies to combine, they drive galaxies toward a constant bulge-to-black hole mass ratio. Provided that GCs are not preferentially destroyed during mergers, this scenario is probably {\it more} applicable for GCs than for black holes: while most GCs are unambiguously old, massive black holes grow both by mergers and by accretion of gas at late times \citep[e.g.][]{Kulier_2015}. 

Indeed, it has been shown observationally that the ratio between total GC mass and black hole mass is also constant, independent of galaxy or halo mass \citep{Burkert_2010, Harris_2014}. This fact is not generally interpreted as indicative of a causal connection between GCs and black holes, as both GC and black hole mass are known independently to scale with halo mass 
\citep[e.g.][]{Gnedin_2014, Kruijssen_2015}. Unsurprisingly, such a correlation is also naturally predicted purely due to hierarchical assembly, as noted by \citet{Jahnke_2011}.

Finally, we note that some other properties of observed GC scaling relations hint that they arise in large part due to hierarchical assembly. The observed GC-to-halo mass relation is tighter for blue GCs than for red GCs \citep{Peng_2006, Harris_2015}. Since blue GCs likely formed in lower-mass halos and at earlier times than red GCs, they are expected to have gone through more mergers by $z=0$ than red GCs, providing more opportunity to linearize their GC-to-halo mass relation. Perhaps relatedly, the fraction of GCs that are red is at fixed halo mass higher for late-type galaxies than for early-type galaxies \citep{Harris_2015}. Since early-type galaxies on average have gone through more mergers, one might expect their GC populations to more closely reflect those of the lower-mass galaxies from which they formed. 

Our fiducial model ignores the effects of GC disruption and mass loss. We show in Appendix~\ref{sec:disrupt} that applying an analytic recipe for GC disruption and mass loss does not substantially change the prediction of a constant GC-to-halo mass relation at high halo masses. {\it If} the disruption efficiency varied strongly with halo mass and disruption occurred primarily at late times, after most mergers, then one might expect disruption to change the shape of the $z=0$ GC-to-halo mass relation. However, we think such a scenario unlikely because the strongest GC disruption is expected to occur in the tidal fields of the gas disk from which GCs form, {\it before} mergers liberate GCs from the galaxies in which they formed and deposit them in the halo \citep[e.g.][]{Kruijssen_2012b, Kruijssen_2015}.

\subsubsection{Other scaling relations resulting from hierarchical assembly}
\label{sec:disc_mstar_mhalo}
\begin{figure}
\includegraphics[width=\columnwidth]{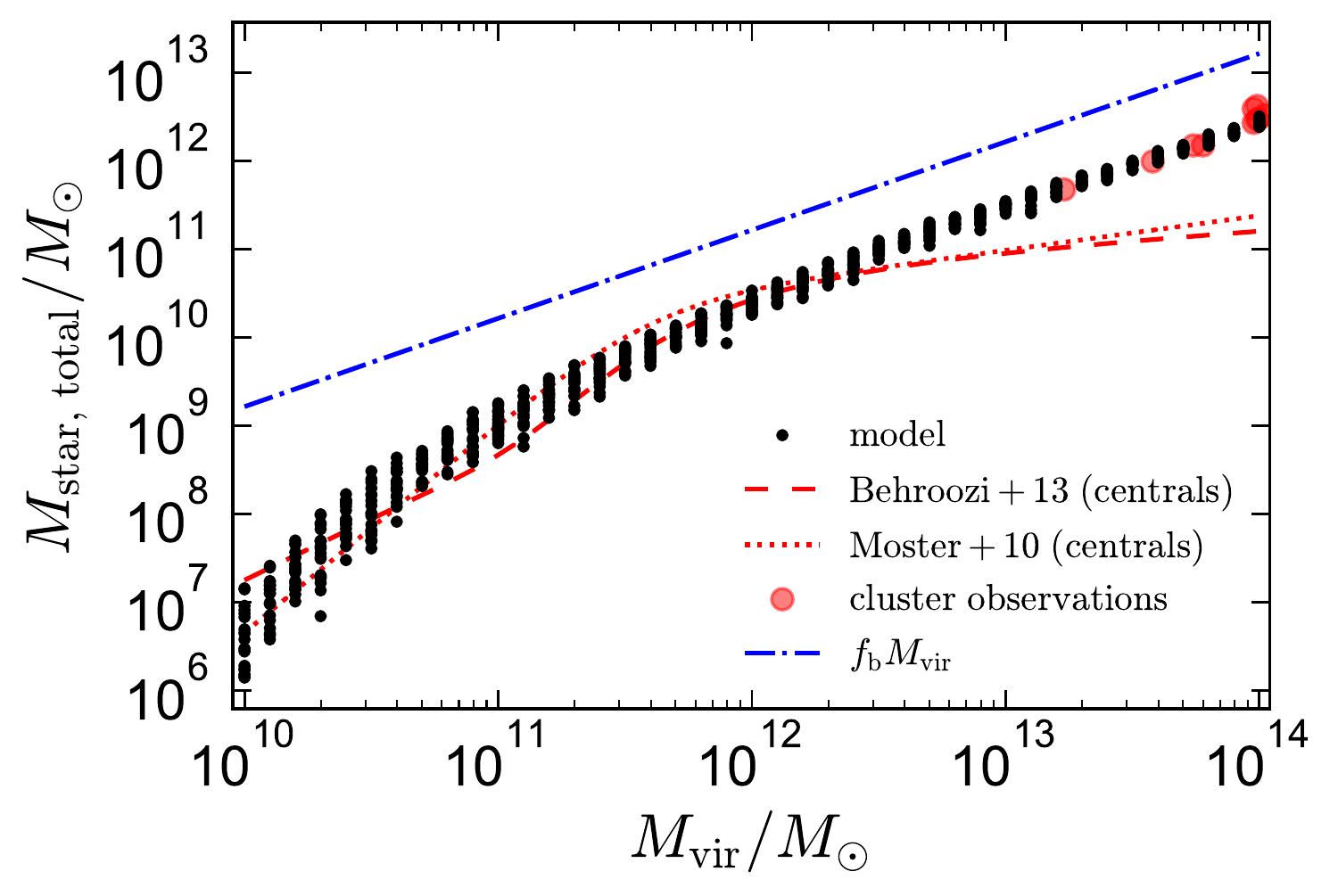}
\caption{Black points show total stellar mass--halo mass relation implied by our model at $z=0$; here $M_{\rm star,\,total}$ represents the sum of all stellar mass within the halo (including satellites and the ICL, which dominate at high halo mass). Red curves show parameterizations of the median stellar mass - halo mass relation for distinct halos (excluding satellites); red points show observations of clusters at $z\sim 0$ (including satellites and the ICL). The relation implied by our simplified model agrees with observational constraints to within a factor of a few. At high halo masses, an almost linear {\it total} stellar-to-halo mass relation is predicted for the same reasons a linear GC-to-halo mass is predicted: most of the stars in clusters are accumulated via mergers.}
\label{fig:mstar_mhalo}
\end{figure}

Although they are not always interpreted as arising due to mergers, other known scaling relations with halo mass may have a similar origin to the GC-to-halo mass relation. The total number of surviving ancient stars in a halo is predicted to scale almost linearly with halo mass (see \citealt{Griffen_2018}; their Figure 10). The same is true for the total stellar mass in groups and clusters (see \citealt{Yang_2007}; their Figure 5). To further illustrate this point, we show in Figure~\ref{fig:mstar_mhalo} the total $z=0$ stellar mass implied by our model as a function of halo mass; i.e., the result of integrating the SFR from Equation~\ref{eq:sfr} over all nodes in the merger tree. Black points correspond to individual merger tree realizations, and the two red lines show stellar-to-halo mass relations {\it for individual galaxies} calculated from abundance matching. 

At the high-mass end, the total stellar mass predicted by our model greatly exceeds the stellar mass predicted by the \citet{Moster_2010} and \citet{Behroozi_2013} relations for central galaxies. This is because the total stellar mass represents not only the stellar mass of the main galaxy, but also the stellar mass of satellite galaxies that have not yet merged with the central galaxy and the mass of stars contributing to the intracluster light (ICL). Because star formation is inefficient in high-mass galaxies and the dynamical friction timescale is long in cluster-mass halos, these components are in fact the dominant contributors to the total stellar mass in massive galaxy clusters, exceeding the mass of the central galaxy by factors of 5-10 \citep{Lin_2004, Yang_2007, Leauthaud_2012,Kravtsov_2018}. To make a fair comparison with our model, we also plot as red hexagons the total stellar mass within a number of intermediate-mass galaxy clusters at low redshift;\footnote{We compile observations from \citet{Leauthaud_2012}, \citet{Gonzalez_2013}, and \citet{Kravtsov_2018}. The points from \citet{Leauthaud_2012} are median values computed for several objects in two mass bins.} these are in good agreement with the predictions of our model.

At the high-mass end, the total stellar-to-halo mass relation in clusters is log-linear for the same reason we predict the GC-to-halo mass relation to be linear. Because a larger fraction of stars than GCs form at late times and star formation is suppressed within massive halos at late times (see Appendix~\ref{sec:prev_fdbk}), the logarithmic slope of the total stellar-to-halo mass relation predicted at high masses is somewhat less than one: we find $M_{{\rm star,\,tot}}\sim M_{{\rm vir}}^{0.9}$. Fitting the data from observed clusters, we find a very similar value. At still-higher masses, $M_{\rm vir} = 10^{14-15} M_{\odot}$, the best-fit exponent is $\sim$0.7 \citep{Vale_2006, Becker_2015, Kravtsov_2018}. 

\subsubsection{The GC-to-halo mass relation in low-mass halos}
\label{sec:low_mass}
\begin{figure}
\includegraphics[width=\columnwidth]{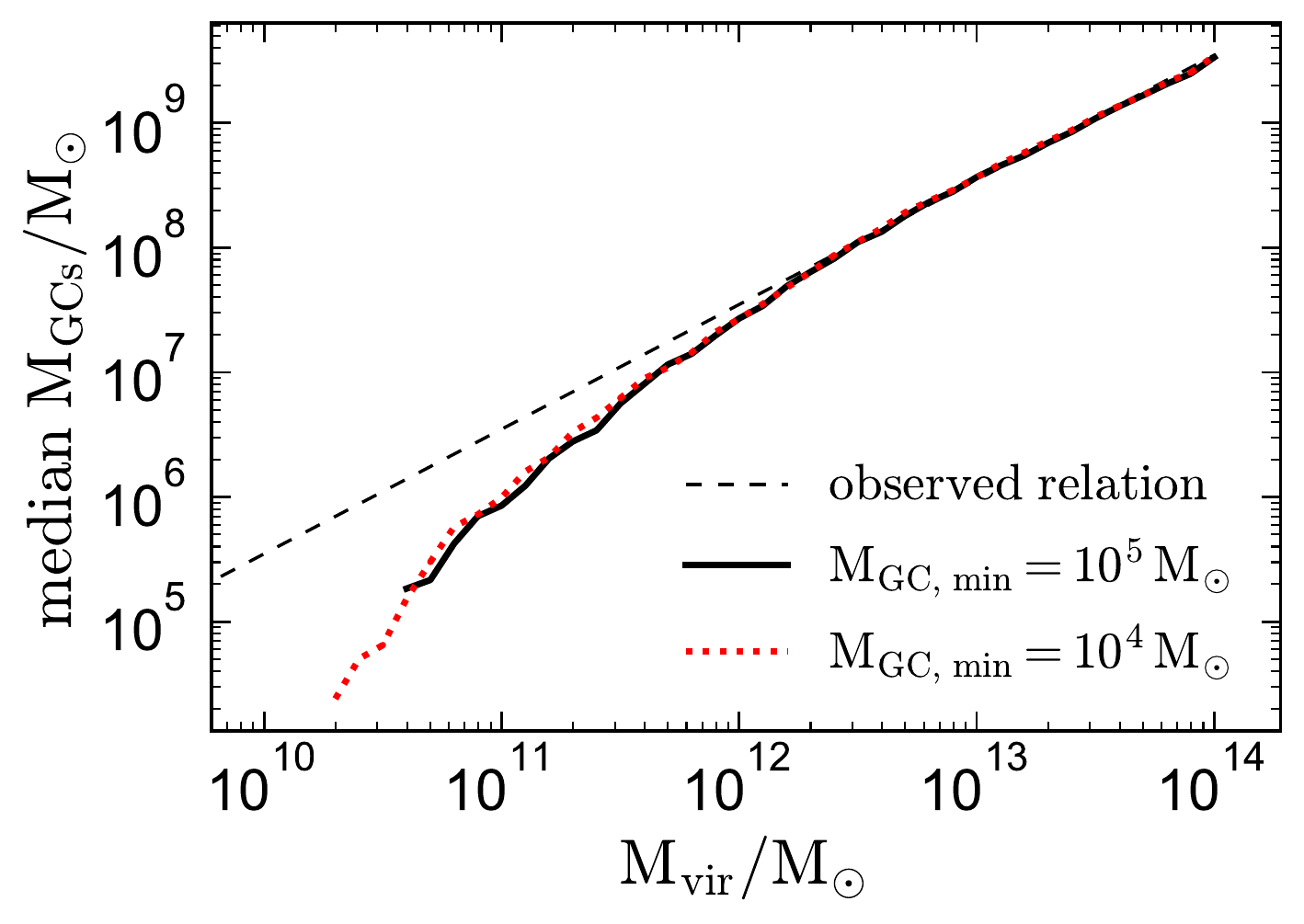}
\caption{Median predicted GC-to-halo mass relation for two choices of the minimum GC mass. Our fiducial model assumes $M_{\rm GC,\,min}=10^5\,M_{\odot}$. Decreasing this value to $10^4 M_{\odot}$ has minimal effects on the GC-to-halo mass relation predicted at high halo masses, but it allows more GCs to form in lower-mass halos, where the fiducial model predicts that most halos will not host any GCs.}
\label{fig:mgc_min}
\end{figure}

A tight, constant GC-to-halo mass ratio cannot be explained purely as a consequence of hierarchical assembly at low halo masses, largely because low-mass halos experience fewer mergers \citep[e.g.][]{Fitts_2018}. The precise scale below which mergers fail to enforce a constant GC-to-halo mass ratio depends on the typical total mass formed per GC formation event, the typical GC formation redshift, and the halo mass limit below which GCs do not form; for both our random and fiducial models, it is $M_{\rm vir} \lesssim 10^{11.5} M_{\odot}$. Below this mass scale, the GC-to-halo mass ratio drops systematically in the fiducial model, and the scatter in the random model increases.

In the fiducial model, most halos with masses below $M_{\rm vir}(z=0) \sim 4 \times 10^{10} M_{\odot}$ do not form any GCs. The gas accretion rates on the progenitors of halos below this mass are sufficiently low, and $\eta$ is sufficiently high, that the implied $\Delta M_{\rm GCs}$ is less than the minimum GC mass we adopt for halos to survive to $z=0$. Some galaxies in halos below this mass {\it are} observed to host GCs \citep{deBoer_2016, Georgiev_2010}, and there are some indications that the observed GC-to-halo mass ratio remains constant down to a few $\times\,10^{10} M_{\odot}$ \citep{Hudson_2014, Zaritsky_2016, Harris_2017}. This may indicate that GC formation is more efficient at early times than is predicted by our model, or that there is a causal origin of the GC-to-halo relation at this mass scale. 

However, substantial uncertainties remain in the observed GC-to-halo mass relation at low masses. At least some dwarf galaxies may deviate strongly from a linear relation \citep[e.g.][]{Lim_2018, Amorisco_2018b, vanDokkum_2018}, and the GC-to-halo mass ratio at $M_{\rm vir} \lesssim 10^{11} M_{\odot}$ appears to be systematically higher-than-average for dwarf ellipticals and lower-than-average for dwarf spheroidals \citep{Spitler_2009, Georgiev_2010}. Measurements of the GC-to-halo mass ratio at low halo masses are also complicated by the fact that most of the lowest-mass observed GC systems are hosted by satellite galaxies, whose halos may have undergone significant tidal stripping. 

Recently, \citet{Forbes_2018} found that the observed mean GC-to-halo mass ratio remains constant down to at least $M_{\rm vir} = 10^9 M_{\odot}$, though the scatter increases substantially at low halo masses. This result is inconsistent with our fiducial model, which predicts a decrease in the GC-to-halo mass ratio below $M_{\rm vir} = 10^{11.5} M_{\odot}$. The increased scatter found at low masses is consistent with what is expected if the constant GC-to-halo mass ratio at higher masses arises from the central limit theorem in hierarchical assembly, though we note that observational measurements of halo mass are also more uncertain at low masses.

A potential concern is that the decrease in $M_{\rm GCs}$ predicted by our model at low masses is an artifact of the minimum GC mass we adopt in sampling GC masses. We test this possibility in Figure~\ref{fig:mgc_min}, which compares the GC-to-halo mass relations predicted for two choices of the minimum GC mass. A lower minimum GC mass allows some GCs to form when the expectation value of the GC mass formed in an accretion event is $\Delta M_{\rm GCs} \ll 10^5 M_{\odot}$. This leads to a lower minimum halo mass for hosting a GC, but does not significantly change the GC-to-halo mass relation at $M_{\rm vir}>5\times 10^{10}\,M_{\odot}$.

The fiducial model's prediction of a drop in the GC-to-halo mass relation at low mases is thus not primarily a consequence of the procedure for sampling GC masses. We have also verified that it is not sensitive to merger tree resolution: it results from the lower gas surface densities and higher mass loading factors predicted for low-mass galaxies in our model. Substantial uncertainties remain in observational measurements of halo mass at the low-mass end. If the observed GC-to-halo mass relation is confirmed to remain constant at low masses, it will imply that our fiducial model assumptions break down at low halo masses. In the context of our fiducial model, a constant GC-to-halo mass ratio at low masses could result from a higher GC formation efficiency, $\Gamma_{\rm GCs}$, in low-mass halos, or a higher GC survival probability in low-mass galaxies.

\subsection{Low GC formation efficiency}
\label{sec:alpha_gamma}
In order for our model to match the normalization of both the GC-to-halo mass relation and the stellar-to-halo mass relations, the coefficient $\alpha_{\Gamma}$ (Equation~\ref{eq:Gamma}), which determines the fraction of star formation that occurs in GC progenitors at asymptotically high surface density, must be quite small, with $\alpha_{\Gamma} \approx 2.1 \times 10^{-3}$ in the fiducial model. Such a low value of $\Gamma_{\rm GCs}$ is somewhat unexpected, since idealized simulations predict the total cluster formation efficiency, $\Gamma$, (which represents the fraction of star formation that occurs in any bound clusters, not only GC progenitors), to asymptote to a value of order unity at high surface density \citepalias{Grudic_2016}. We consider some possible explanations for this discrepancy below. 

\subsubsection{GC disruption}
Our default model does not include GC disruption or stripping. Disruption and mass loss may significantly reduce the total GC mass at $z=0$ relative to the total GC mass formed, increasing the value of $\alpha_{\Gamma}$ required to match observations. However, the fraction of the total GC mass formed that survives until $z=0$ is highly uncertain, as the efficiency of GC disruption depends sensitively both on the redshift at which GCs form \citep[e.g.][]{Katz_2014, Carlberg_2017} and on the spatial distribution of GCs within their host halos \citep{Chernoff_1990, Gieles_2008, Kruijssen_2009, Kruijssen_2012b, Kruijssen_2015, Pfeffer_2018}. 

The simplified model for disruption and evaporation that we test in Appendix~\ref{sec:disrupt} leads to a significant but not overwhelming reduction in the total GC mass, requiring an increase in $\alpha_{\Gamma}$ by a factor of $\sim$2.6. Some other models, particularly those attempting to explain observations of anomalous abundances in GCs as the result of enrichment from a population of stars that was preferentially stripped at late times, invoke much higher disruption efficiencies, typically requiring a reduction in bound GC mass between formation and $z=0$ by factors of 10-100 \citep{DErcole_2008, Conroy_2011b, Conroy_2012}. Combined with the other factors discussed below, such efficient disruption could bring the value of $\alpha_{\Gamma}$ required to match the observed GC-to-halo mass relation into agreement with the expectation from idealized simulations. We note, however, that the trends predicted by self-enrichment models between the fraction of GC stars with anomalous abundances and other cluster properties are generally not observed \citep{Bastian_2015}. If GCs {\it were} much more massive when they formed than they are today, then young GCs likely contribute significantly to the faint end of the high-$z$ luminosity function \citep{Bouwens_2017,BoylanKolchin_2017b}. Future observations with {\it JWST} will be able to test such a scenario. Searches for disrupted GC stars in the MW bulge and stellar halo \citep[e.g.][]{Martell_2016, Schiavon_2017} can also place limits on the efficiency of GC disruption. 

\subsubsection{Contributions from lower-mass clusters}
Because low-mass clusters ($m\lesssim 10^{5} M_{\odot}$ at birth) are expected to be strongly affected by two-body evaporation over a Hubble time \citep{Spitzer_1987, Fall_2001, Prieto_2008, Muratov_2010, Gieles_2011}, we do not consider them as possible progenitors of $z=0$ GCs. However, the total cluster formation efficiency $\Gamma$ {\it does} include bound lower-mass clusters, so we generically expect $\Gamma$ to exceed $\Gamma_{\rm GCs}$. For a ${\rm d}n/{\rm d}m \sim m^{-2}$ initial cluster mass function, each decade of cluster mass contributes equal total mass, so the total mass formed in all clusters is expected to exceed the mass formed in clusters sufficiently massive to survive until $z=0$ by a factor of a few. The shape of the initial cluster mass function is imperfectly understood \citep[e.g.][]{Vesperini_2003, Parmentier_2005, Elmegreen_2010}. If GCs that survive until $z=0$ represent the objects near the high-mass cutoff of a Schechter-type mass function, the mass fraction in since-disrupted lower-mass clusters could be much higher. 

\subsubsection{Additional requirements for bound cluster formation}
In our model, the fraction of star formation that occurs in bound clusters depends only on the local gas surface density, $\Sigma_{\rm GMC}$. The low value of $\alpha_{\Gamma}$ that we find needed to match observations with this model may indicate that other conditions must be met for the formation of GC progenitors, such that our current model based on surface density alone overestimates the fraction of star formation that occurs in massive bound clusters at fixed surface density. For example, the fraction of stars forming in bound clusters likely also depends to some degree on factors such as cloud geometry, the cloud virial parameter, the local tidal field, the amount of shear from neighboring clouds, the Toomre Q parameter, and the gas metallicity \citep{Krumholz_2005, Howard_2016, Kruijssen_2012a}. 

It is also possible that our model underestimates the true critical density for GC formation, $\Sigma_{\rm crit} = 3000\,{\rm M_{\odot}\,pc^{-2}}$, or equivalently, that our merger tree gas calculations overestimate $\Sigma_{\rm GMC}$. Increasing the value of $\Sigma_{\rm crit}$ does increase the value of $\alpha_{\Gamma}$ needed to match the observed GC-to-halo mass ratio for any positive $\beta_{\Gamma}$ (see Appendix~\ref{sec:sig_crit}). However, for the fiducial value of $\beta_{\Gamma} = 1$, it is necessary to increase $\Sigma_{\rm crit}$ by two orders of magnitude in order to produce an order-unity value of $\alpha_{\Gamma}$. Because idealized cloud-collapse calculations find values of $\Sigma_{\rm crit}$ similar to our fiducial value (or lower; e.g. \citealt{Raskutti_2016}), we regard a much-higher value of $\Sigma_{\rm crit}$ as unlikely.

\subsection{Limitations of the model}
\label{sec:limitations}
\subsubsection{Treatment of mergers}
Our model treats GC formation as the high-surface density extension of normal star formation. Because the gas surface density in our model is directly linked to the mass accretion rate calculated from merger trees, a large fraction of GCs form following major mergers  (See Figure~\ref{fig:Sigma_hists}). This prediction is consistent with observations of massive clusters forming in nearby major mergers \citep[e.g.][]{Wilson_2006, Bastian_2009}, and with simulations that find massive bound clusters to form during major mergers \citep[e.g.][]{Bournaud_2008}. Major mergers are explicitly modeled as the dominant site of GC formation in several semi-analytic models \citep{Ashman_1992, Muratov_2010,Li_2014,Choksi_2018}.

In our model, the increased star and GC formation following major mergers is simply a result of the increased inflow rate. However, major mergers are also known to produce large-scale torques that can compress gas and drive it toward the galactic center \citep[e.g.][]{Mihos_1996, Hopkins_2013}, potentially amplifying the subsequent burst of star formation. Our model is designed to approximate this phenomenologically, not to model the details of star or GC formation with high fidelity.

A related limitation is that equilibrium models may be less applicable at high redshift, when galaxies may be in a ``gas accumulation phase'' and cannot process gas as fast as they receive it \citep[e.g.][]{Krumholz_2012, Dave_2012}. Following the arguments of \citet[][their Equation 15]{Dave_2012}, we find that more than 95\% of GCs form after the redshift $z_{\rm eq}$ when galaxies reach equilibrium on average. However, galaxies can  depart from equilibrium at later times during major mergers.\footnote{We estimate that galaxies will depart from equilibrium and temporarily accumulate gas when the SFR implied by the equilibrium model exceeds the gas consumption rate, i.e., when $\dot{M}_{{\rm gas,\,in}}/\left(1+\eta\right)>0.02M_{{\rm gas}}/t_{{\rm dyn}}$, where the gas consumption rate is based on estimates from the local Universe. During such periods, the SFR is likely somewhat lower than predicted by the equilibrium model.} We find that roughly half of the GCs in our model form during such periods. Exploring the effects of a gas accumulation phase on GC formation is a possible avenue for future work. 

\subsubsection{Merger timescale and dynamical friction}
In our model, star and GC formation events corresponding to a particular merger event occur immediately following the merger in the merger tree, on a timescale set by the galaxy dynamical time (Section~\ref{sec:timescale}). In reality, accreted objects will orbit within their host halos as satellites before spiraling to the center of the main halo \citep[e.g.][]{Lacey_1993, Cole_2000, Boylan_Kolchin_2008}. The dynamical friction timescale scales with the dynamical time of the main halo (which is shorter at high redshift) and with the mass ratio of the accreted halo to the primary halo. The majority of GCs in our model form at early times and in major mergers. In these cases, the dynamical friction timescale is short, minimizing the error caused by treating the initial inspiral as instantaneous. A small fraction of GCs in our model do form at late times in halos in which the dynamical friction timescale is of order the Hubble time; in these cases, the  model may overpredict the true SFR and GC formation rate. In a model similar to the one introduced in this work, \citet{Li_2014} tested the effects of including an analytic model for dynamical friction. They found that including dynamical friction typically reduced the number of GCs at $z=0$ by 5-10\%. 

It {\it is} appropriate to account for the GCs of potentially unmerged satellites (which only contribute a large fraction of the total GC mass in cluster-mass halos) when calculating $M_{\rm GCs}$. Observational studies that find a constant GC-to-halo mass ratio at high halo masses also include satellite GC populations, either by direct counting or by extrapolating the GC number density profile to large radii. 

\section*{Acknowledgements}
We are grateful to Yu Lu for making public his implementation of the Parkinson \& Cole merger tree algorithm.
We thank the anonymous referee for constructive comments that improved the paper.
We also thank Mike Fall, Mike Grudic, Charlie Conroy, Chris McKee, Joss Bland-Hawthorn, Aldo Rodriguez-Puebla, and Rohan Naidu for helpful discussions. Ideas for this project were developed in part at the Near/Far Globular Cluster Workshop in Napa in December 2017.
KE and DRW are grateful for the hospitality provided by the MPIA in Heidelberg during the writing of this paper.
KE acknowledges support from an NSF Graduate Research Fellowship. 
EQ and KE are supported by a Simons Investigator Award from the Simons Foundation and by NSF grant AST-1715070.
DRW is supported by fellowships provided by the Alfred P. Sloan Foundation and the Alexander von Humboldt Foundation.
MBK acknowledges support from NSF grant AST-1517226 and CAREER grant AST-1752913 and from NASA grants NNX17AG29G and HST-AR-13888, HST-AR-13896, HST-AR-14282, HST-AR-14554, HST-AR-15006, HST-GO-12914, and HST-GO-14191 from STScI. 
The analysis in this paper relied on the python packages \texttt{NumPy} \citep{vanderwalt_2011}, \texttt{Matplotlib} \citep{Hunter_2007}, and \texttt{AstroPy} \citep{Astropy_2013}.




\bibliographystyle{mnras}

\begin{thebibliography}{}
\makeatletter
\relax
\def\mn@urlcharsother{\let\do\@makeother \do\$\do\&\do\#\do\^\do\_\do\%\do\~}
\def\mn@doi{\begingroup\mn@urlcharsother \@ifnextchar [ {\mn@doi@}
  {\mn@doi@[]}}
\def\mn@doi@[#1]#2{\def\@tempa{#1}\ifx\@tempa\@empty \href
  {http://dx.doi.org/#2} {doi:#2}\else \href {http://dx.doi.org/#2} {#1}\fi
  \endgroup}
\def\mn@eprint#1#2{\mn@eprint@#1:#2::\@nil}
\def\mn@eprint@arXiv#1{\href {http://arxiv.org/abs/#1} {{\tt arXiv:#1}}}
\def\mn@eprint@dblp#1{\href {http://dblp.uni-trier.de/rec/bibtex/#1.xml}
  {dblp:#1}}
\def\mn@eprint@#1:#2:#3:#4\@nil{\def\@tempa {#1}\def\@tempb {#2}\def\@tempc
  {#3}\ifx \@tempc \@empty \let \@tempc \@tempb \let \@tempb \@tempa \fi \ifx
  \@tempb \@empty \def\@tempb {arXiv}\fi \@ifundefined
  {mn@eprint@\@tempb}{\@tempb:\@tempc}{\expandafter \expandafter \csname
  mn@eprint@\@tempb\endcsname \expandafter{\@tempc}}}

\bibitem[\protect\citeauthoryear{{Adamo}, {{\"O}stlin}, {Bastian},
  {Zackrisson}, {Livermore}  \& {Guaita}}{{Adamo} et~al.}{2013}]{Adamo_2013}
{Adamo} A.,  {{\"O}stlin} G.,  {Bastian} N.,  {Zackrisson} E.,  {Livermore}
  R.~C.,   {Guaita} L.,  2013, \mn@doi [\apj] {10.1088/0004-637X/766/2/105},
  \href {http://adsabs.harvard.edu/abs/2013ApJ...766..105A} {766, 105}

\bibitem[\protect\citeauthoryear{{Amorisco}}{{Amorisco}}{2018}]{Amorisco_2018}
{Amorisco} N.~C.,  2018, preprint, \href
  {http://adsabs.harvard.edu/abs/2018arXiv180200812A} {} (\mn@eprint {arXiv}
  {1802.00812})

\bibitem[\protect\citeauthoryear{{Amorisco}, {Monachesi}, {Agnello}  \&
  {White}}{{Amorisco} et~al.}{2018}]{Amorisco_2018b}
{Amorisco} N.~C.,  {Monachesi} A.,  {Agnello} A.,   {White} S.~D.~M.,  2018,
  \mn@doi [\mnras] {10.1093/mnras/sty116}, \href
  {http://adsabs.harvard.edu/abs/2018MNRAS.475.4235A} {475, 4235}

\bibitem[\protect\citeauthoryear{{Ashman} \& {Zepf}}{{Ashman} \&
  {Zepf}}{1992}]{Ashman_1992}
{Ashman} K.~M.,  {Zepf} S.~E.,  1992, \mn@doi [\apj] {10.1086/170850}, \href
  {http://adsabs.harvard.edu/abs/1992ApJ...384...50A} {384, 50}

\bibitem[\protect\citeauthoryear{{Astropy Collaboration} et~al.,}{{Astropy
  Collaboration} et~al.}{2013}]{Astropy_2013}
{Astropy Collaboration} et~al., 2013, \mn@doi [\aap]
  {10.1051/0004-6361/201322068}, \href
  {http://adsabs.harvard.edu/abs/2013A%26A...558A..33A} {558, A33}

\bibitem[\protect\citeauthoryear{{Bailin} \& {Harris}}{{Bailin} \&
  {Harris}}{2009}]{Bailin_2009}
{Bailin} J.,  {Harris} W.~E.,  2009, \mn@doi [\apj]
  {10.1088/0004-637X/695/2/1082}, \href
  {http://adsabs.harvard.edu/abs/2009ApJ...695.1082B} {695, 1082}

\bibitem[\protect\citeauthoryear{{Bastian} \& {Goodwin}}{{Bastian} \&
  {Goodwin}}{2006}]{Bastian_2006}
{Bastian} N.,  {Goodwin} S.~P.,  2006, \mn@doi [\mnras]
  {10.1111/j.1745-3933.2006.00162.x}, \href
  {http://adsabs.harvard.edu/abs/2006MNRAS.369L...9B} {369, L9}

\bibitem[\protect\citeauthoryear{{Bastian} \& {Lardo}}{{Bastian} \&
  {Lardo}}{2015}]{Bastian_2015}
{Bastian} N.,  {Lardo} C.,  2015, \mn@doi [\mnras] {10.1093/mnras/stv1661},
  \href {http://adsabs.harvard.edu/abs/2015MNRAS.453..357B} {453, 357}

\bibitem[\protect\citeauthoryear{{Bastian} \& {Lardo}}{{Bastian} \&
  {Lardo}}{2017}]{Bastian_2017}
{Bastian} N.,  {Lardo} C.,  2017, preprint, \href
  {http://adsabs.harvard.edu/abs/2017arXiv171201286B} {} (\mn@eprint {arXiv}
  {1712.01286})

\bibitem[\protect\citeauthoryear{{Bastian}, {Trancho}, {Konstantopoulos}  \&
  {Miller}}{{Bastian} et~al.}{2009}]{Bastian_2009}
{Bastian} N.,  {Trancho} G.,  {Konstantopoulos} I.~S.,   {Miller} B.~W.,  2009,
  \mn@doi [\apj] {10.1088/0004-637X/701/1/607}, \href
  {http://adsabs.harvard.edu/abs/2009ApJ...701..607B} {701, 607}

\bibitem[\protect\citeauthoryear{{Baumgardt} \& {Kroupa}}{{Baumgardt} \&
  {Kroupa}}{2007}]{Baumgardt_2007}
{Baumgardt} H.,  {Kroupa} P.,  2007, \mn@doi [\mnras]
  {10.1111/j.1365-2966.2007.12209.x}, \href
  {http://adsabs.harvard.edu/abs/2007MNRAS.380.1589B} {380, 1589}

\bibitem[\protect\citeauthoryear{{Baumgardt}, {C{\^o}t{\'e}}, {Hilker},
  {Rejkuba}, {Mieske}, {Djorgovski}  \& {Stetson}}{{Baumgardt}
  et~al.}{2009}]{Baumgardt_2009}
{Baumgardt} H.,  {C{\^o}t{\'e}} P.,  {Hilker} M.,  {Rejkuba} M.,  {Mieske} S.,
  {Djorgovski} S.~G.,   {Stetson} P.,  2009, \mn@doi [\mnras]
  {10.1111/j.1365-2966.2009.14932.x}, \href
  {http://adsabs.harvard.edu/abs/2009MNRAS.396.2051B} {396, 2051}

\bibitem[\protect\citeauthoryear{{Beasley}, {Baugh}, {Forbes}, {Sharples}  \&
  {Frenk}}{{Beasley} et~al.}{2002}]{Beasley_2002}
{Beasley} M.~A.,  {Baugh} C.~M.,  {Forbes} D.~A.,  {Sharples} R.~M.,   {Frenk}
  C.~S.,  2002, \mn@doi [\mnras] {10.1046/j.1365-8711.2002.05402.x}, \href
  {http://adsabs.harvard.edu/abs/2002MNRAS.333..383B} {333, 383}

\bibitem[\protect\citeauthoryear{{Beasley}, {Trujillo}, {Leaman}  \&
  {Montes}}{{Beasley} et~al.}{2018}]{Beasley_2018}
{Beasley} M.~A.,  {Trujillo} I.,  {Leaman} R.,   {Montes} M.,  2018, \mn@doi
  [\nat] {10.1038/nature25756}, \href
  {http://adsabs.harvard.edu/abs/2018Natur.555..483B} {555, 483}

\bibitem[\protect\citeauthoryear{{Becker}}{{Becker}}{2015}]{Becker_2015}
{Becker} M.~R.,  2015, preprint, \href
  {http://adsabs.harvard.edu/abs/2015arXiv150703605B} {} (\mn@eprint {arXiv}
  {1507.03605})

\bibitem[\protect\citeauthoryear{{Behroozi}, {Wechsler}  \&
  {Conroy}}{{Behroozi} et~al.}{2013}]{Behroozi_2013}
{Behroozi} P.~S.,  {Wechsler} R.~H.,   {Conroy} C.,  2013, \mn@doi [\apj]
  {10.1088/0004-637X/770/1/57}, \href
  {http://adsabs.harvard.edu/abs/2013ApJ...770...57B} {770, 57}

\bibitem[\protect\citeauthoryear{{Bekki}}{{Bekki}}{2005}]{Bekki_2005}
{Bekki} K.,  2005, \mn@doi [\apjl] {10.1086/431737}, \href
  {http://adsabs.harvard.edu/abs/2005ApJ...626L..93B} {626, L93}

\bibitem[\protect\citeauthoryear{{Bekki}, {Yahagi}, {Nagashima}  \&
  {Forbes}}{{Bekki} et~al.}{2008}]{Bekki_2008}
{Bekki} K.,  {Yahagi} H.,  {Nagashima} M.,   {Forbes} D.~A.,  2008, \mn@doi
  [\mnras] {10.1111/j.1365-2966.2008.13318.x}, \href
  {http://adsabs.harvard.edu/abs/2008MNRAS.387.1131B} {387, 1131}

\bibitem[\protect\citeauthoryear{{Blakeslee}, {Tonry}  \&
  {Metzger}}{{Blakeslee} et~al.}{1997}]{Blakeslee_1997}
{Blakeslee} J.~P.,  {Tonry} J.~L.,   {Metzger} M.~R.,  1997, \mn@doi [\aj]
  {10.1086/118488}, \href {http://adsabs.harvard.edu/abs/1997AJ....114..482B}
  {114, 482}

\bibitem[\protect\citeauthoryear{{Bolatto}, {Leroy}, {Rosolowsky}, {Walter}  \&
  {Blitz}}{{Bolatto} et~al.}{2008}]{Bolatto_2008}
{Bolatto} A.~D.,  {Leroy} A.~K.,  {Rosolowsky} E.,  {Walter} F.,   {Blitz} L.,
  2008, \mn@doi [\apj] {10.1086/591513}, \href
  {http://adsabs.harvard.edu/abs/2008ApJ...686..948B} {686, 948}

\bibitem[\protect\citeauthoryear{{Boley}, {Lake}, {Read}  \&
  {Teyssier}}{{Boley} et~al.}{2009}]{Boley_2009}
{Boley} A.~C.,  {Lake} G.,  {Read} J.,   {Teyssier} R.,  2009, \mn@doi [\apjl]
  {10.1088/0004-637X/706/1/L192}, \href
  {http://adsabs.harvard.edu/abs/2009ApJ...706L.192B} {706, L192}

\bibitem[\protect\citeauthoryear{{Bond}, {Cole}, {Efstathiou}  \&
  {Kaiser}}{{Bond} et~al.}{1991}]{Bond_1991}
{Bond} J.~R.,  {Cole} S.,  {Efstathiou} G.,   {Kaiser} N.,  1991, \mn@doi
  [\apj] {10.1086/170520}, \href
  {http://adsabs.harvard.edu/abs/1991ApJ...379..440B} {379, 440}

\bibitem[\protect\citeauthoryear{{Bournaud}, {Duc}  \& {Emsellem}}{{Bournaud}
  et~al.}{2008}]{Bournaud_2008}
{Bournaud} F.,  {Duc} P.-A.,   {Emsellem} E.,  2008, \mn@doi [\mnras]
  {10.1111/j.1745-3933.2008.00511.x}, \href
  {http://adsabs.harvard.edu/abs/2008MNRAS.389L...8B} {389, L8}

\bibitem[\protect\citeauthoryear{{Bouwens}, {Illingworth}, {Oesch}, {Caruana},
  {Holwerda}, {Smit}  \& {Wilkins}}{{Bouwens} et~al.}{2015}]{Bouwens_2015b}
{Bouwens} R.~J.,  {Illingworth} G.~D.,  {Oesch} P.~A.,  {Caruana} J.,
  {Holwerda} B.,  {Smit} R.,   {Wilkins} S.,  2015, \mn@doi [\apj]
  {10.1088/0004-637X/811/2/140}, \href
  {http://adsabs.harvard.edu/abs/2015ApJ...811..140B} {811, 140}

\bibitem[\protect\citeauthoryear{{Bouwens}, {van Dokkum}, {Illingworth},
  {Oesch}, {Maseda}, {Ribeiro}, {Stefanon}  \& {Lam}}{{Bouwens}
  et~al.}{2017}]{Bouwens_2017}
{Bouwens} R.~J.,  {van Dokkum} P.~G.,  {Illingworth} G.~D.,  {Oesch} P.~A.,
  {Maseda} M.,  {Ribeiro} B.,  {Stefanon} M.,   {Lam} D.,  2017, preprint,
  \href {http://adsabs.harvard.edu/abs/2017arXiv171102090B} {} (\mn@eprint
  {arXiv} {1711.02090})

\bibitem[\protect\citeauthoryear{{Boylan-Kolchin}}{{Boylan-Kolchin}}{2017a}]{BoylanKolchin_2017b}
{Boylan-Kolchin} M.,  2017a, preprint, \href
  {http://adsabs.harvard.edu/abs/2017arXiv171100009B} {} (\mn@eprint {arXiv}
  {1711.00009})

\bibitem[\protect\citeauthoryear{{Boylan-Kolchin}}{{Boylan-Kolchin}}{2017b}]{BoylanKolchin_2017}
{Boylan-Kolchin} M.,  2017b, \mn@doi [\mnras] {10.1093/mnras/stx2164}, \href
  {http://adsabs.harvard.edu/abs/2017MNRAS.472.3120B} {472, 3120}

\bibitem[\protect\citeauthoryear{{Boylan-Kolchin}, {Ma}  \&
  {Quataert}}{{Boylan-Kolchin} et~al.}{2008}]{Boylan_Kolchin_2008}
{Boylan-Kolchin} M.,  {Ma} C.-P.,   {Quataert} E.,  2008, \mn@doi [\mnras]
  {10.1111/j.1365-2966.2007.12530.x}, \href
  {http://adsabs.harvard.edu/abs/2008MNRAS.383...93B} {383, 93}

\bibitem[\protect\citeauthoryear{{Bressan}, {Marigo}, {Girardi}, {Salasnich},
  {Dal Cero}, {Rubele}  \& {Nanni}}{{Bressan} et~al.}{2012}]{Bressan_2012}
{Bressan} A.,  {Marigo} P.,  {Girardi} L.,  {Salasnich} B.,  {Dal Cero} C.,
  {Rubele} S.,   {Nanni} A.,  2012, \mn@doi [\mnras]
  {10.1111/j.1365-2966.2012.21948.x}, \href
  {http://adsabs.harvard.edu/abs/2012MNRAS.427..127B} {427, 127}

\bibitem[\protect\citeauthoryear{{Brodie} \& {Huchra}}{{Brodie} \&
  {Huchra}}{1991}]{Brodie_1991}
{Brodie} J.~P.,  {Huchra} J.~P.,  1991, \mn@doi [\apj] {10.1086/170492}, \href
  {http://adsabs.harvard.edu/abs/1991ApJ...379..157B} {379, 157}

\bibitem[\protect\citeauthoryear{{Brodie}, {Usher}, {Conroy}, {Strader},
  {Arnold}, {Forbes}  \& {Romanowsky}}{{Brodie} et~al.}{2012}]{Brodie_2012}
{Brodie} J.~P.,  {Usher} C.,  {Conroy} C.,  {Strader} J.,  {Arnold} J.~A.,
  {Forbes} D.~A.,   {Romanowsky} A.~J.,  2012, \mn@doi [\apjl]
  {10.1088/2041-8205/759/2/L33}, \href
  {http://adsabs.harvard.edu/abs/2012ApJ...759L..33B} {759, L33}

\bibitem[\protect\citeauthoryear{{Bryan} \& {Norman}}{{Bryan} \&
  {Norman}}{1998}]{Bryan_1998}
{Bryan} G.~L.,  {Norman} M.~L.,  1998, \mn@doi [\apj] {10.1086/305262}, \href
  {http://adsabs.harvard.edu/abs/1998ApJ...495...80B} {495, 80}

\bibitem[\protect\citeauthoryear{{Bullock}, {Dekel}, {Kolatt}, {Kravtsov},
  {Klypin}, {Porciani}  \& {Primack}}{{Bullock} et~al.}{2001}]{Bullock_2001}
{Bullock} J.~S.,  {Dekel} A.,  {Kolatt} T.~S.,  {Kravtsov} A.~V.,  {Klypin}
  A.~A.,  {Porciani} C.,   {Primack} J.~R.,  2001, \mn@doi [\apj]
  {10.1086/321477}, \href {http://adsabs.harvard.edu/abs/2001ApJ...555..240B}
  {555, 240}

\bibitem[\protect\citeauthoryear{{Burkert} \& {Tremaine}}{{Burkert} \&
  {Tremaine}}{2010}]{Burkert_2010}
{Burkert} A.,  {Tremaine} S.,  2010, \mn@doi [\apj]
  {10.1088/0004-637X/720/1/516}, \href
  {http://adsabs.harvard.edu/abs/2010ApJ...720..516B} {720, 516}

\bibitem[\protect\citeauthoryear{{Carlberg}}{{Carlberg}}{2002}]{Carlberg_2002}
{Carlberg} R.~G.,  2002, \mn@doi [\apj] {10.1086/340500}, \href
  {http://adsabs.harvard.edu/abs/2002ApJ...573...60C} {573, 60}

\bibitem[\protect\citeauthoryear{{Carlberg}}{{Carlberg}}{2017}]{Carlberg_2017}
{Carlberg} R.~G.,  2017, preprint, \href
  {http://adsabs.harvard.edu/abs/2017arXiv170601938C} {} (\mn@eprint {arXiv}
  {1706.01938})

\bibitem[\protect\citeauthoryear{{Chaboyer} et~al.,}{{Chaboyer}
  et~al.}{2017}]{Chaboyer_2017}
{Chaboyer} B.,  et~al., 2017, \mn@doi [\apj] {10.3847/1538-4357/835/2/152},
  \href {http://adsabs.harvard.edu/abs/2017ApJ...835..152C} {835, 152}

\bibitem[\protect\citeauthoryear{{Chen}, {Girardi}, {Bressan}, {Marigo},
  {Barbieri}  \& {Kong}}{{Chen} et~al.}{2014}]{Chen_2014}
{Chen} Y.,  {Girardi} L.,  {Bressan} A.,  {Marigo} P.,  {Barbieri} M.,   {Kong}
  X.,  2014, \mn@doi [\mnras] {10.1093/mnras/stu1605}, \href
  {http://adsabs.harvard.edu/abs/2014MNRAS.444.2525C} {444, 2525}

\bibitem[\protect\citeauthoryear{{Chen}, {Bressan}, {Girardi}, {Marigo}, {Kong}
   \& {Lanza}}{{Chen} et~al.}{2015}]{Chen_2015}
{Chen} Y.,  {Bressan} A.,  {Girardi} L.,  {Marigo} P.,  {Kong} X.,   {Lanza}
  A.,  2015, \mn@doi [\mnras] {10.1093/mnras/stv1281}, \href
  {http://adsabs.harvard.edu/abs/2015MNRAS.452.1068C} {452, 1068}

\bibitem[\protect\citeauthoryear{{Chernoff} \& {Weinberg}}{{Chernoff} \&
  {Weinberg}}{1990}]{Chernoff_1990}
{Chernoff} D.~F.,  {Weinberg} M.~D.,  1990, \mn@doi [\apj] {10.1086/168451},
  \href {http://adsabs.harvard.edu/abs/1990ApJ...351..121C} {351, 121}

\bibitem[\protect\citeauthoryear{{Choksi}, {Gnedin}  \& {Li}}{{Choksi}
  et~al.}{2018}]{Choksi_2018}
{Choksi} N.,  {Gnedin} O.,   {Li} H.,  2018, preprint, \href
  {http://adsabs.harvard.edu/abs/2018arXiv180103515C} {} (\mn@eprint {arXiv}
  {1801.03515})

\bibitem[\protect\citeauthoryear{{Cioffi}, {McKee}  \& {Bertschinger}}{{Cioffi}
  et~al.}{1988}]{Cioffi_1988}
{Cioffi} D.~F.,  {McKee} C.~F.,   {Bertschinger} E.,  1988, \mn@doi [\apj]
  {10.1086/166834}, \href {http://adsabs.harvard.edu/abs/1988ApJ...334..252C}
  {334, 252}

\bibitem[\protect\citeauthoryear{{Cole}, {Lacey}, {Baugh}  \& {Frenk}}{{Cole}
  et~al.}{2000}]{Cole_2000}
{Cole} S.,  {Lacey} C.~G.,  {Baugh} C.~M.,   {Frenk} C.~S.,  2000, \mn@doi
  [\mnras] {10.1046/j.1365-8711.2000.03879.x}, \href
  {http://adsabs.harvard.edu/abs/2000MNRAS.319..168C} {319, 168}

\bibitem[\protect\citeauthoryear{{Conroy}}{{Conroy}}{2012}]{Conroy_2012}
{Conroy} C.,  2012, \mn@doi [\apj] {10.1088/0004-637X/758/1/21}, \href
  {http://adsabs.harvard.edu/abs/2012ApJ...758...21C} {758, 21}

\bibitem[\protect\citeauthoryear{{Conroy} \& {Spergel}}{{Conroy} \&
  {Spergel}}{2011}]{Conroy_2011b}
{Conroy} C.,  {Spergel} D.~N.,  2011, \mn@doi [\apj]
  {10.1088/0004-637X/726/1/36}, \href
  {http://adsabs.harvard.edu/abs/2011ApJ...726...36C} {726, 36}

\bibitem[\protect\citeauthoryear{{Conroy}, {Loeb}  \& {Spergel}}{{Conroy}
  et~al.}{2011}]{Conroy_2011}
{Conroy} C.,  {Loeb} A.,   {Spergel} D.~N.,  2011, \mn@doi [\apj]
  {10.1088/0004-637X/741/2/72}, \href
  {http://adsabs.harvard.edu/abs/2011ApJ...741...72C} {741, 72}

\bibitem[\protect\citeauthoryear{{Corbett Moran}, {Teyssier}  \&
  {Lake}}{{Corbett Moran} et~al.}{2014}]{Moran_2014}
{Corbett Moran} C.,  {Teyssier} R.,   {Lake} G.,  2014, \mn@doi [\mnras]
  {10.1093/mnras/stu1057}, \href
  {http://adsabs.harvard.edu/abs/2014MNRAS.442.2826C} {442, 2826}

\bibitem[\protect\citeauthoryear{{C{\^o}t{\'e}}, {Marzke}  \&
  {West}}{{C{\^o}t{\'e}} et~al.}{1998}]{Cote_1998}
{C{\^o}t{\'e}} P.,  {Marzke} R.~O.,   {West} M.~J.,  1998, \mn@doi [\apj]
  {10.1086/305838}, \href {http://adsabs.harvard.edu/abs/1998ApJ...501..554C}
  {501, 554}

\bibitem[\protect\citeauthoryear{{D'Ercole}, {Vesperini}, {D'Antona},
  {McMillan}  \& {Recchi}}{{D'Ercole} et~al.}{2008}]{DErcole_2008}
{D'Ercole} A.,  {Vesperini} E.,  {D'Antona} F.,  {McMillan} S.~L.~W.,
  {Recchi} S.,  2008, \mn@doi [\mnras] {10.1111/j.1365-2966.2008.13915.x},
  \href {http://adsabs.harvard.edu/abs/2008MNRAS.391..825D} {391, 825}

\bibitem[\protect\citeauthoryear{{Dav{\'e}}, {Oppenheimer}  \&
  {Finlator}}{{Dav{\'e}} et~al.}{2011}]{Dave_2011}
{Dav{\'e}} R.,  {Oppenheimer} B.~D.,   {Finlator} K.,  2011, \mn@doi [\mnras]
  {10.1111/j.1365-2966.2011.18680.x}, \href
  {http://adsabs.harvard.edu/abs/2011MNRAS.415...11D} {415, 11}

\bibitem[\protect\citeauthoryear{{Dav{\'e}}, {Finlator}  \&
  {Oppenheimer}}{{Dav{\'e}} et~al.}{2012}]{Dave_2012}
{Dav{\'e}} R.,  {Finlator} K.,   {Oppenheimer} B.~D.,  2012, \mn@doi [\mnras]
  {10.1111/j.1365-2966.2011.20148.x}, \href
  {http://adsabs.harvard.edu/abs/2012MNRAS.421...98D} {421, 98}

\bibitem[\protect\citeauthoryear{{Dekel} et~al.,}{{Dekel}
  et~al.}{2009}]{Dekel_2009}
{Dekel} A.,  et~al., 2009, \mn@doi [\nat] {10.1038/nature07648}, \href
  {http://adsabs.harvard.edu/abs/2009Natur.457..451D} {457, 451}

\bibitem[\protect\citeauthoryear{{Desmond}, {Mao}, {Wechsler}, {Crain}  \&
  {Schaye}}{{Desmond} et~al.}{2017}]{Desmond_2017}
{Desmond} H.,  {Mao} Y.-Y.,  {Wechsler} R.~H.,  {Crain} R.~A.,   {Schaye} J.,
  2017, \mn@doi [\mnras] {10.1093/mnrasl/slx093}, \href
  {http://adsabs.harvard.edu/abs/2017MNRAS.471L..11D} {471, L11}

\bibitem[\protect\citeauthoryear{{Diemand}, {Madau}  \& {Moore}}{{Diemand}
  et~al.}{2005}]{Diemand_2005}
{Diemand} J.,  {Madau} P.,   {Moore} B.,  2005, \mn@doi [\mnras]
  {10.1111/j.1365-2966.2005.09604.x}, \href
  {http://adsabs.harvard.edu/abs/2005MNRAS.364..367D} {364, 367}

\bibitem[\protect\citeauthoryear{{Dotter}, {Sarajedini}  \&
  {Anderson}}{{Dotter} et~al.}{2011}]{Dotter_2011}
{Dotter} A.,  {Sarajedini} A.,   {Anderson} J.,  2011, \mn@doi [\apj]
  {10.1088/0004-637X/738/1/74}, \href
  {http://adsabs.harvard.edu/abs/2011ApJ...738...74D} {738, 74}

\bibitem[\protect\citeauthoryear{{Durrell} et~al.,}{{Durrell}
  et~al.}{2014}]{Durrell_2014}
{Durrell} P.~R.,  et~al., 2014, \mn@doi [\apj] {10.1088/0004-637X/794/2/103},
  \href {http://adsabs.harvard.edu/abs/2014ApJ...794..103D} {794, 103}

\bibitem[\protect\citeauthoryear{{El-Badry}, {Wetzel}, {Geha}, {Hopkins},
  {Kere{\v s}}, {Chan}  \& {Faucher-Gigu{\`e}re}}{{El-Badry}
  et~al.}{2016}]{ElBadry_2016}
{El-Badry} K.,  {Wetzel} A.,  {Geha} M.,  {Hopkins} P.~F.,  {Kere{\v s}} D.,
  {Chan} T.~K.,   {Faucher-Gigu{\`e}re} C.-A.,  2016, \mn@doi [\apj]
  {10.3847/0004-637X/820/2/131}, \href
  {http://adsabs.harvard.edu/abs/2016ApJ...820..131E} {820, 131}

\bibitem[\protect\citeauthoryear{{El-Badry} et~al.,}{{El-Badry}
  et~al.}{2018a}]{ElBadry_2018b}
{El-Badry} K.,  et~al., 2018a, preprint, \href
  {http://adsabs.harvard.edu/abs/2018arXiv180400659E} {} (\mn@eprint {arXiv}
  {1804.00659})

\bibitem[\protect\citeauthoryear{{El-Badry} et~al.,}{{El-Badry}
  et~al.}{2018b}]{ElBadry_2018}
{El-Badry} K.,  et~al., 2018b, \mn@doi [\mnras] {10.1093/mnras/stx2482}, \href
  {http://adsabs.harvard.edu/abs/2018MNRAS.473.1930E} {473, 1930}

\bibitem[\protect\citeauthoryear{{Elmegreen}}{{Elmegreen}}{2008}]{Elmegreen_2008}
{Elmegreen} B.~G.,  2008, \mn@doi [\apj] {10.1086/523791}, \href
  {http://adsabs.harvard.edu/abs/2008ApJ...672.1006E} {672, 1006}

\bibitem[\protect\citeauthoryear{{Elmegreen}}{{Elmegreen}}{2010}]{Elmegreen_2010}
{Elmegreen} B.~G.,  2010, \mn@doi [\apjl] {10.1088/2041-8205/712/2/L184}, \href
  {http://adsabs.harvard.edu/abs/2010ApJ...712L.184E} {712, L184}

\bibitem[\protect\citeauthoryear{{Elmegreen}}{{Elmegreen}}{2017}]{Elmegreen_2017}
{Elmegreen} B.~G.,  2017, \mn@doi [\apj] {10.3847/1538-4357/836/1/80}, \href
  {http://adsabs.harvard.edu/abs/2017ApJ...836...80E} {836, 80}

\bibitem[\protect\citeauthoryear{{Elmegreen} \& {Efremov}}{{Elmegreen} \&
  {Efremov}}{1997}]{Elmegreen_1997}
{Elmegreen} B.~G.,  {Efremov} Y.~N.,  1997, \mn@doi [\apj] {10.1086/303966},
  \href {http://adsabs.harvard.edu/abs/1997ApJ...480..235E} {480, 235}

\bibitem[\protect\citeauthoryear{{Fall} \& {Chandar}}{{Fall} \&
  {Chandar}}{2012}]{Fall_2012}
{Fall} S.~M.,  {Chandar} R.,  2012, \mn@doi [\apj]
  {10.1088/0004-637X/752/2/96}, \href
  {http://adsabs.harvard.edu/abs/2012ApJ...752...96F} {752, 96}

\bibitem[\protect\citeauthoryear{{Fall} \& {Efstathiou}}{{Fall} \&
  {Efstathiou}}{1980}]{Fall_1980}
{Fall} S.~M.,  {Efstathiou} G.,  1980, \mn@doi [\mnras]
  {10.1093/mnras/193.2.189}, \href
  {http://adsabs.harvard.edu/abs/1980MNRAS.193..189F} {193, 189}

\bibitem[\protect\citeauthoryear{{Fall} \& {Rees}}{{Fall} \&
  {Rees}}{1985}]{Fall_1985}
{Fall} S.~M.,  {Rees} M.~J.,  1985, \mn@doi [\apj] {10.1086/163585}, \href
  {http://adsabs.harvard.edu/abs/1985ApJ...298...18F} {298, 18}

\bibitem[\protect\citeauthoryear{{Fall} \& {Zhang}}{{Fall} \&
  {Zhang}}{2001}]{Fall_2001}
{Fall} S.~M.,  {Zhang} Q.,  2001, \mn@doi [\apj] {10.1086/323358}, \href
  {http://adsabs.harvard.edu/abs/2001ApJ...561..751F} {561, 751}

\bibitem[\protect\citeauthoryear{{Fall}, {Chandar}  \& {Whitmore}}{{Fall}
  et~al.}{2005}]{Fall_2005}
{Fall} S.~M.,  {Chandar} R.,   {Whitmore} B.~C.,  2005, \mn@doi [\apjl]
  {10.1086/496878}, \href {http://adsabs.harvard.edu/abs/2005ApJ...631L.133F}
  {631, L133}

\bibitem[\protect\citeauthoryear{{Fall}, {Krumholz}  \& {Matzner}}{{Fall}
  et~al.}{2010}]{Fall_2010}
{Fall} S.~M.,  {Krumholz} M.~R.,   {Matzner} C.~D.,  2010, \mn@doi [\apjl]
  {10.1088/2041-8205/710/2/L142}, \href
  {http://adsabs.harvard.edu/abs/2010ApJ...710L.142F} {710, L142}

\bibitem[\protect\citeauthoryear{{Faucher-Gigu{\`e}re}, {Kere{\v s}}  \&
  {Ma}}{{Faucher-Gigu{\`e}re} et~al.}{2011}]{FG_2011}
{Faucher-Gigu{\`e}re} C.-A.,  {Kere{\v s}} D.,   {Ma} C.-P.,  2011, \mn@doi
  [\mnras] {10.1111/j.1365-2966.2011.19457.x}, \href
  {http://adsabs.harvard.edu/abs/2011MNRAS.417.2982F} {417, 2982}

\bibitem[\protect\citeauthoryear{{Faucher-Gigu{\`e}re}, {Quataert}  \&
  {Hopkins}}{{Faucher-Gigu{\`e}re} et~al.}{2013}]{FG_2013}
{Faucher-Gigu{\`e}re} C.-A.,  {Quataert} E.,   {Hopkins} P.~F.,  2013, \mn@doi
  [\mnras] {10.1093/mnras/stt866}, \href
  {http://adsabs.harvard.edu/abs/2013MNRAS.433.1970F} {433, 1970}

\bibitem[\protect\citeauthoryear{{Fitts} et~al.,}{{Fitts}
  et~al.}{2018}]{Fitts_2018}
{Fitts} A.,  et~al., 2018, preprint, \href
  {http://adsabs.harvard.edu/abs/2018arXiv180106187F} {} (\mn@eprint {arXiv}
  {1801.06187})

\bibitem[\protect\citeauthoryear{{Forbes}, {Read}, {Gieles}  \&
  {Collins}}{{Forbes} et~al.}{2018}]{Forbes_2018}
{Forbes} D.~A.,  {Read} J.~I.,  {Gieles} M.,   {Collins} M.~L.~M.,  2018,
  \mn@doi [\mnras] {10.1093/mnras/sty2584}, \href
  {http://adsabs.harvard.edu/abs/2018MNRAS.tmp.2463F} {}

\bibitem[\protect\citeauthoryear{{F{\"o}rster Schreiber} et~al.,}{{F{\"o}rster
  Schreiber} et~al.}{2009}]{Forster_2009}
{F{\"o}rster Schreiber} N.~M.,  et~al., 2009, \mn@doi [\apj]
  {10.1088/0004-637X/706/2/1364}, \href
  {http://adsabs.harvard.edu/abs/2009ApJ...706.1364F} {706, 1364}

\bibitem[\protect\citeauthoryear{{F{\"o}rster Schreiber} et~al.,}{{F{\"o}rster
  Schreiber} et~al.}{2011}]{Forster_2011}
{F{\"o}rster Schreiber} N.~M.,  et~al., 2011, \mn@doi [\apj]
  {10.1088/0004-637X/739/1/45}, \href
  {http://adsabs.harvard.edu/abs/2011ApJ...739...45F} {739, 45}

\bibitem[\protect\citeauthoryear{{Garrison-Kimmel} et~al.,}{{Garrison-Kimmel}
  et~al.}{2017}]{GarrisonKimmel_2017}
{Garrison-Kimmel} S.,  et~al., 2017, preprint, \href
  {http://adsabs.harvard.edu/abs/2017arXiv171203966G} {} (\mn@eprint {arXiv}
  {1712.03966})

\bibitem[\protect\citeauthoryear{{Georgiev}, {Puzia}, {Goudfrooij}  \&
  {Hilker}}{{Georgiev} et~al.}{2010}]{Georgiev_2010}
{Georgiev} I.~Y.,  {Puzia} T.~H.,  {Goudfrooij} P.,   {Hilker} M.,  2010,
  \mn@doi [\mnras] {10.1111/j.1365-2966.2010.16802.x}, \href
  {http://adsabs.harvard.edu/abs/2010MNRAS.406.1967G} {406, 1967}

\bibitem[\protect\citeauthoryear{{Geyer} \& {Burkert}}{{Geyer} \&
  {Burkert}}{2001}]{Geyer_2001}
{Geyer} M.~P.,  {Burkert} A.,  2001, \mn@doi [\mnras]
  {10.1046/j.1365-8711.2001.04257.x}, \href
  {http://adsabs.harvard.edu/abs/2001MNRAS.323..988G} {323, 988}

\bibitem[\protect\citeauthoryear{{Gieles} \& {Baumgardt}}{{Gieles} \&
  {Baumgardt}}{2008}]{Gieles_2008}
{Gieles} M.,  {Baumgardt} H.,  2008, \mn@doi [\mnras]
  {10.1111/j.1745-3933.2008.00515.x}, \href
  {http://adsabs.harvard.edu/abs/2008MNRAS.389L..28G} {389, L28}

\bibitem[\protect\citeauthoryear{{Gieles}, {Heggie}  \& {Zhao}}{{Gieles}
  et~al.}{2011}]{Gieles_2011}
{Gieles} M.,  {Heggie} D.~C.,   {Zhao} H.,  2011, \mn@doi [\mnras]
  {10.1111/j.1365-2966.2011.18320.x}, \href
  {http://adsabs.harvard.edu/abs/2011MNRAS.413.2509G} {413, 2509}

\bibitem[\protect\citeauthoryear{{Gnedin}, {Lee}  \& {Ostriker}}{{Gnedin}
  et~al.}{1999}]{Gnedin_1999}
{Gnedin} O.~Y.,  {Lee} H.~M.,   {Ostriker} J.~P.,  1999, \mn@doi [\apj]
  {10.1086/307659}, \href {http://adsabs.harvard.edu/abs/1999ApJ...522..935G}
  {522, 935}

\bibitem[\protect\citeauthoryear{{Gnedin}, {Ostriker}  \& {Tremaine}}{{Gnedin}
  et~al.}{2014}]{Gnedin_2014}
{Gnedin} O.~Y.,  {Ostriker} J.~P.,   {Tremaine} S.,  2014, \mn@doi [\apj]
  {10.1088/0004-637X/785/1/71}, \href
  {http://adsabs.harvard.edu/abs/2014ApJ...785...71G} {785, 71}

\bibitem[\protect\citeauthoryear{{Goddard}, {Bastian}  \&
  {Kennicutt}}{{Goddard} et~al.}{2010}]{Goddard_2010}
{Goddard} Q.~E.,  {Bastian} N.,   {Kennicutt} R.~C.,  2010, \mn@doi [\mnras]
  {10.1111/j.1365-2966.2010.16511.x}, \href
  {http://adsabs.harvard.edu/abs/2010MNRAS.405..857G} {405, 857}

\bibitem[\protect\citeauthoryear{{Gonzalez}, {Sivanandam}, {Zabludoff}  \&
  {Zaritsky}}{{Gonzalez} et~al.}{2013}]{Gonzalez_2013}
{Gonzalez} A.~H.,  {Sivanandam} S.,  {Zabludoff} A.~I.,   {Zaritsky} D.,  2013,
  \mn@doi [\apj] {10.1088/0004-637X/778/1/14}, \href
  {http://adsabs.harvard.edu/abs/2013ApJ...778...14G} {778, 14}

\bibitem[\protect\citeauthoryear{{Griffen}, {Dooley}, {Ji}, {O'Shea},
  {G{\'o}mez}  \& {Frebel}}{{Griffen} et~al.}{2018}]{Griffen_2018}
{Griffen} B.~F.,  {Dooley} G.~A.,  {Ji} A.~P.,  {O'Shea} B.~W.,  {G{\'o}mez}
  F.~A.,   {Frebel} A.,  2018, \mn@doi [\mnras] {10.1093/mnras/stx2749}, \href
  {http://adsabs.harvard.edu/abs/2018MNRAS.474..443G} {474, 443}

\bibitem[\protect\citeauthoryear{{Grudi{\'c}}, {Hopkins}, {Quataert}  \&
  {Murray}}{{Grudi{\'c}} et~al.}{2018a}]{Grudic_2018}
{Grudi{\'c}} M.~Y.,  {Hopkins} P.~F.,  {Quataert} E.,   {Murray} N.,  2018a,
  preprint, \href {http://adsabs.harvard.edu/abs/2018arXiv180404137G} {}
  (\mn@eprint {arXiv} {1804.04137})

\bibitem[\protect\citeauthoryear{{Grudi{\'c}}, {Hopkins},
  {Faucher-Gigu{\`e}re}, {Quataert}, {Murray}  \& {Kere{\v s}}}{{Grudi{\'c}}
  et~al.}{2018b}]{Grudic_2016}
{Grudi{\'c}} M.~Y.,  {Hopkins} P.~F.,  {Faucher-Gigu{\`e}re} C.-A.,  {Quataert}
  E.,  {Murray} N.,   {Kere{\v s}} D.,  2018b, \mn@doi [\mnras]
  {10.1093/mnras/sty035}, \href
  {http://adsabs.harvard.edu/abs/2018MNRAS.475.3511G} {475, 3511}

\bibitem[\protect\citeauthoryear{{Harris}}{{Harris}}{1996}]{Harris_1996}
{Harris} W.~E.,  1996, \mn@doi [\aj] {10.1086/118116}, \href
  {http://adsabs.harvard.edu/abs/1996AJ....112.1487H} {112, 1487}

\bibitem[\protect\citeauthoryear{{Harris}, {Whitmore}, {Karakla}, {Oko{\'n}},
  {Baum}, {Hanes}  \& {Kavelaars}}{{Harris} et~al.}{2006}]{Harris_2006}
{Harris} W.~E.,  {Whitmore} B.~C.,  {Karakla} D.,  {Oko{\'n}} W.,  {Baum}
  W.~A.,  {Hanes} D.~A.,   {Kavelaars} J.~J.,  2006, \mn@doi [\apj]
  {10.1086/498058}, \href {http://adsabs.harvard.edu/abs/2006ApJ...636...90H}
  {636, 90}

\bibitem[\protect\citeauthoryear{{Harris}, {Harris}  \& {Alessi}}{{Harris}
  et~al.}{2013}]{Harris_2013}
{Harris} W.~E.,  {Harris} G.~L.~H.,   {Alessi} M.,  2013, \mn@doi [\apj]
  {10.1088/0004-637X/772/2/82}, \href
  {http://adsabs.harvard.edu/abs/2013ApJ...772...82H} {772, 82}

\bibitem[\protect\citeauthoryear{{Harris}, {Poole}  \& {Harris}}{{Harris}
  et~al.}{2014}]{Harris_2014}
{Harris} G.~L.~H.,  {Poole} G.~B.,   {Harris} W.~E.,  2014, \mn@doi [\mnras]
  {10.1093/mnras/stt2337}, \href
  {http://adsabs.harvard.edu/abs/2014MNRAS.438.2117H} {438, 2117}

\bibitem[\protect\citeauthoryear{{Harris}, {Harris}  \& {Hudson}}{{Harris}
  et~al.}{2015}]{Harris_2015}
{Harris} W.~E.,  {Harris} G.~L.,   {Hudson} M.~J.,  2015, \mn@doi [\apj]
  {10.1088/0004-637X/806/1/36}, \href
  {http://adsabs.harvard.edu/abs/2015ApJ...806...36H} {806, 36}

\bibitem[\protect\citeauthoryear{{Harris}, {Blakeslee}, {Whitmore}, {Gnedin},
  {Geisler}  \& {Rothberg}}{{Harris} et~al.}{2016}]{Harris_2016}
{Harris} W.~E.,  {Blakeslee} J.~P.,  {Whitmore} B.~C.,  {Gnedin} O.~Y.,
  {Geisler} D.,   {Rothberg} B.,  2016, \mn@doi [\apj]
  {10.3847/0004-637X/817/1/58}, \href
  {http://adsabs.harvard.edu/abs/2016ApJ...817...58H} {817, 58}

\bibitem[\protect\citeauthoryear{{Harris}, {Ciccone}, {Eadie}, {Gnedin},
  {Geisler}, {Rothberg}  \& {Bailin}}{{Harris} et~al.}{2017a}]{Harris_2017b}
{Harris} W.~E.,  {Ciccone} S.~M.,  {Eadie} G.~M.,  {Gnedin} O.~Y.,  {Geisler}
  D.,  {Rothberg} B.,   {Bailin} J.,  2017a, \mn@doi [\apj]
  {10.3847/1538-4357/835/1/101}, \href
  {http://adsabs.harvard.edu/abs/2017ApJ...835..101H} {835, 101}

\bibitem[\protect\citeauthoryear{{Harris}, {Blakeslee}  \& {Harris}}{{Harris}
  et~al.}{2017b}]{Harris_2017}
{Harris} W.~E.,  {Blakeslee} J.~P.,   {Harris} G.~L.~H.,  2017b, \mn@doi [\apj]
  {10.3847/1538-4357/836/1/67}, \href
  {http://adsabs.harvard.edu/abs/2017ApJ...836...67H} {836, 67}

\bibitem[\protect\citeauthoryear{{Hirschmann}, {Khochfar}, {Burkert}, {Naab},
  {Genel}  \& {Somerville}}{{Hirschmann} et~al.}{2010}]{Hirschmann_2010}
{Hirschmann} M.,  {Khochfar} S.,  {Burkert} A.,  {Naab} T.,  {Genel} S.,
  {Somerville} R.~S.,  2010, \mn@doi [\mnras]
  {10.1111/j.1365-2966.2010.17006.x}, \href
  {http://adsabs.harvard.edu/abs/2010MNRAS.407.1016H} {407, 1016}

\bibitem[\protect\citeauthoryear{{Hopkins}, {Quataert}  \& {Murray}}{{Hopkins}
  et~al.}{2012}]{Hopkins_2012}
{Hopkins} P.~F.,  {Quataert} E.,   {Murray} N.,  2012, \mn@doi [\mnras]
  {10.1111/j.1365-2966.2012.20578.x}, \href
  {http://adsabs.harvard.edu/abs/2012MNRAS.421.3488H} {421, 3488}

\bibitem[\protect\citeauthoryear{{Hopkins}, {Cox}, {Hernquist}, {Narayanan},
  {Hayward}  \& {Murray}}{{Hopkins} et~al.}{2013}]{Hopkins_2013}
{Hopkins} P.~F.,  {Cox} T.~J.,  {Hernquist} L.,  {Narayanan} D.,  {Hayward}
  C.~C.,   {Murray} N.,  2013, \mn@doi [\mnras] {10.1093/mnras/stt017}, \href
  {http://adsabs.harvard.edu/abs/2013MNRAS.430.1901H} {430, 1901}

\bibitem[\protect\citeauthoryear{{Howard}, {Pudritz}  \& {Harris}}{{Howard}
  et~al.}{2016}]{Howard_2016}
{Howard} C.~S.,  {Pudritz} R.~E.,   {Harris} W.~E.,  2016, \mn@doi [\mnras]
  {10.1093/mnras/stw1476}, \href
  {http://adsabs.harvard.edu/abs/2016MNRAS.461.2953H} {461, 2953}

\bibitem[\protect\citeauthoryear{{Huang} et~al.,}{{Huang}
  et~al.}{2017}]{Huang_2017}
{Huang} K.-H.,  et~al., 2017, \mn@doi [\apj] {10.3847/1538-4357/aa62a6}, \href
  {http://adsabs.harvard.edu/abs/2017ApJ...838....6H} {838, 6}

\bibitem[\protect\citeauthoryear{{Hudson}, {Harris}  \& {Harris}}{{Hudson}
  et~al.}{2014}]{Hudson_2014}
{Hudson} M.~J.,  {Harris} G.~L.,   {Harris} W.~E.,  2014, \mn@doi [\apjl]
  {10.1088/2041-8205/787/1/L5}, \href
  {http://adsabs.harvard.edu/abs/2014ApJ...787L...5H} {787, L5}

\bibitem[\protect\citeauthoryear{Hunter}{Hunter}{2007}]{Hunter_2007}
Hunter J.~D.,  2007, \mn@doi [Computing In Science \& Engineering]
  {10.1109/MCSE.2007.55}, 9, 90

\bibitem[\protect\citeauthoryear{{Ibata}, {Nipoti}, {Sollima}, {Bellazzini},
  {Chapman}  \& {Dalessandro}}{{Ibata} et~al.}{2013}]{Ibata_2013}
{Ibata} R.,  {Nipoti} C.,  {Sollima} A.,  {Bellazzini} M.,  {Chapman} S.~C.,
  {Dalessandro} E.,  2013, \mn@doi [\mnras] {10.1093/mnras/sts302}, \href
  {http://adsabs.harvard.edu/abs/2013MNRAS.428.3648I} {428, 3648}

\bibitem[\protect\citeauthoryear{{Jahnke} \& {Macci{\`o}}}{{Jahnke} \&
  {Macci{\`o}}}{2011}]{Jahnke_2011}
{Jahnke} K.,  {Macci{\`o}} A.~V.,  2011, \mn@doi [\apj]
  {10.1088/0004-637X/734/2/92}, \href
  {http://adsabs.harvard.edu/abs/2011ApJ...734...92J} {734, 92}

\bibitem[\protect\citeauthoryear{{Jiang} \& {van den Bosch}}{{Jiang} \& {van
  den Bosch}}{2014}]{Jiang_2014}
{Jiang} F.,  {van den Bosch} F.~C.,  2014, \mn@doi [\mnras]
  {10.1093/mnras/stu280}, \href
  {http://adsabs.harvard.edu/abs/2014MNRAS.440..193J} {440, 193}

\bibitem[\protect\citeauthoryear{{Johnson} et~al.,}{{Johnson}
  et~al.}{2016}]{Johnson_2016}
{Johnson} L.~C.,  et~al., 2016, \mn@doi [\apj] {10.3847/0004-637X/827/1/33},
  \href {http://adsabs.harvard.edu/abs/2016ApJ...827...33J} {827, 33}

\bibitem[\protect\citeauthoryear{{Katz} \& {Ricotti}}{{Katz} \&
  {Ricotti}}{2013}]{Katz_2013}
{Katz} H.,  {Ricotti} M.,  2013, \mn@doi [\mnras] {10.1093/mnras/stt676}, \href
  {http://adsabs.harvard.edu/abs/2013MNRAS.432.3250K} {432, 3250}

\bibitem[\protect\citeauthoryear{{Katz} \& {Ricotti}}{{Katz} \&
  {Ricotti}}{2014}]{Katz_2014}
{Katz} H.,  {Ricotti} M.,  2014, \mn@doi [\mnras] {10.1093/mnras/stu1489},
  \href {http://adsabs.harvard.edu/abs/2014MNRAS.444.2377K} {444, 2377}

\bibitem[\protect\citeauthoryear{{Kavelaars}}{{Kavelaars}}{1999}]{Kavelaars_1999}
{Kavelaars} J.~J.,  1999, in {Merritt} D.~R.,  {Valluri} M.,   {Sellwood}
  J.~A.,  eds,  Astronomical Society of the Pacific Conference Series Vol. 182,
  Galaxy Dynamics - A Rutgers Symposium.  (\mn@eprint {} {astro-ph/9806094})

\bibitem[\protect\citeauthoryear{{Keto}, {Ho}  \& {Lo}}{{Keto}
  et~al.}{2005}]{Keto_2005}
{Keto} E.,  {Ho} L.~C.,   {Lo} K.-Y.,  2005, \mn@doi [\apj] {10.1086/497575},
  \href {http://adsabs.harvard.edu/abs/2005ApJ...635.1062K} {635, 1062}

\bibitem[\protect\citeauthoryear{{Kim}, {Kim}  \& {Ostriker}}{{Kim}
  et~al.}{2016}]{Kim_2016}
{Kim} J.-G.,  {Kim} W.-T.,   {Ostriker} E.~C.,  2016, \mn@doi [\apj]
  {10.3847/0004-637X/819/2/137}, \href
  {http://adsabs.harvard.edu/abs/2016ApJ...819..137K} {819, 137}

\bibitem[\protect\citeauthoryear{{Kimm}, {Cen}, {Rosdahl}  \& {Yi}}{{Kimm}
  et~al.}{2016}]{Kimm_2016}
{Kimm} T.,  {Cen} R.,  {Rosdahl} J.,   {Yi} S.~K.,  2016, \mn@doi [\apj]
  {10.3847/0004-637X/823/1/52}, \href
  {http://adsabs.harvard.edu/abs/2016ApJ...823...52K} {823, 52}

\bibitem[\protect\citeauthoryear{{Kirby}, {Cohen}, {Guhathakurta}, {Cheng},
  {Bullock}  \& {Gallazzi}}{{Kirby} et~al.}{2013}]{Kirby_2013}
{Kirby} E.~N.,  {Cohen} J.~G.,  {Guhathakurta} P.,  {Cheng} L.,  {Bullock}
  J.~S.,   {Gallazzi} A.,  2013, \mn@doi [\apj] {10.1088/0004-637X/779/2/102},
  \href {http://adsabs.harvard.edu/abs/2013ApJ...779..102K} {779, 102}

\bibitem[\protect\citeauthoryear{{Kravtsov}}{{Kravtsov}}{2013}]{Kravtsov_2013}
{Kravtsov} A.~V.,  2013, \mn@doi [\apjl] {10.1088/2041-8205/764/2/L31}, \href
  {http://adsabs.harvard.edu/abs/2013ApJ...764L..31K} {764, L31}

\bibitem[\protect\citeauthoryear{{Kravtsov} \& {Gnedin}}{{Kravtsov} \&
  {Gnedin}}{2005}]{Kravtsov_2005}
{Kravtsov} A.~V.,  {Gnedin} O.~Y.,  2005, \mn@doi [\apj] {10.1086/428636},
  \href {http://adsabs.harvard.edu/abs/2005ApJ...623..650K} {623, 650}

\bibitem[\protect\citeauthoryear{{Kravtsov}, {Vikhlinin}  \&
  {Meshcheryakov}}{{Kravtsov} et~al.}{2018}]{Kravtsov_2018}
{Kravtsov} A.~V.,  {Vikhlinin} A.~A.,   {Meshcheryakov} A.~V.,  2018, \mn@doi
  [Astronomy Letters] {10.1134/S1063773717120015}, \href
  {http://adsabs.harvard.edu/abs/2018AstL...44....8K} {44, 8}

\bibitem[\protect\citeauthoryear{{Kroupa}}{{Kroupa}}{2001}]{Kroupa_2001}
{Kroupa} P.,  2001, \mn@doi [\mnras] {10.1046/j.1365-8711.2001.04022.x}, \href
  {http://adsabs.harvard.edu/abs/2001MNRAS.322..231K} {322, 231}

\bibitem[\protect\citeauthoryear{{Kroupa}, {Weidner}, {Pflamm-Altenburg},
  {Thies}, {Dabringhausen}, {Marks}  \& {Maschberger}}{{Kroupa}
  et~al.}{2013}]{Kroupa_2013}
{Kroupa} P.,  {Weidner} C.,  {Pflamm-Altenburg} J.,  {Thies} I.,
  {Dabringhausen} J.,  {Marks} M.,   {Maschberger} T.,  2013, {The Stellar and
  Sub-Stellar Initial Mass Function of Simple and Composite Populations}.
p.~115, \mn@doi{10.1007/978-94-007-5612-0_4}

\bibitem[\protect\citeauthoryear{{Kruijssen}}{{Kruijssen}}{2012}]{Kruijssen_2012a}
{Kruijssen} J.~M.~D.,  2012, \mn@doi [\mnras]
  {10.1111/j.1365-2966.2012.21923.x}, \href
  {http://adsabs.harvard.edu/abs/2012MNRAS.426.3008K} {426, 3008}

\bibitem[\protect\citeauthoryear{{Kruijssen}}{{Kruijssen}}{2015}]{Kruijssen_2015}
{Kruijssen} J.~M.~D.,  2015, \mn@doi [\mnras] {10.1093/mnras/stv2026}, \href
  {http://adsabs.harvard.edu/abs/2015MNRAS.454.1658K} {454, 1658}

\bibitem[\protect\citeauthoryear{{Kruijssen} \& {Mieske}}{{Kruijssen} \&
  {Mieske}}{2009}]{Kruijssen_2009}
{Kruijssen} J.~M.~D.,  {Mieske} S.,  2009, \mn@doi [\aap]
  {10.1051/0004-6361/200811453}, \href
  {http://adsabs.harvard.edu/abs/2009A%26A...500..785K} {500, 785}

\bibitem[\protect\citeauthoryear{{Kruijssen}, {Maschberger}, {Moeckel},
  {Clarke}, {Bastian}  \& {Bonnell}}{{Kruijssen}
  et~al.}{2012}]{Kruijssen_2012b}
{Kruijssen} J.~M.~D.,  {Maschberger} T.,  {Moeckel} N.,  {Clarke} C.~J.,
  {Bastian} N.,   {Bonnell} I.~A.,  2012, \mn@doi [\mnras]
  {10.1111/j.1365-2966.2011.19748.x}, \href
  {http://adsabs.harvard.edu/abs/2012MNRAS.419..841K} {419, 841}

\bibitem[\protect\citeauthoryear{{Krumholz}}{{Krumholz}}{2014}]{Krumholz_2014}
{Krumholz} M.~R.,  2014, \mn@doi [\physrep] {10.1016/j.physrep.2014.02.001},
  \href {http://adsabs.harvard.edu/abs/2014PhR...539...49K} {539, 49}

\bibitem[\protect\citeauthoryear{{Krumholz} \& {Dekel}}{{Krumholz} \&
  {Dekel}}{2012}]{Krumholz_2012}
{Krumholz} M.~R.,  {Dekel} A.,  2012, \mn@doi [\apj]
  {10.1088/0004-637X/753/1/16}, \href
  {http://adsabs.harvard.edu/abs/2012ApJ...753...16K} {753, 16}

\bibitem[\protect\citeauthoryear{{Krumholz} \& {McKee}}{{Krumholz} \&
  {McKee}}{2005}]{Krumholz_2005}
{Krumholz} M.~R.,  {McKee} C.~F.,  2005, \mn@doi [\apj] {10.1086/431734}, \href
  {http://adsabs.harvard.edu/abs/2005ApJ...630..250K} {630, 250}

\bibitem[\protect\citeauthoryear{{Krumholz} \& {Ting}}{{Krumholz} \&
  {Ting}}{2018}]{Krumholz_2018}
{Krumholz} M.~R.,  {Ting} Y.-S.,  2018, \mn@doi [\mnras]
  {10.1093/mnras/stx3286}, \href
  {http://adsabs.harvard.edu/abs/2018MNRAS.475.2236K} {475, 2236}

\bibitem[\protect\citeauthoryear{{Kulier}, {Ostriker}, {Natarajan}, {Lackner}
  \& {Cen}}{{Kulier} et~al.}{2015}]{Kulier_2015}
{Kulier} A.,  {Ostriker} J.~P.,  {Natarajan} P.,  {Lackner} C.~N.,   {Cen} R.,
  2015, \mn@doi [\apj] {10.1088/0004-637X/799/2/178}, \href
  {http://adsabs.harvard.edu/abs/2015ApJ...799..178K} {799, 178}

\bibitem[\protect\citeauthoryear{{Lacey} \& {Cole}}{{Lacey} \&
  {Cole}}{1993}]{Lacey_1993}
{Lacey} C.,  {Cole} S.,  1993, \mn@doi [\mnras] {10.1093/mnras/262.3.627},
  \href {http://adsabs.harvard.edu/abs/1993MNRAS.262..627L} {262, 627}

\bibitem[\protect\citeauthoryear{{Lada} \& {Lada}}{{Lada} \&
  {Lada}}{2003}]{Lada_2003}
{Lada} C.~J.,  {Lada} E.~A.,  2003, \mn@doi [\araa]
  {10.1146/annurev.astro.41.011802.094844}, \href
  {http://adsabs.harvard.edu/abs/2003ARA%26A..41...57L} {41, 57}

\bibitem[\protect\citeauthoryear{{Larsen} \& {Richtler}}{{Larsen} \&
  {Richtler}}{2000}]{Larsen_2000}
{Larsen} S.~S.,  {Richtler} T.,  2000, \aap, \href
  {http://adsabs.harvard.edu/abs/2000A%26A...354..836L} {354, 836}

\bibitem[\protect\citeauthoryear{{Larsen}, {Brodie}, {Huchra}, {Forbes}  \&
  {Grillmair}}{{Larsen} et~al.}{2001}]{Larsen_2001}
{Larsen} S.~S.,  {Brodie} J.~P.,  {Huchra} J.~P.,  {Forbes} D.~A.,
  {Grillmair} C.~J.,  2001, \mn@doi [\aj] {10.1086/321081}, \href
  {http://adsabs.harvard.edu/abs/2001AJ....121.2974L} {121, 2974}

\bibitem[\protect\citeauthoryear{{Leaman}, {VandenBerg}  \& {Mendel}}{{Leaman}
  et~al.}{2013}]{Leaman_2013}
{Leaman} R.,  {VandenBerg} D.~A.,   {Mendel} J.~T.,  2013, \mn@doi [\mnras]
  {10.1093/mnras/stt1540}, \href
  {http://adsabs.harvard.edu/abs/2013MNRAS.436..122L} {436, 122}

\bibitem[\protect\citeauthoryear{{Leauthaud} et~al.,}{{Leauthaud}
  et~al.}{2012}]{Leauthaud_2012}
{Leauthaud} A.,  et~al., 2012, \mn@doi [\apj] {10.1088/0004-637X/746/1/95},
  \href {http://adsabs.harvard.edu/abs/2012ApJ...746...95L} {746, 95}

\bibitem[\protect\citeauthoryear{{Li} \& {Gnedin}}{{Li} \&
  {Gnedin}}{2014}]{Li_2014}
{Li} H.,  {Gnedin} O.~Y.,  2014, \mn@doi [\apj] {10.1088/0004-637X/796/1/10},
  \href {http://adsabs.harvard.edu/abs/2014ApJ...796...10L} {796, 10}

\bibitem[\protect\citeauthoryear{{Li}, {Gnedin}, {Gnedin}, {Meng}, {Semenov}
  \& {Kravtsov}}{{Li} et~al.}{2017}]{Li_2017}
{Li} H.,  {Gnedin} O.~Y.,  {Gnedin} N.~Y.,  {Meng} X.,  {Semenov} V.~A.,
  {Kravtsov} A.~V.,  2017, \mn@doi [\apj] {10.3847/1538-4357/834/1/69}, \href
  {http://adsabs.harvard.edu/abs/2017ApJ...834...69L} {834, 69}

\bibitem[\protect\citeauthoryear{{Lilly}, {Carollo}, {Pipino}, {Renzini}  \&
  {Peng}}{{Lilly} et~al.}{2013}]{Lilly_2013}
{Lilly} S.~J.,  {Carollo} C.~M.,  {Pipino} A.,  {Renzini} A.,   {Peng} Y.,
  2013, \mn@doi [\apj] {10.1088/0004-637X/772/2/119}, \href
  {http://adsabs.harvard.edu/abs/2013ApJ...772..119L} {772, 119}

\bibitem[\protect\citeauthoryear{{Lim}, {Peng}, {C{\^o}t{\'e}}, {Sales}, {den
  Brok}, {Blakeslee}  \& {Guhathakurta}}{{Lim} et~al.}{2018}]{Lim_2018}
{Lim} S.,  {Peng} E.~W.,  {C{\^o}t{\'e}} P.,  {Sales} L.~V.,  {den Brok} M.,
  {Blakeslee} J.~P.,   {Guhathakurta} P.,  2018, \mn@doi [\apj]
  {10.3847/1538-4357/aacb81}, \href
  {http://adsabs.harvard.edu/abs/2018ApJ...862...82L} {862, 82}

\bibitem[\protect\citeauthoryear{{Lin}, {Mohr}  \& {Stanford}}{{Lin}
  et~al.}{2004}]{Lin_2004}
{Lin} Y.-T.,  {Mohr} J.~J.,   {Stanford} S.~A.,  2004, \mn@doi [\apj]
  {10.1086/421714}, \href {http://adsabs.harvard.edu/abs/2004ApJ...610..745L}
  {610, 745}

\bibitem[\protect\citeauthoryear{{Ma}, {Hopkins}, {Faucher-Gigu{\`e}re},
  {Zolman}, {Muratov}, {Kere{\v s}}  \& {Quataert}}{{Ma}
  et~al.}{2016}]{Ma_2016}
{Ma} X.,  {Hopkins} P.~F.,  {Faucher-Gigu{\`e}re} C.-A.,  {Zolman} N.,
  {Muratov} A.~L.,  {Kere{\v s}} D.,   {Quataert} E.,  2016, \mn@doi [\mnras]
  {10.1093/mnras/stv2659}, \href
  {http://adsabs.harvard.edu/abs/2016MNRAS.456.2140M} {456, 2140}

\bibitem[\protect\citeauthoryear{{Mandelker}, {Dekel}, {Ceverino}, {DeGraf},
  {Guo}  \& {Primack}}{{Mandelker} et~al.}{2017}]{Mandelker_2017}
{Mandelker} N.,  {Dekel} A.,  {Ceverino} D.,  {DeGraf} C.,  {Guo} Y.,
  {Primack} J.,  2017, \mn@doi [\mnras] {10.1093/mnras/stw2358}, \href
  {http://adsabs.harvard.edu/abs/2017MNRAS.464..635M} {464, 635}

\bibitem[\protect\citeauthoryear{{Martell} et~al.,}{{Martell}
  et~al.}{2016}]{Martell_2016}
{Martell} S.~L.,  et~al., 2016, \mn@doi [\apj] {10.3847/0004-637X/825/2/146},
  \href {http://adsabs.harvard.edu/abs/2016ApJ...825..146M} {825, 146}

\bibitem[\protect\citeauthoryear{{Mashchenko} \& {Sills}}{{Mashchenko} \&
  {Sills}}{2005}]{Mashchenko_2005}
{Mashchenko} S.,  {Sills} A.,  2005, \mn@doi [\apj] {10.1086/426132}, \href
  {http://adsabs.harvard.edu/abs/2005ApJ...619..243M} {619, 243}

\bibitem[\protect\citeauthoryear{{McKee} \& {Ostriker}}{{McKee} \&
  {Ostriker}}{2007}]{McKee_2007}
{McKee} C.~F.,  {Ostriker} E.~C.,  2007, \mn@doi [\araa]
  {10.1146/annurev.astro.45.051806.110602}, \href
  {http://adsabs.harvard.edu/abs/2007ARA%26A..45..565M} {45, 565}

\bibitem[\protect\citeauthoryear{{McLaughlin} \& {Fall}}{{McLaughlin} \&
  {Fall}}{2008}]{McLaughlin_2008}
{McLaughlin} D.~E.,  {Fall} S.~M.,  2008, \mn@doi [\apj] {10.1086/533485},
  \href {http://adsabs.harvard.edu/abs/2008ApJ...679.1272M} {679, 1272}

\bibitem[\protect\citeauthoryear{{Mihos} \& {Hernquist}}{{Mihos} \&
  {Hernquist}}{1996}]{Mihos_1996}
{Mihos} J.~C.,  {Hernquist} L.,  1996, \mn@doi [\apj] {10.1086/177353}, \href
  {http://adsabs.harvard.edu/abs/1996ApJ...464..641M} {464, 641}

\bibitem[\protect\citeauthoryear{{Mo}, {Mao}  \& {White}}{{Mo}
  et~al.}{1998}]{Mo_1998}
{Mo} H.~J.,  {Mao} S.,   {White} S.~D.~M.,  1998, \mn@doi [\mnras]
  {10.1046/j.1365-8711.1998.01227.x}, \href
  {http://adsabs.harvard.edu/abs/1998MNRAS.295..319M} {295, 319}

\bibitem[\protect\citeauthoryear{{Moore}}{{Moore}}{1996}]{Moore_1996}
{Moore} B.,  1996, \mn@doi [\apjl] {10.1086/309998}, \href
  {http://adsabs.harvard.edu/abs/1996ApJ...461L..13M} {461, L13}

\bibitem[\protect\citeauthoryear{{Moore}, {Diemand}, {Madau}, {Zemp}  \&
  {Stadel}}{{Moore} et~al.}{2006}]{Moore_2006}
{Moore} B.,  {Diemand} J.,  {Madau} P.,  {Zemp} M.,   {Stadel} J.,  2006,
  \mn@doi [\mnras] {10.1111/j.1365-2966.2006.10116.x}, \href
  {http://adsabs.harvard.edu/abs/2006MNRAS.368..563M} {368, 563}

\bibitem[\protect\citeauthoryear{{Moster}, {Somerville}, {Maulbetsch}, {van den
  Bosch}, {Macci{\`o}}, {Naab}  \& {Oser}}{{Moster} et~al.}{2010}]{Moster_2010}
{Moster} B.~P.,  {Somerville} R.~S.,  {Maulbetsch} C.,  {van den Bosch} F.~C.,
  {Macci{\`o}} A.~V.,  {Naab} T.,   {Oser} L.,  2010, \mn@doi [\apj]
  {10.1088/0004-637X/710/2/903}, \href
  {http://adsabs.harvard.edu/abs/2010ApJ...710..903M} {710, 903}

\bibitem[\protect\citeauthoryear{{Muratov} \& {Gnedin}}{{Muratov} \&
  {Gnedin}}{2010}]{Muratov_2010}
{Muratov} A.~L.,  {Gnedin} O.~Y.,  2010, \mn@doi [\apj]
  {10.1088/0004-637X/718/2/1266}, \href
  {http://adsabs.harvard.edu/abs/2010ApJ...718.1266M} {718, 1266}

\bibitem[\protect\citeauthoryear{{Murray}, {Quataert}  \& {Thompson}}{{Murray}
  et~al.}{2005}]{Murray_2005}
{Murray} N.,  {Quataert} E.,   {Thompson} T.~A.,  2005, \mn@doi [\apj]
  {10.1086/426067}, \href {http://adsabs.harvard.edu/abs/2005ApJ...618..569M}
  {618, 569}

\bibitem[\protect\citeauthoryear{{Murray}, {Quataert}  \& {Thompson}}{{Murray}
  et~al.}{2010}]{Murray_2010}
{Murray} N.,  {Quataert} E.,   {Thompson} T.~A.,  2010, \mn@doi [\apj]
  {10.1088/0004-637X/709/1/191}, \href
  {http://adsabs.harvard.edu/abs/2010ApJ...709..191M} {709, 191}

\bibitem[\protect\citeauthoryear{{O'Shea}, {Wise}, {Xu}  \& {Norman}}{{O'Shea}
  et~al.}{2015}]{OShea_2015}
{O'Shea} B.~W.,  {Wise} J.~H.,  {Xu} H.,   {Norman} M.~L.,  2015, \mn@doi
  [\apjl] {10.1088/2041-8205/807/1/L12}, \href
  {http://adsabs.harvard.edu/abs/2015ApJ...807L..12O} {807, L12}

\bibitem[\protect\citeauthoryear{{Okamoto}, {Gao}  \& {Theuns}}{{Okamoto}
  et~al.}{2008}]{Okamoto_2008}
{Okamoto} T.,  {Gao} L.,   {Theuns} T.,  2008, \mn@doi [\mnras]
  {10.1111/j.1365-2966.2008.13830.x}, \href
  {http://adsabs.harvard.edu/abs/2008MNRAS.390..920O} {390, 920}

\bibitem[\protect\citeauthoryear{{Ostriker} \& {Shetty}}{{Ostriker} \&
  {Shetty}}{2011}]{Ostriker_2011}
{Ostriker} E.~C.,  {Shetty} R.,  2011, \mn@doi [\apj]
  {10.1088/0004-637X/731/1/41}, \href
  {http://adsabs.harvard.edu/abs/2011ApJ...731...41O} {731, 41}

\bibitem[\protect\citeauthoryear{{Parkinson}, {Cole}  \& {Helly}}{{Parkinson}
  et~al.}{2008}]{Parkinson_2008}
{Parkinson} H.,  {Cole} S.,   {Helly} J.,  2008, \mn@doi [\mnras]
  {10.1111/j.1365-2966.2007.12517.x}, \href
  {http://adsabs.harvard.edu/abs/2008MNRAS.383..557P} {383, 557}

\bibitem[\protect\citeauthoryear{{Parmentier} \& {Gilmore}}{{Parmentier} \&
  {Gilmore}}{2005}]{Parmentier_2005}
{Parmentier} G.,  {Gilmore} G.,  2005, \mn@doi [\mnras]
  {10.1111/j.1365-2966.2005.09455.x}, \href
  {http://adsabs.harvard.edu/abs/2005MNRAS.363..326P} {363, 326}

\bibitem[\protect\citeauthoryear{{Peebles}}{{Peebles}}{1984}]{Peebles_1984}
{Peebles} P.~J.~E.,  1984, \mn@doi [\apj] {10.1086/161714}, \href
  {http://adsabs.harvard.edu/abs/1984ApJ...277..470P} {277, 470}

\bibitem[\protect\citeauthoryear{{Peebles} \& {Dicke}}{{Peebles} \&
  {Dicke}}{1968}]{Peebles_1968}
{Peebles} P.~J.~E.,  {Dicke} R.~H.,  1968, \mn@doi [\apj] {10.1086/149811},
  \href {http://adsabs.harvard.edu/abs/1968ApJ...154..891P} {154, 891}

\bibitem[\protect\citeauthoryear{{Peng}}{{Peng}}{2007}]{Peng_2007}
{Peng} C.~Y.,  2007, \mn@doi [\apj] {10.1086/522774}, \href
  {http://adsabs.harvard.edu/abs/2007ApJ...671.1098P} {671, 1098}

\bibitem[\protect\citeauthoryear{{Peng} et~al.,}{{Peng}
  et~al.}{2006}]{Peng_2006}
{Peng} E.~W.,  et~al., 2006, \mn@doi [\apj] {10.1086/498210}, \href
  {http://adsabs.harvard.edu/abs/2006ApJ...639...95P} {639, 95}

\bibitem[\protect\citeauthoryear{{Peng} et~al.,}{{Peng}
  et~al.}{2008}]{Peng_2008}
{Peng} E.~W.,  et~al., 2008, \mn@doi [\apj] {10.1086/587951}, \href
  {http://adsabs.harvard.edu/abs/2008ApJ...681..197P} {681, 197}

\bibitem[\protect\citeauthoryear{{Peng} et~al.,}{{Peng}
  et~al.}{2011}]{Peng_2011}
{Peng} E.~W.,  et~al., 2011, \mn@doi [\apj] {10.1088/0004-637X/730/1/23}, \href
  {http://adsabs.harvard.edu/abs/2011ApJ...730...23P} {730, 23}

\bibitem[\protect\citeauthoryear{{Pfeffer}, {Kruijssen}, {Crain}  \&
  {Bastian}}{{Pfeffer} et~al.}{2018}]{Pfeffer_2018}
{Pfeffer} J.,  {Kruijssen} J.~M.~D.,  {Crain} R.~A.,   {Bastian} N.,  2018,
  \mn@doi [\mnras] {10.1093/mnras/stx3124}, \href
  {http://adsabs.harvard.edu/abs/2018MNRAS.475.4309P} {475, 4309}

\bibitem[\protect\citeauthoryear{{Piotto} et~al.,}{{Piotto}
  et~al.}{2015}]{Piotto_2015}
{Piotto} G.,  et~al., 2015, \mn@doi [\aj] {10.1088/0004-6256/149/3/91}, \href
  {http://adsabs.harvard.edu/abs/2015AJ....149...91P} {149, 91}

\bibitem[\protect\citeauthoryear{{Planck Collaboration} et~al.,}{{Planck
  Collaboration} et~al.}{2016}]{Plank_2016}
{Planck Collaboration} et~al., 2016, \mn@doi [\aap]
  {10.1051/0004-6361/201525830}, \href
  {http://adsabs.harvard.edu/abs/2016A%26A...594A..13P} {594, A13}

\bibitem[\protect\citeauthoryear{{Portegies Zwart}, {McMillan}  \&
  {Gieles}}{{Portegies Zwart} et~al.}{2010}]{Zwart_2010}
{Portegies Zwart} S.~F.,  {McMillan} S.~L.~W.,   {Gieles} M.,  2010, \mn@doi
  [\araa] {10.1146/annurev-astro-081309-130834}, \href
  {http://adsabs.harvard.edu/abs/2010ARA%26A..48..431P} {48, 431}

\bibitem[\protect\citeauthoryear{{Prieto} \& {Gnedin}}{{Prieto} \&
  {Gnedin}}{2008}]{Prieto_2008}
{Prieto} J.~L.,  {Gnedin} O.~Y.,  2008, \mn@doi [\apj] {10.1086/591777}, \href
  {http://adsabs.harvard.edu/abs/2008ApJ...689..919P} {689, 919}

\bibitem[\protect\citeauthoryear{{Raskutti}, {Ostriker}  \&
  {Skinner}}{{Raskutti} et~al.}{2016}]{Raskutti_2016}
{Raskutti} S.,  {Ostriker} E.~C.,   {Skinner} M.~A.,  2016, \mn@doi [\apj]
  {10.3847/0004-637X/829/2/130}, \href
  {http://adsabs.harvard.edu/abs/2016ApJ...829..130R} {829, 130}

\bibitem[\protect\citeauthoryear{{Renzini}}{{Renzini}}{2017}]{Renzini_2017}
{Renzini} A.,  2017, \mn@doi [\mnras] {10.1093/mnrasl/slx057}, \href
  {http://adsabs.harvard.edu/abs/2017MNRAS.469L..63R} {469, L63}

\bibitem[\protect\citeauthoryear{{Richtler}}{{Richtler}}{2006}]{Richtler_2006}
{Richtler} T.,  2006, Bulletin of the Astronomical Society of India, \href
  {http://adsabs.harvard.edu/abs/2006BASI...34...83R} {34, 83}

\bibitem[\protect\citeauthoryear{{Rodr{\'{\i}}guez-Puebla}, {Primack},
  {Behroozi}  \& {Faber}}{{Rodr{\'{\i}}guez-Puebla}
  et~al.}{2016a}]{Rodriguez_2016b}
{Rodr{\'{\i}}guez-Puebla} A.,  {Primack} J.~R.,  {Behroozi} P.,   {Faber}
  S.~M.,  2016a, \mn@doi [\mnras] {10.1093/mnras/stv2513}, \href
  {http://adsabs.harvard.edu/abs/2016MNRAS.455.2592R} {455, 2592}

\bibitem[\protect\citeauthoryear{{Rodr{\'{\i}}guez-Puebla}, {Behroozi},
  {Primack}, {Klypin}, {Lee}  \& {Hellinger}}{{Rodr{\'{\i}}guez-Puebla}
  et~al.}{2016b}]{Rodriguez_2016}
{Rodr{\'{\i}}guez-Puebla} A.,  {Behroozi} P.,  {Primack} J.,  {Klypin} A.,
  {Lee} C.,   {Hellinger} D.,  2016b, \mn@doi [\mnras] {10.1093/mnras/stw1705},
  \href {http://adsabs.harvard.edu/abs/2016MNRAS.462..893R} {462, 893}

\bibitem[\protect\citeauthoryear{{Rosenblatt}, {Faber}  \&
  {Blumenthal}}{{Rosenblatt} et~al.}{1988}]{Rosenblatt_1988}
{Rosenblatt} E.~I.,  {Faber} S.~M.,   {Blumenthal} G.~R.,  1988, \mn@doi [\apj]
  {10.1086/166466}, \href {http://adsabs.harvard.edu/abs/1988ApJ...330..191R}
  {330, 191}

\bibitem[\protect\citeauthoryear{{Sanders} et~al.,}{{Sanders}
  et~al.}{2012}]{Sanders_2012}
{Sanders} N.~E.,  et~al., 2012, \mn@doi [\apj] {10.1088/0004-637X/758/2/132},
  \href {http://adsabs.harvard.edu/abs/2012ApJ...758..132S} {758, 132}

\bibitem[\protect\citeauthoryear{{Santos}}{{Santos}}{2003}]{Santos_2003}
{Santos} M.~R.,  2003, in {Kissler-Patig} M.,  ed., Extragalactic Globular
  Cluster Systems. p.~348, \mn@doi{10.1007/10857603_53}

\bibitem[\protect\citeauthoryear{{Schiavon} et~al.,}{{Schiavon}
  et~al.}{2017}]{Schiavon_2017}
{Schiavon} R.~P.,  et~al., 2017, \mn@doi [\mnras] {10.1093/mnras/stw2162},
  \href {http://adsabs.harvard.edu/abs/2017MNRAS.465..501S} {465, 501}

\bibitem[\protect\citeauthoryear{{Schroetter}, {Bouch{\'e}}, {P{\'e}roux},
  {Murphy}, {Contini}  \& {Finley}}{{Schroetter}
  et~al.}{2015}]{Schroetter_2015}
{Schroetter} I.,  {Bouch{\'e}} N.,  {P{\'e}roux} C.,  {Murphy} M.~T.,
  {Contini} T.,   {Finley} H.,  2015, \mn@doi [\apj]
  {10.1088/0004-637X/804/2/83}, \href
  {http://adsabs.harvard.edu/abs/2015ApJ...804...83S} {804, 83}

\bibitem[\protect\citeauthoryear{{Shapiro}, {Genzel}  \& {F{\"o}rster
  Schreiber}}{{Shapiro} et~al.}{2010}]{Shapiro_2010}
{Shapiro} K.~L.,  {Genzel} R.,   {F{\"o}rster Schreiber} N.~M.,  2010, \mn@doi
  [\mnras] {10.1111/j.1745-3933.2010.00810.x}, \href
  {http://adsabs.harvard.edu/abs/2010MNRAS.403L..36S} {403, L36}

\bibitem[\protect\citeauthoryear{{Shibuya}, {Ouchi}  \& {Harikane}}{{Shibuya}
  et~al.}{2015}]{Shibuya_2015}
{Shibuya} T.,  {Ouchi} M.,   {Harikane} Y.,  2015, \mn@doi [\apjs]
  {10.1088/0067-0049/219/2/15}, \href
  {http://adsabs.harvard.edu/abs/2015ApJS..219...15S} {219, 15}

\bibitem[\protect\citeauthoryear{{Skinner} \& {Ostriker}}{{Skinner} \&
  {Ostriker}}{2015}]{Skinner_2015}
{Skinner} M.~A.,  {Ostriker} E.~C.,  2015, \mn@doi [\apj]
  {10.1088/0004-637X/809/2/187}, \href
  {http://adsabs.harvard.edu/abs/2015ApJ...809..187S} {809, 187}

\bibitem[\protect\citeauthoryear{{Spitler} \& {Forbes}}{{Spitler} \&
  {Forbes}}{2009}]{Spitler_2009}
{Spitler} L.~R.,  {Forbes} D.~A.,  2009, \mn@doi [\mnras]
  {10.1111/j.1745-3933.2008.00567.x}, \href
  {http://adsabs.harvard.edu/abs/2009MNRAS.392L...1S} {392, L1}

\bibitem[\protect\citeauthoryear{{Spitler}, {Forbes}, {Strader}, {Brodie}  \&
  {Gallagher}}{{Spitler} et~al.}{2008}]{Spitler_2008}
{Spitler} L.~R.,  {Forbes} D.~A.,  {Strader} J.,  {Brodie} J.~P.,   {Gallagher}
  J.~S.,  2008, \mn@doi [\mnras] {10.1111/j.1365-2966.2007.12823.x}, \href
  {http://adsabs.harvard.edu/abs/2008MNRAS.385..361S} {385, 361}

\bibitem[\protect\citeauthoryear{{Spitzer}}{{Spitzer}}{1987}]{Spitzer_1987}
{Spitzer} L.,  1987, {Dynamical evolution of globular clusters}

\bibitem[\protect\citeauthoryear{{Strader}, {Brodie}, {Cenarro}, {Beasley}  \&
  {Forbes}}{{Strader} et~al.}{2005}]{Strader_2005}
{Strader} J.,  {Brodie} J.~P.,  {Cenarro} A.~J.,  {Beasley} M.~A.,   {Forbes}
  D.~A.,  2005, \mn@doi [\aj] {10.1086/432717}, \href
  {http://adsabs.harvard.edu/abs/2005AJ....130.1315S} {130, 1315}

\bibitem[\protect\citeauthoryear{{Strader}, {Brodie}, {Spitler}  \&
  {Beasley}}{{Strader} et~al.}{2006}]{Strader_2006}
{Strader} J.,  {Brodie} J.~P.,  {Spitler} L.,   {Beasley} M.~A.,  2006, \mn@doi
  [\aj] {10.1086/509124}, \href
  {http://adsabs.harvard.edu/abs/2006AJ....132.2333S} {132, 2333}

\bibitem[\protect\citeauthoryear{{Strader} et~al.,}{{Strader}
  et~al.}{2011}]{Strader_2011}
{Strader} J.,  et~al., 2011, \mn@doi [\apjs] {10.1088/0067-0049/197/2/33},
  \href {http://adsabs.harvard.edu/abs/2011ApJS..197...33S} {197, 33}

\bibitem[\protect\citeauthoryear{{Tang}, {Bressan}, {Rosenfield}, {Slemer},
  {Marigo}, {Girardi}  \& {Bianchi}}{{Tang} et~al.}{2014}]{Tang_2014}
{Tang} J.,  {Bressan} A.,  {Rosenfield} P.,  {Slemer} A.,  {Marigo} P.,
  {Girardi} L.,   {Bianchi} L.,  2014, \mn@doi [\mnras]
  {10.1093/mnras/stu2029}, \href
  {http://adsabs.harvard.edu/abs/2014MNRAS.445.4287T} {445, 4287}

\bibitem[\protect\citeauthoryear{{Thompson} \& {Krumholz}}{{Thompson} \&
  {Krumholz}}{2016}]{Thompson_2016}
{Thompson} T.~A.,  {Krumholz} M.~R.,  2016, \mn@doi [\mnras]
  {10.1093/mnras/stv2331}, \href
  {http://adsabs.harvard.edu/abs/2016MNRAS.455..334T} {455, 334}

\bibitem[\protect\citeauthoryear{{Tonini}}{{Tonini}}{2013}]{Tonini_2013}
{Tonini} C.,  2013, \mn@doi [\apj] {10.1088/0004-637X/762/1/39}, \href
  {http://adsabs.harvard.edu/abs/2013ApJ...762...39T} {762, 39}

\bibitem[\protect\citeauthoryear{{Tremonti} et~al.,}{{Tremonti}
  et~al.}{2004}]{Tremonti_2004}
{Tremonti} C.~A.,  et~al., 2004, \mn@doi [\apj] {10.1086/423264}, \href
  {http://adsabs.harvard.edu/abs/2004ApJ...613..898T} {613, 898}

\bibitem[\protect\citeauthoryear{{Trenti}, {Padoan}  \& {Jimenez}}{{Trenti}
  et~al.}{2015}]{Trenti_2015}
{Trenti} M.,  {Padoan} P.,   {Jimenez} R.,  2015, \mn@doi [\apjl]
  {10.1088/2041-8205/808/2/L35}, \href
  {http://adsabs.harvard.edu/abs/2015ApJ...808L..35T} {808, L35}

\bibitem[\protect\citeauthoryear{{Tsang} \& {Milosavljevic}}{{Tsang} \&
  {Milosavljevic}}{2017}]{Tsang_2017}
{Tsang} B.~T.-H.,  {Milosavljevic} M.,  2017, preprint, \href
  {http://adsabs.harvard.edu/abs/2017arXiv170907539T} {} (\mn@eprint {arXiv}
  {1709.07539})

\bibitem[\protect\citeauthoryear{{Tutukov}}{{Tutukov}}{1978}]{Tutukov_1978}
{Tutukov} A.~V.,  1978, \aap, \href
  {http://adsabs.harvard.edu/abs/1978A%26A....70...57T} {70, 57}

\bibitem[\protect\citeauthoryear{{Vale} \& {Ostriker}}{{Vale} \&
  {Ostriker}}{2006}]{Vale_2006}
{Vale} A.,  {Ostriker} J.~P.,  2006, \mn@doi [\mnras]
  {10.1111/j.1365-2966.2006.10605.x}, \href
  {http://adsabs.harvard.edu/abs/2006MNRAS.371.1173V} {371, 1173}

\bibitem[\protect\citeauthoryear{{Van Der Walt}, {Colbert}  \&
  {Varoquaux}}{{Van Der Walt} et~al.}{2011}]{vanderwalt_2011}
{Van Der Walt} S.,  {Colbert} S.~C.,   {Varoquaux} G.,  2011, preprint, \href
  {http://adsabs.harvard.edu/abs/2011arXiv1102.1523V} {} (\mn@eprint {arXiv}
  {1102.1523})

\bibitem[\protect\citeauthoryear{{VandenBerg}, {Brogaard}, {Leaman}  \&
  {Casagrande}}{{VandenBerg} et~al.}{2013}]{VandenBerg_2013}
{VandenBerg} D.~A.,  {Brogaard} K.,  {Leaman} R.,   {Casagrande} L.,  2013,
  \mn@doi [\apj] {10.1088/0004-637X/775/2/134}, \href
  {http://adsabs.harvard.edu/abs/2013ApJ...775..134V} {775, 134}

\bibitem[\protect\citeauthoryear{{Vanzella} et~al.,}{{Vanzella}
  et~al.}{2017}]{Vanzella_2017}
{Vanzella} E.,  et~al., 2017, \mn@doi [\mnras] {10.1093/mnras/stx351}, \href
  {http://adsabs.harvard.edu/abs/2017MNRAS.467.4304V} {467, 4304}

\bibitem[\protect\citeauthoryear{{Vesperini} \& {Zepf}}{{Vesperini} \&
  {Zepf}}{2003}]{Vesperini_2003}
{Vesperini} E.,  {Zepf} S.~E.,  2003, \mn@doi [\apjl] {10.1086/375313}, \href
  {http://adsabs.harvard.edu/abs/2003ApJ...587L..97V} {587, L97}

\bibitem[\protect\citeauthoryear{{Wilson}, {Harris}, {Longden}  \&
  {Scoville}}{{Wilson} et~al.}{2006}]{Wilson_2006}
{Wilson} C.~D.,  {Harris} W.~E.,  {Longden} R.,   {Scoville} N.~Z.,  2006,
  \mn@doi [\apj] {10.1086/500577}, \href
  {http://adsabs.harvard.edu/abs/2006ApJ...641..763W} {641, 763}

\bibitem[\protect\citeauthoryear{{Wise}, {Turk}, {Norman}  \& {Abel}}{{Wise}
  et~al.}{2012}]{Wise_2012}
{Wise} J.~H.,  {Turk} M.~J.,  {Norman} M.~L.,   {Abel} T.,  2012, \mn@doi
  [\apj] {10.1088/0004-637X/745/1/50}, \href
  {http://adsabs.harvard.edu/abs/2012ApJ...745...50W} {745, 50}

\bibitem[\protect\citeauthoryear{{Woodley}, {Harris}, {Puzia}, {G{\'o}mez},
  {Harris}  \& {Geisler}}{{Woodley} et~al.}{2010}]{Woodley_2010}
{Woodley} K.~A.,  {Harris} W.~E.,  {Puzia} T.~H.,  {G{\'o}mez} M.,  {Harris}
  G.~L.~H.,   {Geisler} D.,  2010, \mn@doi [\apj]
  {10.1088/0004-637X/708/2/1335}, \href
  {http://adsabs.harvard.edu/abs/2010ApJ...708.1335W} {708, 1335}

\bibitem[\protect\citeauthoryear{{Yang}, {Mo}, {van den Bosch}, {Pasquali},
  {Li}  \& {Barden}}{{Yang} et~al.}{2007}]{Yang_2007}
{Yang} X.,  {Mo} H.~J.,  {van den Bosch} F.~C.,  {Pasquali} A.,  {Li} C.,
  {Barden} M.,  2007, \mn@doi [\apj] {10.1086/522027}, \href
  {http://adsabs.harvard.edu/abs/2007ApJ...671..153Y} {671, 153}

\bibitem[\protect\citeauthoryear{{Yoon}, {Yi}  \& {Lee}}{{Yoon}
  et~al.}{2006}]{Yoon_2006}
{Yoon} S.-J.,  {Yi} S.~K.,   {Lee} Y.-W.,  2006, \mn@doi [Science]
  {10.1126/science.1122294}, \href
  {http://adsabs.harvard.edu/abs/2006Sci...311.1129Y} {311, 1129}

\bibitem[\protect\citeauthoryear{{Zaritsky}, {Crnojevi{\'c}}  \&
  {Sand}}{{Zaritsky} et~al.}{2016}]{Zaritsky_2016}
{Zaritsky} D.,  {Crnojevi{\'c}} D.,   {Sand} D.~J.,  2016, \mn@doi [\apjl]
  {10.3847/2041-8205/826/1/L9}, \href
  {http://adsabs.harvard.edu/abs/2016ApJ...826L...9Z} {826, L9}

\bibitem[\protect\citeauthoryear{{Zepf} \& {Ashman}}{{Zepf} \&
  {Ashman}}{1993}]{Zepf_1993}
{Zepf} S.~E.,  {Ashman} K.~M.,  1993, \mn@doi [\mnras]
  {10.1093/mnras/264.3.611}, \href
  {http://adsabs.harvard.edu/abs/1993MNRAS.264..611Z} {264, 611}

\bibitem[\protect\citeauthoryear{{Zhao}}{{Zhao}}{2005}]{Zhao_2005}
{Zhao} H.,  2005, ArXiv Astrophysics e-prints, \href
  {http://adsabs.harvard.edu/abs/2005astro.ph..8635Z} {}

\bibitem[\protect\citeauthoryear{{Zick}, {Weisz}  \& {Boylan-Kolchin}}{{Zick}
  et~al.}{2018}]{Zick_2018}
{Zick} T.~O.,  {Weisz} D.~R.,   {Boylan-Kolchin} M.,  2018, preprint, \href
  {http://adsabs.harvard.edu/abs/2018arXiv180206801Z} {} (\mn@eprint {arXiv}
  {1802.06801})

\bibitem[\protect\citeauthoryear{{de Boer} \& {Fraser}}{{de Boer} \&
  {Fraser}}{2016}]{deBoer_2016}
{de Boer} T.~J.~L.,  {Fraser} M.,  2016, \mn@doi [\aap]
  {10.1051/0004-6361/201527580}, \href
  {http://adsabs.harvard.edu/abs/2016A%26A...590A..35D} {590, A35}

\bibitem[\protect\citeauthoryear{{van Dokkum} et~al.,}{{van Dokkum}
  et~al.}{2018}]{vanDokkum_2018}
{van Dokkum} P.,  et~al., 2018, \mn@doi [\apjl] {10.3847/2041-8213/aab60b},
  \href {http://adsabs.harvard.edu/abs/2018ApJ...856L..30V} {856, L30}

\makeatother
\end{thebibliography}



\appendix


\section{Cold gas accretion rate from total accretion rate}
\label{sec:prev_fdbk}

\begin{figure*}
\includegraphics[width=\textwidth]{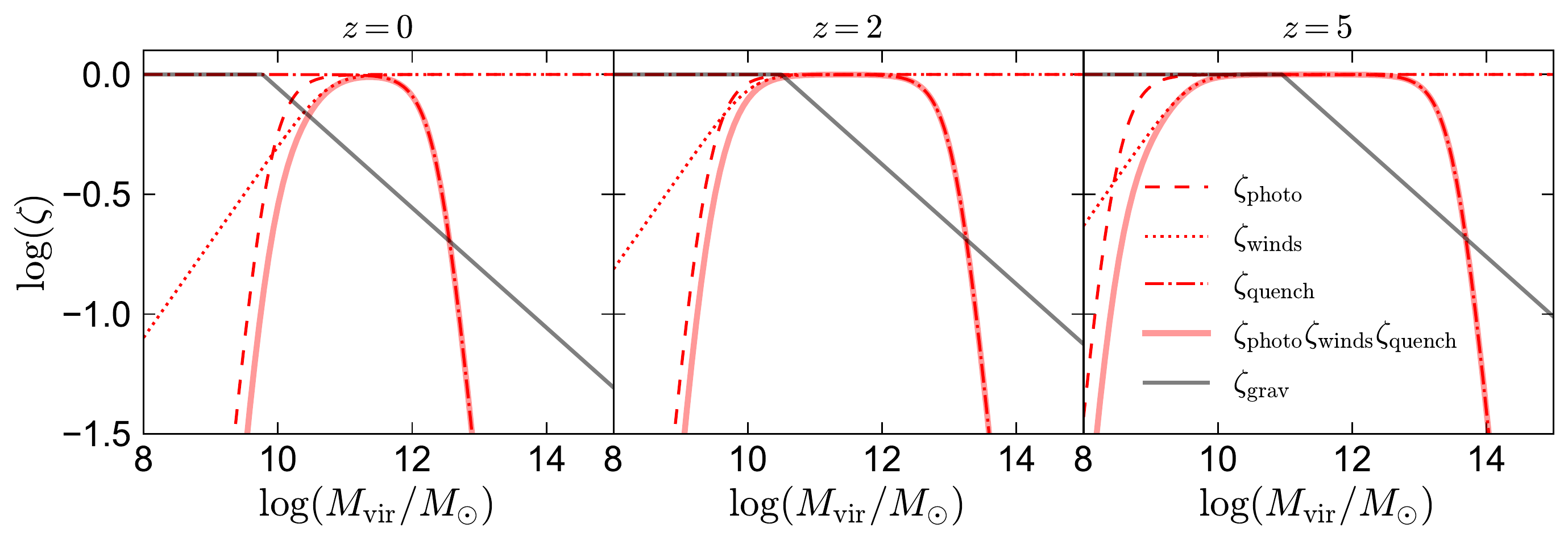}
\caption{Components of the feedback parameter $\zeta$ that determines the cold gas mass accreted at each accretion event relative to the cosmic baryon fraction. Solid red curve represents the total reduction in the cold gas mass of a halo of mass $M_{\rm vir}$ being accreted at redshift $z$; dashed and dotted red lines show the contributions of individual feedback sources. Solid gray curve represents the additional suppression of cold gas accretion due to hot halo gas during an accretion event {\it into} a halo of mass $M_{\rm vir}$. Equations~\ref{eq:zeta}-\ref{eq:zeta_grav} give parameterizations of our adopted $\zeta$. }
\label{fig:zeta2D}
\end{figure*}

As discussed in Section~\ref{sec:implementation}, the equilibrium model that we use to estimate the SFR and gas surface density throughout a merger tree depends on the inflow rate of cold gas, $\dot{M}_{\rm gas,\,in}$. We calculate $\dot{M}_{\rm gas,\,in}$ from the total accretion rate using a suppression function $\zeta\left(M_{{\rm vir,1}},M_{{\rm vir,2}},z\right)$ (see Equation~\ref{eq:mdot_gas_in}) to crudely account for the effects of gas heating by the UV background, stellar feedback-driven winds, AGN quenching, and ambient hot gas in the primary halo. Here $M_{\rm vir,1}$ and $M_{\rm vir,2}$ represent the mass of the primary and accreted halo, respectively. 

Following \citet{Dave_2012}, we model $\zeta$ as a product of several different terms accounting for different processes: 
\begin{align}
\label{eq:zeta}
\zeta=\zeta_{{\rm photo}}\times\zeta_{{\rm winds}}\times\zeta_{{\rm quench}}\times\zeta_{{\rm grav}}.
\end{align}
$\zeta_{\rm photo}$, $\zeta_{\rm winds}$, and $\zeta_{\rm quench}$ represent the reduction in the cold gas content, relative to the cosmic baryon fraction, of a galaxy being accreted; they are functions of $M_{\rm vir,2}$. On the other hand, $\zeta_{\rm grav}$ represents the suppression of cold gas accretion due to the hot gas in the primary halo; it is a function of $M_{\rm vir,1}$.

$\zeta_{\rm photo}$ represents the decrease in the cold gas mass of low-mass halos due to photoionization heating after the epoch of reionization. It drops to 0 below the ``photosuppression mass,'' $M_{\gamma}(z)$, which increases from $\sim 10^{8}M_{\odot}$ during reionization to a few $\times\,10^{9}M_{\odot}$ at $z=0$. Following \citet[][their Equation 1]{Okamoto_2008}, this is parameterized as 
\begin{align}
\label{eq:zeta_phot}
\zeta_{{\rm photo}}=\left\{ 1+\left[2^{\alpha/\beta}-1\right]\left(\frac{M_{{\rm vir,2}}}{M_{\gamma}\left(z\right)}\right)^{-\alpha}\right\} ^{-\beta/\alpha},
\end{align}
where $\alpha = 2$ and $\beta = 3$. We calculate $M_{\gamma}(z)$ by interpolating on the results of the simulations presented in \citet[][their Figure 5]{Okamoto_2008}.

$\zeta_{\rm winds}$ represents the removal of cold gas from the accreted galaxy by winds prior to its accretion. The precise form of $\zeta_{\rm winds}$ is highly uncertain; but in general, winds are expected to affect low-mass galaxies more strongly than massive galaxies and to reduce the cold gas content more at late times than at early times. $\zeta_{\rm winds}$ is parameterized analogously to $\zeta_{\rm photo}$: 
\begin{align}
\label{eq:zeta_winds}
\zeta_{{\rm winds}}=\left\{ 1+\left[2^{\alpha/\beta}-1\right]\left(\frac{M_{{\rm vir,2}}}{M_{{\rm w}}\left(z\right)}\right)^{-\alpha}\right\} ^{-\beta/\alpha},
\end{align}
with $\alpha = 2$, $\beta = 0.4$, and $M_{{\rm w}}\left(z\right)=10^{10}\left(1+z\right)^{-1.5}M_{\odot}$. 

$\zeta_{\rm quench}$ represents the effects of whatever processes heat gas in high-mass halos, likely connected to AGN. It drops to 0 above a ``quenching mass,'' $M_{\rm q}(z)$, which increases at higher redshift. It is parameterized as: 
\begin{align}
\label{eq:zeta_quench}
\zeta_{{\rm quench}}=\left\{ 1+\left[2^{\alpha/\beta}-1\right]\left(\frac{M_{{\rm vir,2}}}{M_{{\rm q}}\left(z\right)}\right)^{\alpha}\right\} ^{-\beta/\alpha},
\end{align} 
where $M_{{\rm q}}\left(z\right)=10^{12.3}\left(1+z\right)^{1.47}M_{\odot}$ and $\alpha = 2$ and $\beta=3$. 

Finally, $\zeta_{\rm grav}$ represents the suppression of cold gas accretion due to hot halo gas, which is heated by virial shocks. Following \citet{FG_2011} and \citet{Dave_2012}, we parameterize it as 
\begin{align}
\label{eq:zeta_grav}
\zeta_{{\rm grav}}=0.47\left(\frac{1+z}{4}\right)^{0.38}\left(\frac{M_{{\rm vir,1}}}{10^{12}M_{\odot}}\right)^{-0.25}.
\end{align}
$\zeta_{\rm grav}$ suppresses cooling into high-mass halos, especially at lower redshifts. When Equation~\ref{eq:zeta_grav} exceeds unity, $\zeta_{\rm grav}$ is set to 1. 

The combined effects of our parameterization of $\zeta$ are shown in Figure~\ref{fig:zeta2D}. We emphasize that this model is largely phenomenological and is not expected to hold in detail for any galaxy. The main point of the model is to capture the facts that (a) cold gas fractions are higher at high redshift and in intermediate mass halos, and (b) gas accretion and cooling are suppressed in high-mass halos.

\section{Varying the critical density for cluster formation}
\label{sec:sig_crit}
\begin{figure}
\includegraphics[width=\columnwidth]{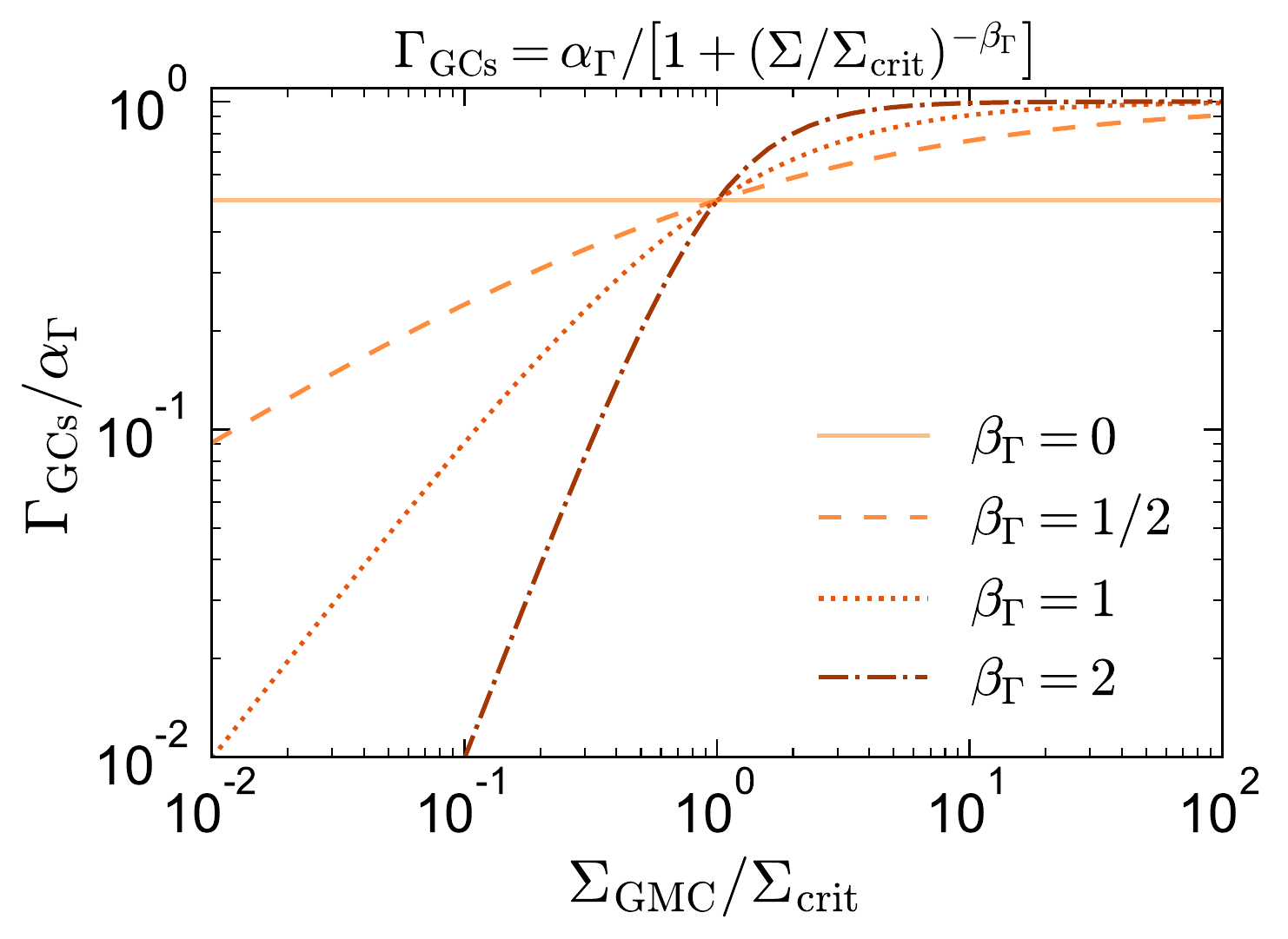}
\caption{Parametric form of the GC formation efficiency, $\Gamma_{\rm GCs}=\dot{M}_{\rm GCs}/{\rm SFR}$, assumed in our model (Equation~\ref{eq:Gamma}). $\Gamma_{\rm GCs}$ dictates the fraction of star formation that occurs in massive bound clusters that survive until $z=0$. In our model, $\Gamma_{\rm GCs}$ goes to 0 at $\Sigma_{\rm GMC} \ll \Sigma_{\rm crit}$ and plateaus at a value $\alpha_{\Gamma}$ at $\Sigma_{\rm GMC} \gg \Sigma_{\rm crit}$; the free parameter $\beta_{\Gamma}$ determines the steepness with which $\Gamma_{\rm GCs}$ falls off at low surface densities.}
\label{fig:sup_funcs}
\end{figure}

Figure~\ref{fig:sup_funcs} shows examples of the GC formation efficiency parameterization assumed in our model for several values of $\beta_{\Gamma}$. 

\begin{figure}
\includegraphics[width=\columnwidth]{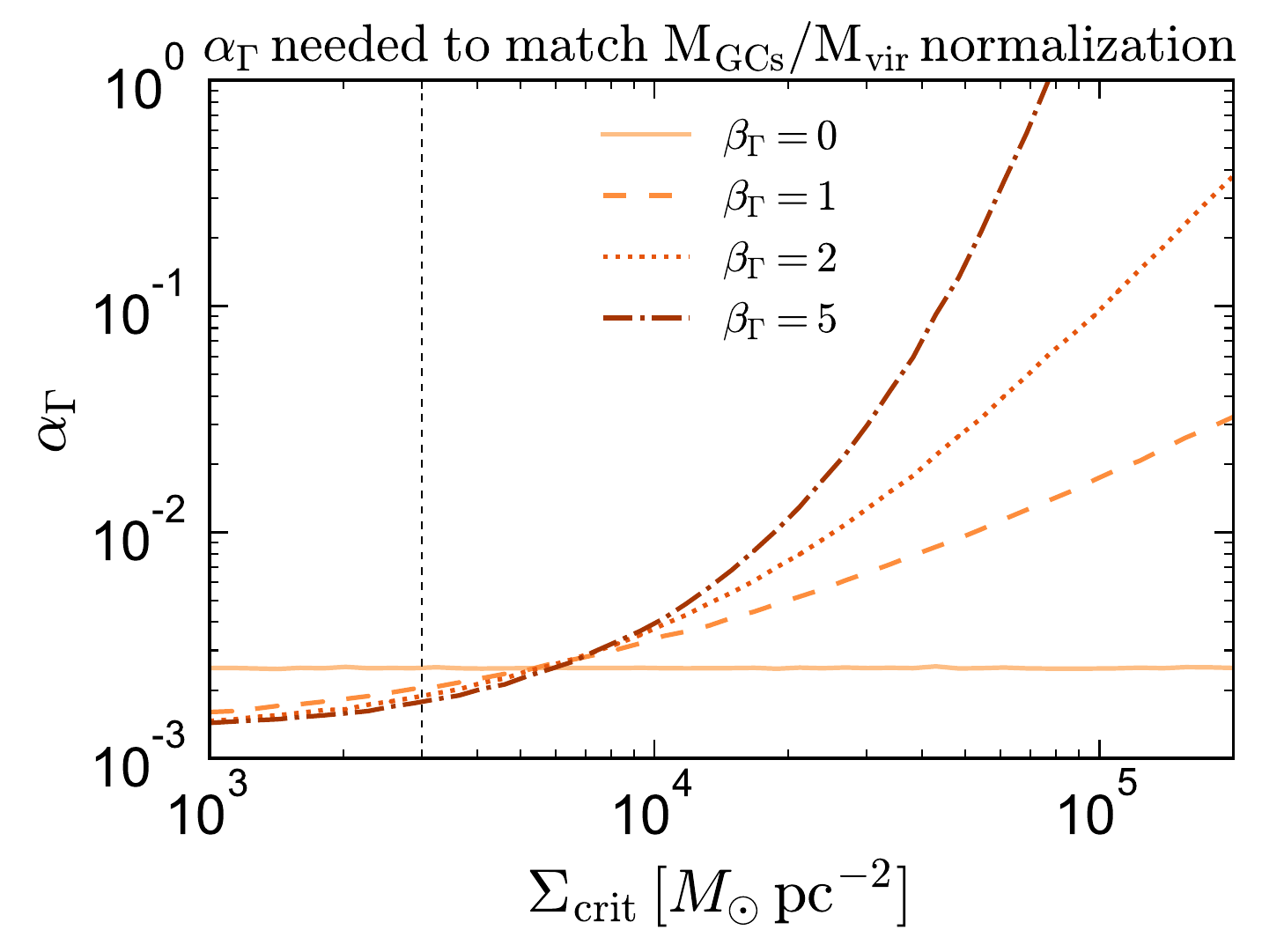}
\caption{Normalization of the cluster formation efficiency, $\alpha_{\Gamma}$ (Equation~\ref{eq:Gamma}), vs. $\Sigma_{\rm crit}$, the critical density above which the cluster formation efficiency $\Gamma_{\rm GCs}$ plateaus. For each $\beta_{\Gamma}$ and $\Sigma_{\rm crit}$, we plot the value of $\alpha_{\Gamma}$ that matches the  normalization of the observed GC-to-halo mass relation at the high-mass end (Figure~\ref{fig:mgc_mhalo}). Choosing a much higher value of $\Sigma_{\rm crit}$ than our default value of $3000\,M_{\odot}\,{\rm pc^{-2}}$ increases the value of $\alpha_{\Gamma}$. }
\label{fig:alpha_vs_sigcrit}
\end{figure}

Throughout our analysis, we fixed the value of $\Sigma_{\rm crit}$, the critical surface density above which the GC formation efficiency plateaus, to $3000\,M_{\odot}\,{\rm pc^{-2}}$. This is approximately the value predicted by analytic theory and found in idealized cloud-collapse simulations \citep[e.g.][]{Murray_2010, Fall_2010, Kim_2016, Grudic_2016, Grudic_2018}. As discussed in Section~\ref{sec:alpha_gamma}, this results in a low value of $\alpha_{\Gamma}$, implying that even at asymptotically high surface densities, the fraction of stars forming in proto-GCs is low. In Figure~\ref{fig:alpha_vs_sigcrit}, we investigate how changing the value of $\Sigma_{\rm crit}$ changes the value of $\alpha_{\Gamma}$ that is required to match the observed GC-to-halo mass relation at the high-mass end. We fix $\beta_{\eta} = 1/3$.

As expected, a higher value of $\Sigma_{\rm crit}$ requires a higher value of $\alpha_{\Gamma}$, because a smaller fraction of star forming events have $\Sigma_{\rm GMC} \gtrsim \Sigma_{\rm crit}$. However, even for large values of $\beta_{\Gamma}$ (i.e., a sharp truncation of GC formation at $\Sigma_{\rm GMC} < \Sigma_{\rm crit}$), it is necessary to increase $\Sigma_{\rm crit}$ by more than an order of magnitude in order to bring $\alpha_{\Gamma}$ to order unity. Such high values of $\Sigma_{\rm crit}$ significantly exceed those predicted by idealized cloud-collapse simulations. 

\begin{figure}
\includegraphics[width=\columnwidth]{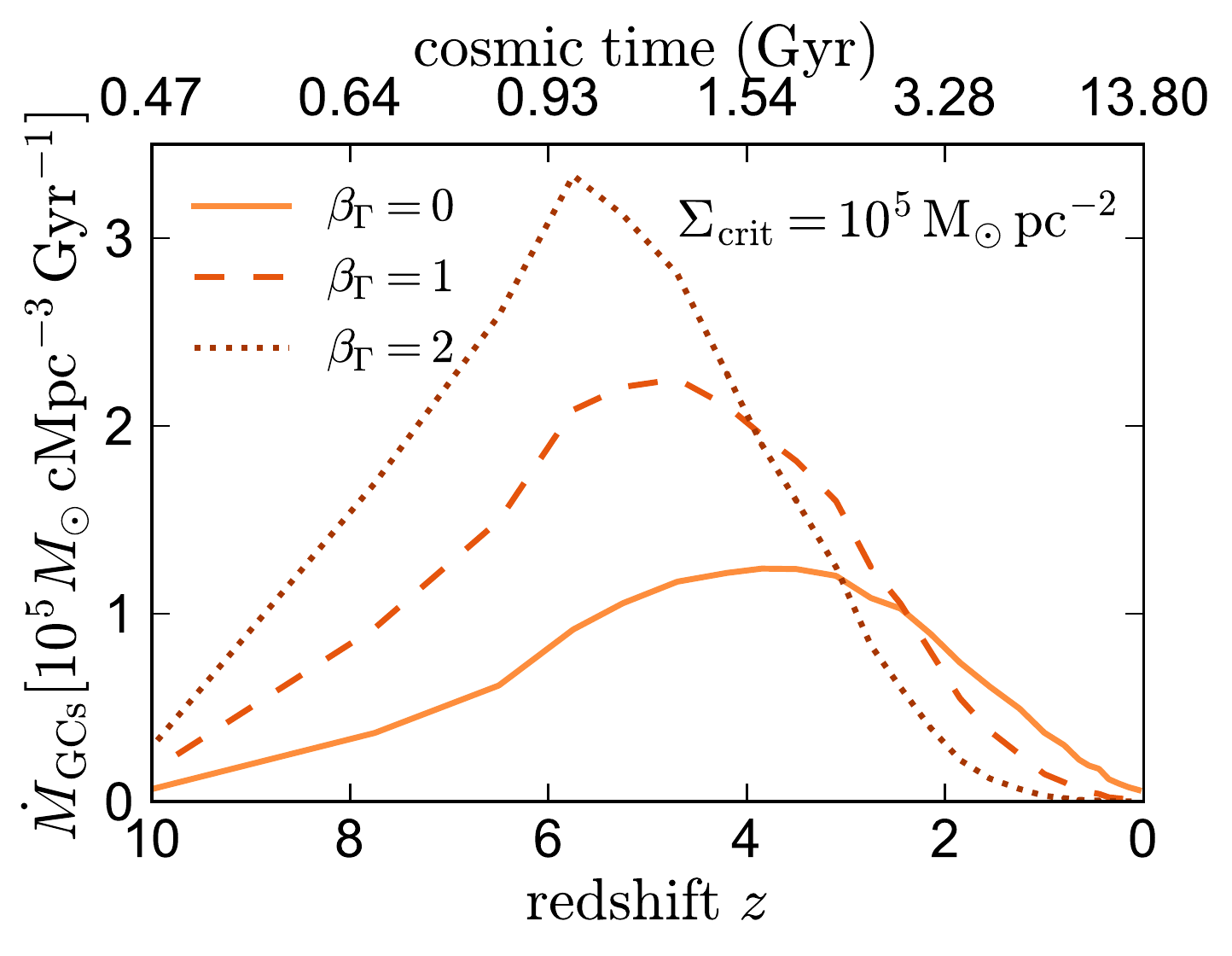}
\caption{Cosmic GC formation rate (similar to the top panel of Figure~\ref{fig:analytic}) using a higher critical density for GC formation. We fix $\beta_{\eta} = 1/3$. Increasing $\Sigma_{\rm crit}$ causes GCs to form earlier and makes the epoch of GC formation more sensitive to $\beta_{\Gamma}$.}
\label{fig:cosmic_SFGC_sigma}
\end{figure}

In Figure~\ref{fig:cosmic_SFGC_sigma}, we show the effects of increasing $\Sigma_{\rm crit}$ on the predicted cosmic GC formation rate. The solid line is the same as the solid line in the top panel of Figure~\ref{fig:analytic}, since the cluster formation efficiency is independent of $\Sigma_{\rm crit}$ for $\beta_{\Gamma} = 0$.  With $\Sigma_{\rm crit} = 10^5 M_{\odot}\,\rm pc^{-2}$, the cosmic GC formation rate varies more strongly with $\beta_{\Gamma}$, and increasing $\beta_{\Gamma}$ causes the epoch of GC formation to move to earlier times. This makes GCs contribute more to the UV luminosity density during reionzation, though still only at the $\sim$5\% level. Although we consider a value of $\Sigma_{\rm crit}$ as high as $10^5 M_{\odot}\,\rm pc^{-2}$ unlikely, it is possible that approximations in our model, such as the adopted merger timescale or the assumption of $\Sigma_{\rm GMC} = 5\times \Sigma_{\rm gas}$, overestimate the true value of $\Sigma_{\rm GMC}$. Using a higher value of $\Sigma_{\rm crit}$ has exactly the same effect on the predicted GC population as using a longer $\tau_{\rm merger}$ (Equation~\ref{eq:tau_merger}) or a lower value of $\Sigma_{\rm GMC}$ relative to $\Sigma_{\rm gas}$. 

\section{Stellar metallicity distribution}
\label{sec:stellar_mdf}
\begin{figure}
\includegraphics[width=\columnwidth]{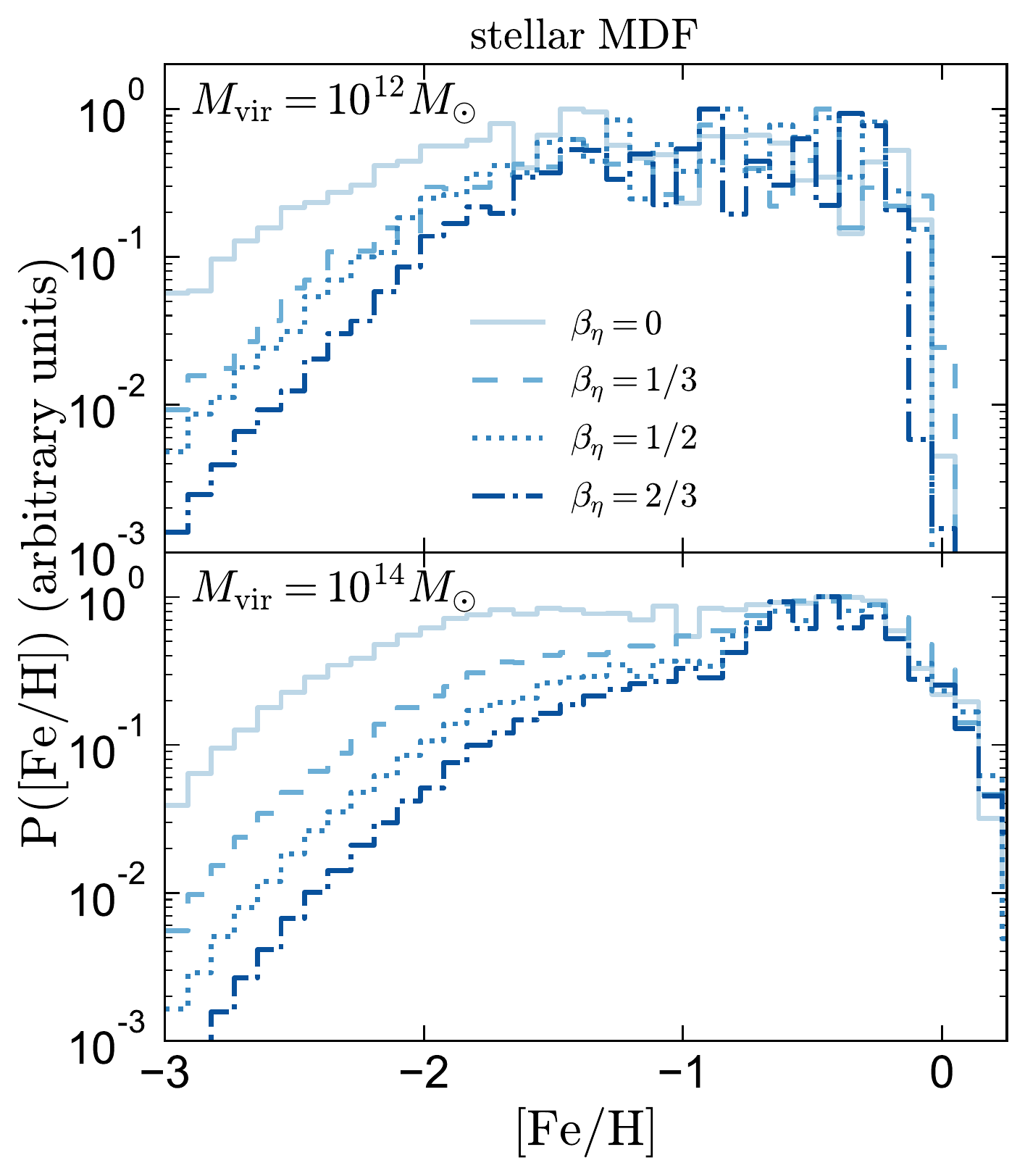}
\caption{Metallicity distribution of all stars in a halo with $M_{\rm vir}=10^{12}M_{\odot}$ (top) and $M_{\rm vir}=10^{14}M_{\odot}$ (bottom) at $z=0$, for different scalings of $\eta$ with halo mass. Increasing $\beta_{\eta}$ decreases the fraction of low-metallicity stars. Beyond the GC color distribution (Figure~\ref{fig:bimodality_surface}), the implied metallicity distribution for all stars provides a secondary constraint on $\beta_{\eta}$. The fiducial model with $\beta_{\eta} =1/3$ somewhat underestimates the mean metallicity of field stars.}
\label{fig:stellar_mdfs}
\end{figure}

Changing the value of $\beta_{\eta}$ changes the metallicity distribution for both GCs and field stars. The effect of varying $\beta_{\eta}$ on the metallicity of all stars in a halo (i.e., the result of integrating Equation~\ref{eq:sfr} over all nodes in the merger tree) is shown in Figure~\ref{fig:stellar_mdfs}. Increasing the value of $\beta_{\eta}$ suppresses star formation in low-mass halos and thus reduces the fraction of low-metallicity stars. This serves as an additional constraint on $\beta_{\eta}$: although models with $\beta_{\eta}=0$ can produce plausible GC metallicities, ages, and color distributions (Figures~\ref{fig:GC_mmr} and~\ref{fig:bimodality_surface}), $\beta_{\eta} \gtrsim 1/3$ is required for a plausible stellar metallicity distribution. 

The fiducial model somewhat underestimes the typical metallicity of field stars. At $M_{\rm vir} = 10^{12} M_{\odot}$, it predicts a mean metallicity for all stars in the halo of $\rm [Fe/H] \approx -1$; we find the same value for the simulated MW-mass galaxies studied in \citet{ElBadry_2018} to be $\rm [Fe/H] \approx -0.4$. Perhaps relatedly, the fiducial model predicts the cosmic star formation rate to peak at $z\sim 3.5$ (Figure~\ref{fig:analytic}), which is earlier than the value $z\sim 2$ found observationally. This may indicate that our approximations for the suppression of cold gas accretion (Appendix~\ref{sec:prev_fdbk}) prevent accretion at late times too strongly, or that the accretion-based model overestimates the SFR at early times due to a gas accumulate phase (see also \citealt{Rodriguez_2016b}).

\section{Effects of cluster disruption}
\label{sec:disrupt}

\begin{figure}
\includegraphics[width=\columnwidth]{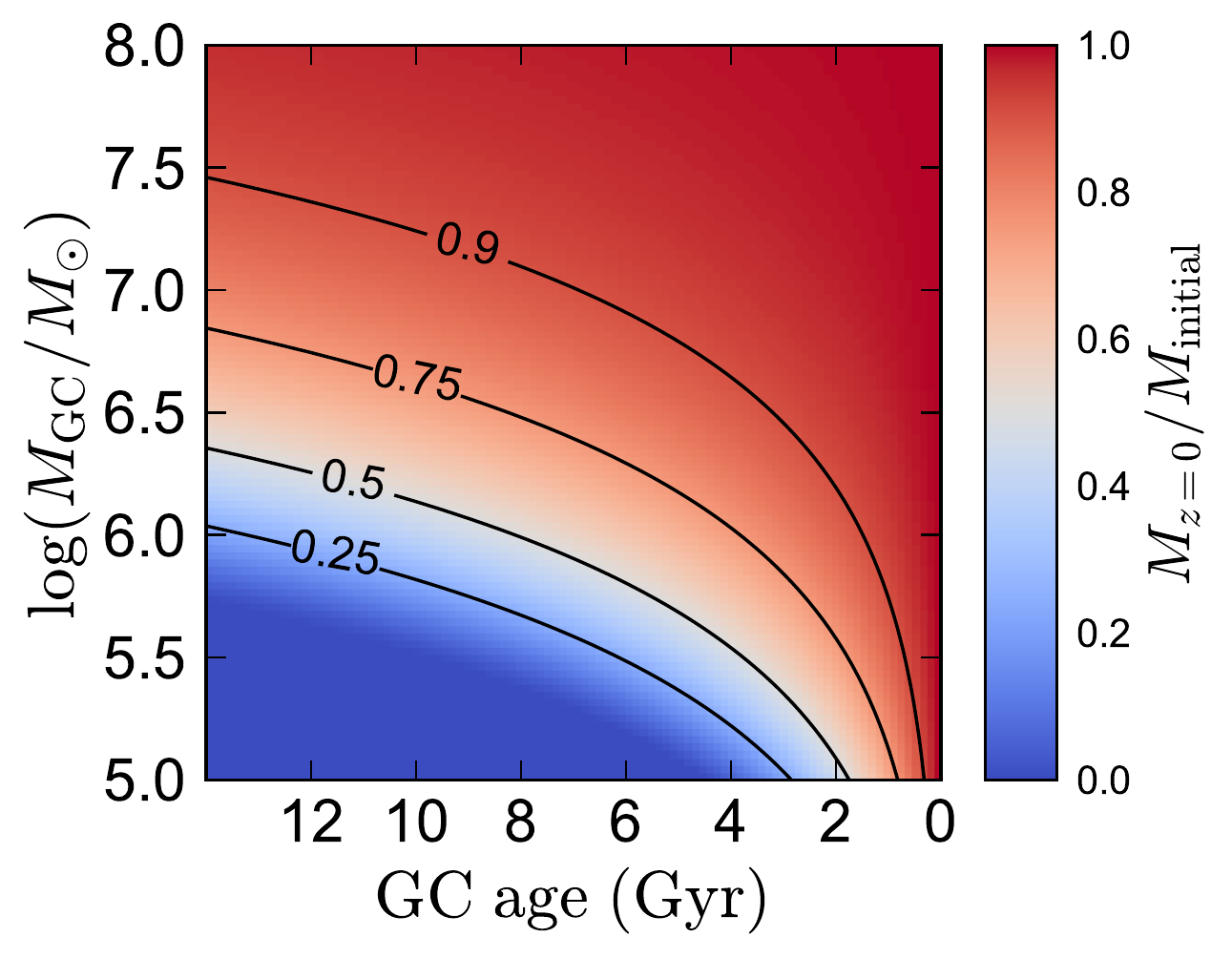}
\caption{Effective disruption model taken from \citet{Choksi_2018} and implemented in Figure~\ref{fig:disrupt_effect}. Color scale shows the fraction of a single GC's initial mass that survives at $z=0$; GCs in dark blue regions of parameter space are disrupted completely.}
\label{fig:disrupt_summary}
\end{figure}

Here we test the effects of a simplified model for GC mass loss and disruption due to both tidal fields and two-body evaporation. We use the model from \citet[][their Equation 9]{Choksi_2018}. The model attempts to account both for disruption due to tidal fields (averaging over all spatial distributions) and two-body evaporation; tidal effects are dominant for all but the least massive clusters. In this model, the mass of a GC at $z=0$ depends only on its age and initial mass. 

\begin{figure}
\includegraphics[width=\columnwidth]{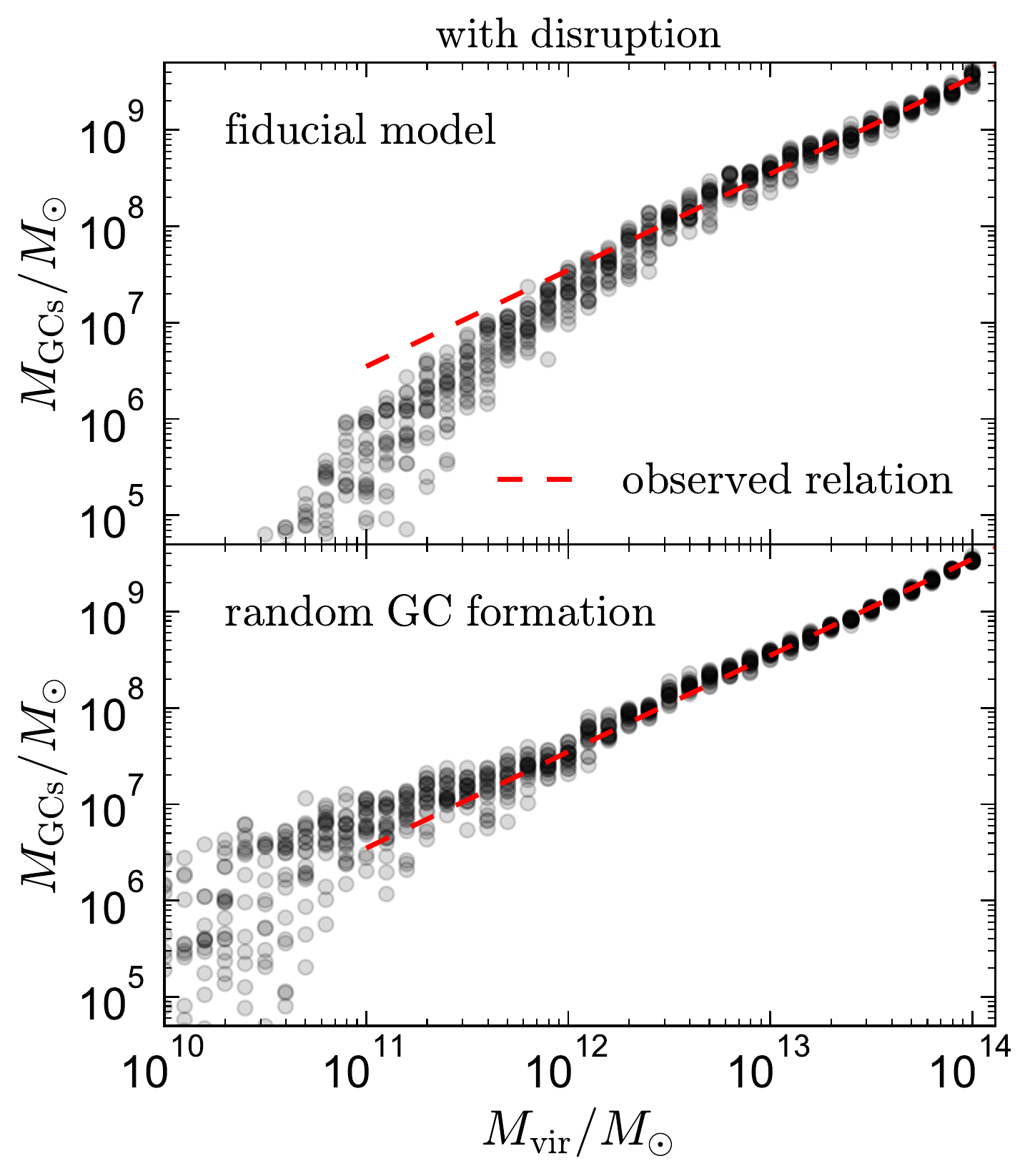}
\caption{GC-to-halo mass relation for our fiducial model (top) and random GC formation (bottom), but now including the age- and mass-dependent GC disruption model from \citet{Choksi_2018} (see Figure~\ref{fig:disrupt_summary}). Including disruption has little effect on the linearity of the GC-to-halo mass relation (compare to Figure~\ref{fig:mgc_mhalo}).}
\label{fig:disrupt_effect}
\end{figure}

The combined total effects of these processes as we implement them are illustrated in Figure~\ref{fig:disrupt_summary}. In this model, old GCs with initial masses $m\lesssim 10^{6} M_{\odot}$ are disrupted or lose a dominant fraction of their mass by $z=0$; GCs that form at later times or are more massive lose a smaller fraction of their initial mass. The disruption model was calibrated by \citet{Choksi_2018} to produce a realistic $z=0$ cluster mass function. \citet{Choksi_2018} also showed that the combination of this disruption model and their sampling procedure for drawing GCs masses (which we also adopt) reproduces the observed ``blue tilt''; i.e., the trend of more massive blue GCs to be more metal rich on average \citep[e.g.][]{Strader_2006}.

Figure~\ref{fig:disrupt_effect} shows the GC-to-halo mass relation predicted by our model after implementing the disruption model. Applying the GC disruption model causes the normalization of the GC-to-halo mass ratio at the high mass end to drop by a factor of 2.6 for the fiducial model and a factor of 4 for the random model, so we increase the free parameter $\alpha_{\Gamma}$ by a factor of 2.6 and increase the probability of each halo hosting a GC formation event in the random model by a factor of 4. After these adjustments are made, both the fiducial and random GC formation models produce the a similar constant GC-to-halo mass relation as when no disruption is included. Disruption primarily affects old and low-mass GCs, but the fraction of GCs in a halo at $z=0$ that are old or low mass is not a strong function of halo mass, so in this model disruption has little effect beyond changing the overall normalization of the total GC mass formed. 

Although we do not explore the other effects of this GC disruption model in detail, we find that applying it also does not change our conclusion from Section~\ref{sec:bimodality} that bimodal GC color and metallicity distributions can be produced for a wide range of model parameters. The model preferentially disrupts old GCs, which are bluer than average, so when $\beta_{\Gamma}$ and $\beta_{\eta}$ are held fixed, including disruption tends to increase the mass fraction of GCs that are red. 

\section{Distributions of GC properties}
\label{sec:dists}
\begin{figure*}
\includegraphics[width=\textwidth]{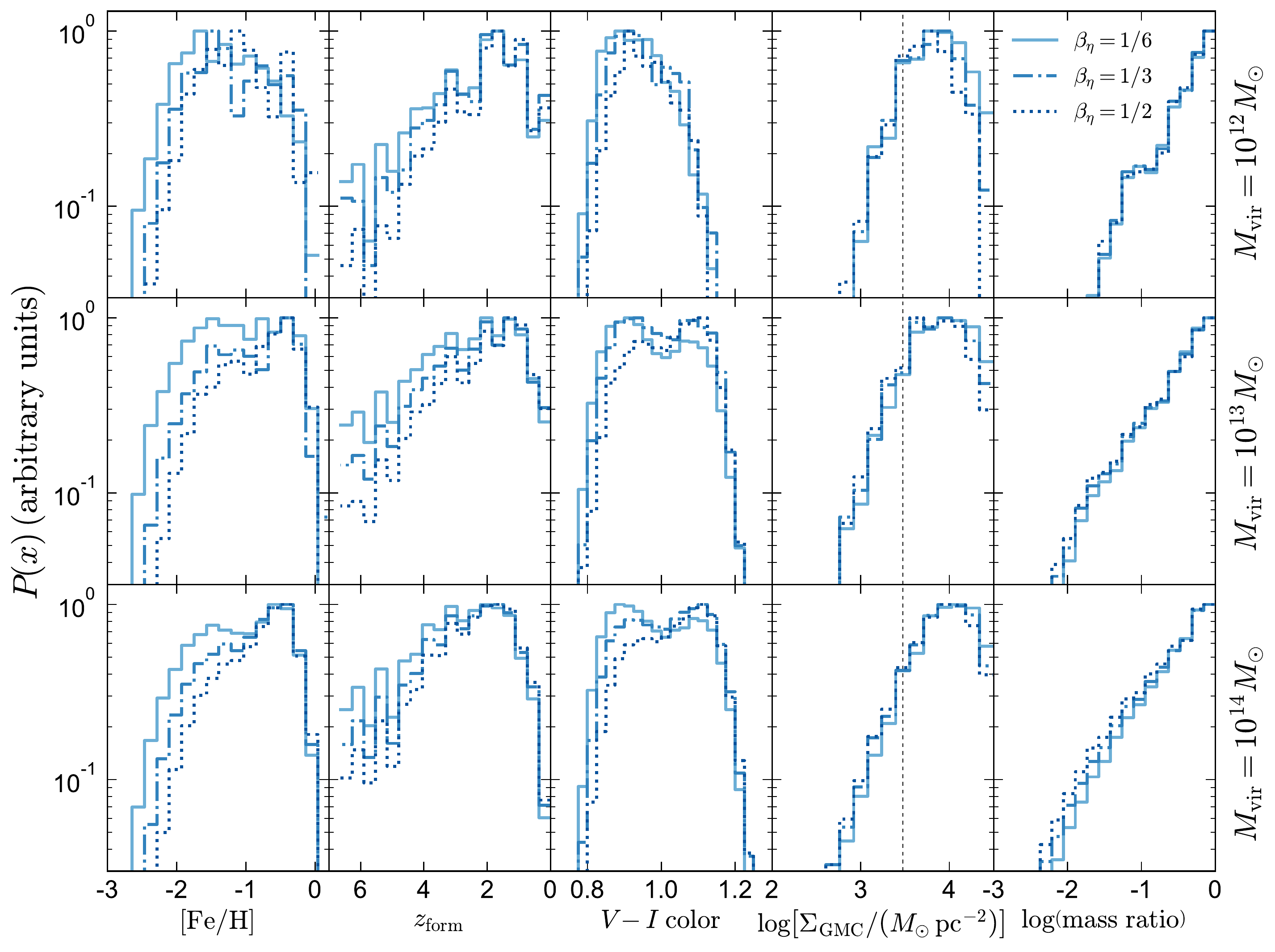}
\caption{Mass-weighted distributions of GC metallicity, formation redshift, color, GMC surface density at the time of formation, and the mass ratio of the merger event in which the GC formed. Different histograms show three values of $\beta_{\eta}$ (Equation~\ref{eq:eta}); different rows show different halo masses at $z=0$. All models assume $\beta_{\Gamma}=1$ (Equation~\ref{eq:Gamma}).}
\label{fig:Sigma_hists}
\end{figure*}
Figure~\ref{fig:Sigma_hists} shows distributions of several GC properties for the GC populations of halos at three different mass scales. Each distribution is an ensemble for the GC populations of 20 merger tree realizations. We show predictions for three different values of $\beta_{\eta}$.

The distributions of most GC properties are similar in halos of different $z=0$ masses. However, the GC color and metallicity distributions evolve with halo mass due to the adopted mass-metallicity relation. At high halo masses, GC color distributions are often more strongly bimodal than GC metallicity distributions. This is a result of the coincidental alignment of GCs in the age-metallicity plane along lines of constant color. Increasing $\beta_{\eta}$ makes GC formation more efficient in higher-mass halos and thus increases the fraction of the GC population that is in the metal-rich, red mode.

Most GCs form in disks with $\Sigma_{\rm GMC} \sim 10^4\,\rm M_{\odot}\,pc^{-2}$, higher than our adopted critical density $\Sigma_{\rm crit} = 3000\,\rm M_{\odot}\,pc^{-2}$. 
The majority of GCs form in major mergers. This is not an explicit requirement of our GC formation model, but is simply a consequence of the fact that the gas accretion rate, SFR, and gas surface density are all highest during major mergers. For this reason, many of the predictions of our model are similar to those of models in which GC formation is explicitly tied to major mergers \citep[e.g.][]{Muratov_2010, Li_2014, Choksi_2018}.

\bsp	
\label{lastpage}
\end{document}